\documentclass[12pt]{article}
\newlength{\pubnumber} 

\catcode`\@=11
\@addtoreset{equation}{section}

\def\section{\@startsection{section}{1}{\z@}{3.5ex plus 1ex minus .2ex}
 {2.3ex plus .2ex}{\large\bf}}
\def\subsection{\@startsection{subsection}{2}{\z@}{2.3ex plus .2ex}
 {2.3ex plus .2ex}{\bf}}

\usepackage{latexsym}
\usepackage{caption}
\usepackage{ragged2e}
\usepackage[justification=RaggedRight]{subfig}
\usepackage{amsmath}

\begin{document}

\begin{titlepage}
\samepage{
\setcounter{page}{0}
\rightline{April 2012}
\rightline{NPAC 11-15}
\vfill
\begin{center}
    {\Large \bf Charged Current Universality and the MSSM}
\vfill
   {\large
      Sky Bauman$^a$\footnote{
     E-mail address:  sbauman@physics.wisc.edu}
     $,$ Jens Erler$^b$\footnote{E-mail address:  erler@fisica.unam.mx}
        $\,$and$\,$ Michael J. Ramsey-Musolf$^{a,c}$\footnote{
     E-mail address:  mjrm@physics.wisc.edu}
    \\}
\vspace{.10in}
 
 {\it $^a$Department of Physics, University of Wisconsin, Madison, WI  53706  USA\\}
 {\it $^b$Departamento de F\'isica Te\'orica, Instituto de F\'isica, \\
              Universidad Nacional Aut\'onoma de M\'exico, 04510 M\'exico D.F., M\'exico\\}
 {\it $^c$California Institute of Technology, Pasadena, CA 91125 USA}
\end{center}
\vfill
\begin{abstract}
We analyze the prospective impact of supersymmetric radiative corrections on tests of charged current universality involving light quarks and leptons. Working within the R-parity conserving Minimal Supersymmetric Standard Model, we compute the corresponding one-loop corrections that enter the extraction of the Cabibbo-Kobayashi-Maskawa matrix element $V_{ud}$ from a comparison of the muon-decay Fermi constant with the vector coupling constant determined from nuclear and neutron $\beta$-decay. We also revisit earlier studies of the corrections to the ratio $R_{e/\mu}$  of pion leptonic decay rates $\Gamma[\pi^+\to e^+\nu (\gamma)]$ and $\Gamma[\pi^+\to \mu^+\nu (\gamma)]$. In both cases, we observe that the magnitude of the corrections can be on the order of $10^{-3}$. We show that a comparison of the  first row CKM unitarity tests with measurements of $R_{e/\mu}$  can provide unique probes of the spectrum of first generation squarks and first and second generation sleptons.
\end{abstract}
\vfill
\smallskip}
\end{titlepage}

\setcounter{footnote}{0}

\def\beq{\begin{equation}}
\def\eeq{\end{equation}}
\def\beqn{\begin{eqnarray}}
\def\eeqn{\end{eqnarray}}
\def\half{{\textstyle{1\over 2}}}
\def\quarter{{\textstyle{1\over 4}}}

\def\calO{{\cal O}}
\def\calE{{\cal E}}
\def\calT{{\cal T}}
\def\calM{{\cal M}}
\def\calF{{\cal F}}
\def\calS{{\cal S}}
\def\calY{{\cal Y}}
\def\calV{{\cal V}}
\def\ibar{{\overline{\imath}}}
\def\chibar{{\overline{\chi}}}
\def\ttwo{{\vartheta_2}}
\def\tthree{{\vartheta_3}}
\def\tfour{{\vartheta_4}}
\def\ttwob{{\overline{\vartheta}_2}}
\def\tthreeb{{\overline{\vartheta}_3}}
\def\tfourb{{\overline{\vartheta}_4}}

\def\qbar{{\overline{q}}}
\def\mm{{\tilde m}}
\def\nn{{\tilde n}}
\def\rep#1{{\bf {#1}}}
\def\ie{{\it i.e.}\/}
\def\eg{{\it e.g.}\/}

\newcommand{\newc}{\newcommand}
\newc{\gsim}{\lower.7ex\hbox{$\;\stackrel{\textstyle>}{\sim}\;$}}
\newc{\lsim}{\lower.7ex\hbox{$\;\stackrel{\textstyle<}{\sim}\;$}}

\hyphenation{su-per-sym-met-ric non-su-per-sym-met-ric}
\hyphenation{space-time-super-sym-met-ric}
\hyphenation{mod-u-lar mod-u-lar--in-var-i-ant}


\def\inbar{\,\vrule height1.5ex width.4pt depth0pt}

\def\IC{\relax\hbox{$\inbar\kern-.3em{\rm C}$}}
\def\IQ{\relax\hbox{$\inbar\kern-.3em{\rm Q}$}}
\def\IR{\relax{\rm I\kern-.18em R}}
 \font\cmss=cmss10 \font\cmsss=cmss10 at 7pt
\def\IZ{\relax\ifmmode\mathchoice
 {\hbox{\cmss Z\kern-.4em Z}}{\hbox{\cmss Z\kern-.4em Z}}
 {\lower.9pt\hbox{\cmsss Z\kern-.4em Z}}
 {\lower1.2pt\hbox{\cmsss Z\kern-.4em Z}}\else{\cmss Z\kern-.4em Z}\fi}

\long\def\@caption#1[#2]#3{\par\addcontentsline{\csname
  ext@#1\endcsname}{#1}{\protect\numberline{\csname
  the#1\endcsname}{\ignorespaces #2}}\begingroup
    \small
    \@parboxrestore
    \@makecaption{\csname fnum@#1\endcsname}{\ignorespaces #3}\par
  \endgroup}
\catcode`@=12

\input epsf

\section{Introduction
\label{intro}
}
\setcounter{footnote}{0}

New physics beyond the Standard Model (BSM) is widely expected to be discovered at the Large Hadron Collider (LHC). If so, a key challenge will be to identify the scenario that best accounts for the collider signatures and to determine the parameters of the corresponding Lagrangian. In this respect, high precision measurements of electroweak precision observables (EWPOs), such as the muon anomalous magnetic moment, may provide crucial input. During the first decade of LHC operations, much of the effort at the \lq\lq intensity frontier" or \lq\lq precision frontier" will involve low-energy studies involving hadronic, nuclear, and atomic systems (for recent reviews, see \eg, Refs.~\cite{mjrm_su,Erler:2004cx}). In this paper, we consider one such class of observables that involve the weak decays of light quarks and leptons.

Historically, such studies played a crucial role in testing and confirming the universality of the Standard Model (SM) charged current (CC) interaction. The comparison of Fermi constants extracted from the muon lifetime and neutron/nuclear $\beta$-decays, respectively, indicated that the underlying universality of CC interactions of leptons and quarks is obscured by the mismatch between quark flavor and mass eigenstates -- leading ultimately to the Cabibbo-Kobayashi-Maskawa (CKM) matrix --  but is otherwise intact. Today, the most stringent tests of lepton-quark universality involve the first-row CKM unitarity relation,
\beq
|V_{ud}|^2 + |V_{us}|^2 + |V_{ub}|^2 = 1~.
\label{unitary}
\eeq

The largest and most precisely known entry in this relation, $V_{ud}$ is obtained from a comparison of the muon decay Fermi constant, $G_\mu$ with the corresponding $\beta$-decay Fermi (or vector coupling) constant $G_V^\beta$ extracted from superallowed $0^+\to 0^+$ nuclear $\beta$-decays~\cite{hardy_towner}. The value of $V_{us}$ is obtained from $K_{e3}$ decay branching ratios~\cite{data_group}. For both the nuclear and kaon decays, extraction of the corresponding CKM matrix element requires theoretical input (see \eg, Refs.~\cite{hardy_towner,data_group,Towner:2010zz,marc_sirlin}). Given the overall resulting uncertainty and the much smaller magnitude of $V_{ub}$, the latter can be ignored in testing Eq.~(\ref{unitary}). A measure of this test is given by the quantity 
\beq
\Delta_\mathrm{CKM}=\left(|V_{ud}|^2 + |V_{us}|^2 + |V_{ub}|^2\right)_\mathrm{exp} - 1~,
\label{eq:Delta}
\eeq
where the \lq\lq exp" subscript indicates the value extracted from experiment with the corresponding theoretical input. Currently,
\beq
\Delta_\mathrm{CKM} = -0.0001\pm 0.0006~,
\label{eq:Deltaexp}
\eeq
with comparable uncertainties coming from $V_{ud}$ and $V_{us}$~\cite{Towner:2010zz}. This agreement with the SM places stringent constraints on a variety of BSM scenarios.

A similarly powerful test of CC universality involves the ratio of pion decay branching ratios
\beq
R_{e/\mu} = \frac{\Gamma[\pi^+\to e^+\nu (\gamma)]}{\Gamma[\pi^+\to \mu^+\nu (\gamma)]}~.
\label{eq:Remudef}
\eeq
The theoretical interpretation of this ratio in terms of BSM physics is remarkably clean, as many hadronic theory uncertainties that affect the individual branching ratios cancel from the ratio. Recent work using chiral perturbation theory puts the overall relative error bar at the $10^{-4}$ level~\cite{Cirigliano:2007ga}, leading to a present error bar dominated by the experimental uncertainty:
\beq
\Delta_{e/\mu}\equiv\frac{\Delta R_{e/\mu} }{R_{e/\mu} }\equiv \frac{R_{e/\mu}^\mathrm{exp}- R_{e/\mu}^\mathrm{SM}}{R_{e/\mu}^\mathrm{SM} } = -0.0034\pm 0.0030\pm 0.0001~,
\label{eq:DeltaRemuexp}
\eeq
where the \lq\lq SM" superscript indicates the theoretical SM prediction~\cite{Cirigliano:2007ga}. The first error is the experimental and the second is the theoretical error in the SM prediction for $R_{e/\mu}^\mathrm{SM}$.

In what follows, we analyze the sensitivity of $\Delta_\mathrm{CKM}$ and $\Delta_{e/\mu}$ to supersymmetric radiative corrections in the R-parity conserving Minimal Supersymmetric Standard Model (MSSM). (For a discussion of the effects of R-parity violation, see {\it e.g.}, Ref.~\cite{RP_mjrm}.) Supersymmetry (SUSY) is one of the most widely considered and strongly motivated BSM scenarios, and the MSSM represents the natural starting point for any study of  SUSY effects on EWPOs. Our focus on CC universality tests is motivated by the prospects of significant improvements in experimental and theoretical precision in both $\Delta_\mathrm{CKM}$ and $\Delta_{e/\mu}$. Experiments presently underway at TRIUMF~\cite{arXiv:1001.3121} and PSI~\cite{865126}  aim to reduce the experimental uncertainty in $\Delta_{e/\mu}$ to the level of $5\times 10^{-4}$, and one hopes that future generation experiments will lead to additional significant reductions. Similarly, new measurements of the neutron decay correlation parameters using the PERC~\cite{PERC} detector may lead to an overall uncertainty in $|V_{ud}|^2$ of a few times $10^{-4}$, while progress in computing the ratio of pseudoscalar decay constants $F_K/F_\pi$ using lattice QCD may yield a similar improvement in the error bar on $|V_{us}|^2$. Since radiative corrections involving weak scale particles generally have the scale $\alpha/\pi\sim 10^{-3}$, it is interesting to analyze the prospective sensitivity of these observables to weak scale SUSY. 

Previous analyses of CC universality in the R-parity conserving MSSM have appeared in Refs.~\cite{kurylov} and~\cite{sean}. The authors of Ref.~\cite{ckm_operator} performed a model-independent analysis of first row CKM unitarity violation in an effective operator framework. The author of Ref.~\cite{right_current} analyzed the effects of right-handed currents in the determinations of $|V_{ub}|$ and $|V_{cb}|$, specifically in the MSSM. At the time Ref.~\cite{kurylov} appeared, there existed a long-standing $\sim 2\sigma$ deviation of $\Delta_\mathrm{CKM}$ from zero. The authors of Ref.~\cite{kurylov} showed -- using a semi-analytical exploration of the MSSM parameter space --  that the sign of the discrepancy was at odds with the implications of conventional models for SUSY-breaking mediation. Subsequently, the re-measurement of kaon decay branching ratios has lead to agreement with CKM unitarity, assuming the values of the kaon form factor $f_K^+(0)$ is taken from lattice QCD computations~\cite{Boyle:2007qe}. Alternately, one may on the ratio of $K_{\ell 2}$ and $\pi_{\ell 2}$ decay widths and lattice QCD computations of the decay constant ratio $F_K/F_\pi$~\cite{kaon}. Thus, it is interesting to revisit the analysis of Ref.~\cite{kurylov}. In doing so, we carry out a more general investigation of the relevant MSSM parameters using a numerical scan that takes into account relevant experimental constraints. At the same time, we consider the behavior of $\Delta_{e/\mu}$ in the same scan, reproducing the results of Ref.~\cite{sean} but uncovering a novel correlation with $\Delta_\mathrm{CKM}$. We show that the correlation of these two EWPOs may provide unique diagnostic of the first and second generation squark and slepton spectrum that may ultimately be compared with the results of LHC searches if the latter discover superpartners.

We summarize our findings here:
\begin{itemize}
\item[(i)] The generic magnitude of the SUSY corrections is of order $10^{-3}$ or smaller.
\item[(ii)] The corrections entering the determination of $V_{ud}$ -- and thus $\Delta_\mathrm{CKM}$ -- are largest for relatively light charginos and either light first generation squarks or second generation sleptons. Current collider bounds on squark masses~\cite{lhc_bound, lepton_photon}, together with the deviation of the muon anomalous magnetic moment from the SM expectation, suggest that the scenario with relatively light charginos and second generation sleptons is most likely. The effects of first generation slepton loops on $V_{ud}$ are suppressed.
\item[(iii)] SUSY loop contributions to $\Delta_{e/\mu}$ are largest in magnitude in the presence of light charginos and a relatively large mass splitting between the first and second generation left-handed sleptons. In this case, the sign of the correction indicates which of the two slepton generations is lightest. 
\item[(iv)] Global constraints on SUSY contributions to $\Delta_\mathrm{CKM}$ and $\Delta_{e/\mu}$ from EW precision data are relatively weak, since the corrections to gauge boson propagators cancel in both cases. Our results illustrate are the more general insensitivity of low-energy CC observables to the $\rho$ parameter that is otherwise strongly constrained by $Z$-pole observables. 
\item[(v)] There exist strong correlations between the SUSY loop corrections to $\Delta_\mathrm{CKM}$ and $\Delta_{e/\mu}$ for various representative slepton and squark spectra for light charginos. These are given in Table~\ref{table1}.
\end{itemize}

\begin{table}[h]
\caption{Correlations between $\Delta_{CKM}$ and $\Delta_{e/\mu}$ (see also Fig.~\ref{fig10c}).}
\begin{tabular}{|l|c|c|}
\hline
Case & $|\Delta_{CKM}|$ & $|\Delta_{e/\mu}|$ \\
\hline
$$ & $$ & $$ \\
light $\tilde{\ell}_2$, heavy $\tilde{\ell}_1$ and $\tilde{q}_1$ & large & large \\
\hline
$$ & $$ & $$ \\
heavy and  nearly degenerate $\tilde{\ell}_1$ and $\tilde{\ell}_2$, light $\tilde{q}_1$ & large & small \\
\hline
$$ & $$ & $$ \\
light $\tilde{\ell}_1$, heavy and nearly degenerate $\tilde{\ell}_2$ and $\tilde{q}_1$ & small & large \\
\hline
$$ & $$ & $$ \\
light and nearly degenerate $\tilde{\ell}_1$, $\tilde{\ell}_2$ and $\tilde{q}_1$ & small & small \\
\hline
$$ & $$ & $$ \\
heavy $\tilde{\ell}_1$, $\tilde{\ell}_2$ and $\tilde{q}_1$ & small & small \\
\hline
\end{tabular}
\label{table1}
\end{table}

Our discussion of the calculations and analysis leading to these findings are organized in the remainder of the paper as follows. In Sect.~\ref{beta_decay}, we briefly review $\beta$-decay and $\pi_{\ell 2}$ decays. In Sect.~\ref{radiate}, we set up our computation and summarize current constraints on the parameters of the MSSM. In Sect.~\ref{scan}, we numerically evaluate the MSSM corrections to $\Delta_\mathrm{CKM}$ and $\Delta_{e/\mu}$ for a large space of MSSM parameters by performing scans over the relevant parameters. We also study the corrections to both quantities as functions of a single mass parameter with all others held fixed in order to derive insight into regions where effects become largest.   In Sect.~\ref{conclude}, we give our conclusions.  For the benefit of readers wishing to carry out their own numerical studies, we provide expressions for the individual loop corrections in the Appendix.

\section{An Overview of $\beta$- and $\pi_{\ell 2}$-Decays}
\label{beta_decay}

The most precise value of $V_{ud}$ is obtained from an analysis of $Q$-values, branching ratios, and corrected half lives or \lq\lq $ft$" values from a series of 13 $0^+\to 0^+$ \lq\lq superallowed"  nuclear decays. For a general nuclear or hadronic decay, the $ft$ values can be expressed in terms of the $\beta$-decay vector ($V$) and axial vector ($A$) coupling constants, $G^{\beta}_{V,A}$ as
\beqn
ft &=& \frac{K}{(G_V^\beta)^2 M_F^2 + (G_A^\beta)^2 M_{GT}^2} \\
K & = & \hbar (2\pi^2 \ln 2) (\hbar c)^6/(m_ec^2)^5~,
\eeqn
where $M_F$ and $M_{GT}$ denote the Fermi and Gamow-Teller transition matrix elements, respectively. For the superallowed decays of interest, $M_F=\sqrt{2}$ and $M_{GT} = 0$, while for neutron decay $M_F=1$ and $M_{GT}=\sqrt{3}$. 

One obtains $V_{ud}$ by expressing it in terms of $G^{\beta}_V$,  the muon decay Fermi constant $G_\mu$, and electroweak radiative corrections to both processes:
\beq
G^{\beta}_V = G_{\mu}V_{ud}[1 + \Delta r^{(V)}_{\beta} - \Delta r_{\mu}]g_V (0)~.
\label{G_V}
\eeq
Here, $\Delta r^{(V)}_{\beta}$ is the correction to tree-level four-fermion semileptonic amplitude for 
 for $\beta$-decay and $\Delta r_{\mu}$ is the corresponding correction for muon decay, while $g_V(0)$ is an appropriate hadronic form factor evaluated at zero momentum transfer. Note that the corrections $\Delta r^{(V)}_{\beta}$ and $\Delta r_{\mu}$ do not include pure QED corrections to the effective four fermion interaction. The latter are conventionally computed separately and combined with the corresponding real photon corrections to the decay rates before extracting the corresponding Fermi constants. This procedure ensures the appropriate cancellation of infrared (IR) divergences. The effects of BSM physics, including contributions from superpartner loops that we consider here, are incorporated in the difference
\beq
\label{eq:betarcsusy}
\left[\Delta r^{(V)}_{\beta} - \Delta r_{\mu}\right]^\mathrm{BSM}~.
\eeq
Since superpartners are all massive and since our assumption of $R$-parity precludes the presence of any massless particles (photons or gluons) in the one-loop SUSY graphs, our calculation introduces no new IR singularities. 

In order to relate the quantity in Eq.~(\ref{eq:betarcsusy}) to $\Delta_\mathrm{CKM}$, we invert Eq.~(\ref{G_V}) to solve for $V_{ud}$ in terms of $\Delta r^{(V)}_{\beta} - \Delta r_{\mu}$. The resulting shift in $\Delta_\mathrm{CKM}$ due to BSM physics is, thus, given by
\beq
\delta \Delta_\mathrm{CKM} = -2|V_{ud}|^2 \left[\Delta r^{(V)}_{\beta} - \Delta r_{\mu}\right]^\mathrm{BSM}~.
\label{eq:shiftCKM}
\eeq
The meaning of this quantity is as follows: if for example $[\Delta r^{(V)}_{\beta} - \Delta r_{\mu}]^\mathrm{BSM}$ were positive (negative), then the value of $V_{ud}$ extracted from $G_V^\beta$ would be decreased (increased) relative to the value obtained using only SM radiative corrections, thereby decreasing (increasing) $\Delta_\mathrm{CKM}$ by twice the magnitude of the BSM correction (due to squaring of $V_{ud}$). Conversely, the lower (upper) end of the range in Eq.~(\ref{eq:Deltaexp}) implies an upper (lower) bound on $[\Delta r^{(V)}_{\beta} - \Delta r_{\mu}]^\mathrm{BSM}$ at a given level of confidence. As we discuss below, the present error on $\Delta_\mathrm{CKM}$ is on the verge of allowing one to infer new constraints on the MSSM parameter space, but further reductions in both experimental and theoretical uncertainties would be needed in order to do so. 

These uncertainties have several sources. Here, we concentrate on those associated with the determination of $V_{ud}$. For the case of neutron-decay, the form factor $g_V$ is given by,
\beq
<f| \bar{u}\gamma_{\lambda}d |i> = \bar{U}_p (P^{\prime})\left[ g_V (q^2) \gamma_{\lambda} + \frac{ig_M (q^2)}{2m_N}\sigma_{\lambda \nu}q^{\nu} \right] U_n (P)~,
\label{vect_form}
\eeq
where $|i>$ is the state of the initial neutron having momentum $P$;   $|f>$ and $P^{\prime}$ refer to the final state proton; and $q = P^{\prime} - P$. The conserved vector current (CVC) property of the SM implies that $g_V(0)=1$. Small corrections due to isospin breaking have been in Ref.~\cite{Kaiser:2001yc} using chiral perturbation theory. The magnitude is less than $10^{-4}$ and can be neglected for present purposes. For the superallowed decays, additional isospin-breaking corrections are incorporated into \lq\lq corrected" $ft$ values:
\beq
\mathcal{F}t= ft\, (1+\delta_R)(1+\delta_C)~,
\eeq
where $\delta_C$ is a nucleus-dependent isospin-breaking correction and $\delta_R$ is an additional nucleus-dependent correction to the $\mathcal{O}(\alpha)$ electroweak radiative corrections. 

The current world average for the thirteen most precisely-known corrected $ft$ values is~\cite{hardy_towner,Towner:2010zz} 
\beq
\overline{\mathcal{F}t} = 3071.87\pm  0.83s~,
\eeq
where the nuclear shell model (NSM) computations of Towner and Hardy (TH)  have been used to evaluate the corrections $\delta_{R,C}$ and where the error bar has been increased to include slight differences with a result obtained using Hartree-Fock methods. As a result, one obtains
\beq
\label{eq:vudvalue}
V_{ud} = 0.97425(14)(19) \qquad \mathrm {(superallowed)}~.
\eeq
The first error is the combined experimental and nuclear theory error, while the second arises from hadronic uncertainties in the SM contribution to $\Delta r^{(V)}_{\beta}$~\cite{marc_sirlin}. The  combined fractional uncertainty of $0.024\%$ is dominated by the hadronic theory error in $\Delta r^{(V)}_{\beta}$ that is common to both the nuclear and neutron decays.  

In contrast to superallowed decays, whose spin-parity quantum numbers select only the vector current transition, the neutron lifetime ($\tau_n$) also depends on the axial vector coupling
\beq
G_A^\beta = G_{\mu}V_{ud}[1 + \Delta r^{(A)}_{\beta} - \Delta r_{\mu}]g_A (0)~,
\eeq
where $\Delta r^{(A)}_\beta$ can in principle differ from $\Delta r^{(V)}_\beta$ due to BSM physics and where $g_A$ is the nucleon axial vector form factor. The value of the latter is not protected from strong interaction renormalization of the underlying quark axial current. At present, it is not feasible to compute $g_A(0)$ from first principles in the SM with the precision needed for probes of new physics. Consequently, an additional neutron decay observable -- having a different relative dependence on $G_V^\beta$ and $G_A^\beta$ than $\tau_n$ -- must be measured in order to extract $G_V^\beta$ with sufficient precision. What currently is the most precise value of $G_V^\beta$ from neutron $\beta$-decay was obtained by measuring the neutron lifetime $\tau_n$ and angular correlations in the decay. (See Ref.~\cite{perkeo_liu}.) The angular correlations relate to the ratio,
\beq
\lambda=\frac{G_A^\beta}{G_V^\beta} \approx \frac{g_A(0)}{g_V(0)} \left(1+\Delta r^{(A)}_{\beta}-\Delta r^{(V)}_{\beta}      \right)~.
\eeq
At present, the value of $V_{ud}$ derived from neutron decay has a larger uncertainty than given in Eq.~(\ref{eq:vudvalue}), owing largely to the experimental uncertainties in $\tau_n$ and $\lambda$. Improvements in the precision of $\lambda$ are expected with measurements of other neutron decay parameters at the Fundamental Neutron Physics Beamline at the Oak Ridge Spallation Neutron Source and with the PERC detector under construction in Vienna and Heidelberg.

A detailed discussion of the determination of $V_{us}$ can be found in Ref.~\cite{data_group}. The result, which we use below, is
\beq
V_{us} = 0.2252\pm 0.0009~.
\eeq
Combining the latter with Eq.~(\ref{eq:vudvalue}) leads to the result for $\Delta_\mathrm{CKM}$ quoted in Eq.~(\ref{eq:Deltaexp}). 

We turn now to pion leptonic decays. Theoretically, we will denote the corrections to the $\pi_{\ell 2}$ decay widths as $\Delta r^{(A)}_{\pi}(\ell)-\Delta r_{\mu} $ with
\beqn
\Gamma[\pi^+\to\ell^+ \nu_\ell (\gamma)] & = & \frac{G_\mu^2 |V_{ud}|^2}{4\pi} F_\pi^2 m_\pi m_\ell^2 \left[1-\frac{m_\ell^2}{m_\pi^2}\right]^2 \nonumber \\
&& \left\{
1+2\left[\Delta r^{(A)}_{\pi}(\ell)-\Delta r_{\mu}\right]+\ \mathrm{brem}\right\}~,
\label{eq:pidecay}
\eeqn
where $F_\pi=92.4$ MeV is the pion decay constant and where the \lq\lq $(\gamma)$" and \lq\lq $+\ \mathrm{brem}$" indicate the inclusion of real radiation as needed to cancel infrared divergences in the Standard Model contributions to $\Delta r^{(A)}_{\pi}(\ell)$. The subscript \lq\lq $A$" appears since only matrix elements of the hadronic axial vector current contribute to pion decays, in contrast to the determination of $V_{ud}$ for which we are interested in the hadronic vector current. The ratio $R_{e/\mu}$ is then insensitive to any quantities that are lepton-species independent, such as $F_\pi$, $V_{ud}$, and $\Delta r_{\mu}$. The resulting dependence of $R_{e/\mu}$ and $\Delta_{e/\mu}$ on BSM physics is then encoded in the difference
\beq
\label{eq:pircsusy}
2\left[\Delta r^{(A)}_{\pi}(e)-\Delta r^{(A)}_{\pi}(\mu)\right]^\mathrm{BSM}~.
\eeq
As in the case of Eq.~(\ref{eq:betarcsusy}), the superpartner loop contributions to the difference (\ref{eq:pircsusy}) are free from IR divergences.

The result for $\Delta_{e/\mu}$ given in Eq.~(\ref{eq:DeltaRemuexp}) has been obtained from a comparison of the average of separate measurements of $R_{e/\mu}$ carried out at TRIUMF~\cite{TRI-PP-92-15} and PSI~\cite{355366} and with theoretical SM prediction~\cite{Cirigliano:2007ga}. 
Two new measurements are underway at these laboratories~\cite{arXiv:1001.3121,865126} that plan for an experimental error of $\sim 0.0005$, comparable to the previous and longstanding value for the theoretical uncertainty in the SM prediction. The smaller theoretical error quoted above is given in a recent two-loop chiral perturbation theory computation. Part of the reduction in the theory error results from matching the low-energy constants, or counterterms, to low-energy QCD in the large $N_C$ limit. Assuming both experiments achieve their planned precision, the resulting uncertainty in $\Delta_{e/\mu}$ will be comparable in magnitude to, but slightly smaller than, the error in $\Delta_\mathrm{CKM}$. 

Looking further to the future, we observe that further reductions in the uncertainties in both $\Delta_\mathrm{CKM}$ and $\Delta_{e/\mu}$ would be needed if these observables are to probe significant portions of the MSSM parameter space. In the case of $\Delta_{e/\mu}$, the challenge will be entirely experimental as the present theory error is smaller than the expected magnitude of SUSY corrections, particularly in the regime of light superpartners. For $\Delta_\mathrm{CKM}$, one would require progress from a combination of experiment and hadronic physics theory. In what follows, we use the magnitude of the SUSY corrections to set a benchmark for these future improvements.

\section{MSSM Radiative Corrections
\label{radiate}}

One-loop corrections come in the form of vertex, propagator and box diagrams for both muon decay ($\Delta r_{\mu}$) and light quark-decay ($\Delta r^{(V)}_{\beta}$, $\Delta r^{(A)}_{\pi}(\ell)$).  Graphs for the supersymmetric contributions are shown in Figs.~\ref{fig13}--\ref{fig17} of Appendix~\ref{individual}. (Note that we do not show corrections to the $W$-boson propagators, since these cancel from the differences in Eqs.~(\ref{eq:betarcsusy}) and~(\ref{eq:pircsusy}).) Explicit results for individual graphs are given in Appendix~\ref{individual}. Here, we outline the general framework for the computation and comment on some general characteristics. 

In the R-parity conserving MSSM, all of the internal lines involve superpartners. As noted above, the masses of the latter are much greater than the scale of external momenta in the decay process,  and we encounter no infrared (IR) divergences whose effect would have to be compensated by inclusion of real radiation. This situation contrasts with that of the SM corrections, where the presence of internal photon and charged lepton lines lead to soft and collinear IR divergences. As a result, our computation can be simplified by neglecting external masses and momenta in the loop integrals. 

The SUSY contributions do, however, lead to ultraviolet (UV) divergences in the vertex and external leg corrections. We regularize these UV divergences using dimensional reduction, working in $d=4-2\epsilon$ spacetime dimensions for the momenta and $d=4$  dimensions for the Clifford algebra. Use of the latter is needed to preserve supersymmetry at one-loop order. Renormalization is carried out by subtracting all terms proportional to $1/\epsilon-\gamma+\ln 4\pi$ with appropriate counterterms, a procedure known is \lq\lq $\overline{\mathrm{DR}}$ renormalization" \footnote{It is common to denote all $\overline{\mathrm{DR}}$-renormalized quantities with a hat, {\em viz}, $\widehat\Delta r_{\mu}$. We will not do so here, however, to avoid cumbersome notation.}. All divergences cancel in the differences $\Delta r^{(V)}_{\beta}-\Delta r_{\mu}$ and $\Delta r^{(A)}_{\pi}(e)-\Delta r^{(A)}_{\pi}(\mu)$ (see Appendix~\ref{cancel}); the one-loop corrections are finite and insensitive to the UV regulator.

In addition to the cancellation of $W$-boson propagator corrections from the differences (\ref{eq:betarcsusy}) and (\ref{eq:pircsusy}), 
there exist additional cancellations that simplify the analysis of the SUSY corrections on the underlying MSSM parameters. In the case of $\Delta r^{(V)}_{\beta}$  and $\Delta r_{\mu}$, the residual finite corrections to the $e\nu W$ vertex and electron propagators are identical, so these corrections cancel from the difference (\ref{eq:betarcsusy}) . Specifically, denoting the external leg, vertex, and box corrections as $\Delta_{\rm leg}$, $\Delta_{\rm vertex}$, $\Delta_{\rm box}$ respectively  and writing
\beq
\Delta r^{(V)}_{\beta} - \Delta r_{\mu} = \Delta_{\rm leg} + \Delta_{\rm vertex} + \Delta_{\rm box}~,
\label{Delta_contr}
\eeq
this cancellation implies that
\beq
\Delta_{\rm leg} \equiv \Delta_{\beta-{\rm leg}} - \Delta_{\mu-{\rm leg}}~,
\label{Delta_leg}
\eeq
is independent of any corrections to the electron propagator, while all corrections to the $e\nu W$ vertex cancel from
\beq
\Delta_{\rm vertex} \equiv \Delta_{\beta-{\rm vertex}} -  \Delta_{\mu-{\rm vertex}}~.
\label{Delta_vertex}
\eeq

The remaining first generation slepton mass-dependence enters via the box graphs, whose contributions are numerically suppressed~\cite{kurylov}. Thus, we expect the SUSY corrections to $\Delta_\mathrm{CKM}$ to be largely independent of first generation slepton masses -- an expectation that we confirm numerically below. Similarly, corrections to the $udW$ vertex and external light quark fermion propagators cancel from (\ref{eq:pircsusy}), thereby desensitizing $\Delta_{e/\mu}$ to first generation squark masses. Thus, we expect 
\begin{itemize}
\item[(i)] $\Delta r^{(V)}_{\beta}-\Delta r_{\mu}$ will be most sensitive to details of the electroweak gaugino, first generation squark, and second generation slepton spectrum, becoming largest when the latter classes of sfermions are non-degenerate with one set being relatively light,
\item[(ii)] $\Delta r^{(A)}_{\pi}(e)-\Delta r^{(A)}_{\pi}(\mu)$ will be most sensitive to spectra of electroweak gauginos plus those of the first and second generation sleptons, becoming largest when the latter are non-generate with, again, one set being relatively light. 
\end{itemize}
In short, $\Delta_\mathrm{CKM}$ and $\Delta_{e/\mu}$ probe, respectively, slepton-squark and slepton universality in the MSSM. 

Given the cancellation of the $udW$ vertex and external leg corrections from (\ref{eq:pircsusy}), $\Delta_{e/\mu}$ carries no dependence on the gluino mass at $\mathcal{O}(\alpha_s)$. In principle, (\ref{eq:betarcsusy}) will display some gluino mass sensitivity since the hadronic CC vertex and external leg corrections do not cancel. In practice, this dependence is typically negligible, due to conservation of the vector current, or CVC. In the case of SM corrections, CVC implies that in the limit of exact isospin symmetry, the hadronic vector charged current receives no strong interaction renormalization. Tiny corrections associated with the light quark mass differences and electromagnetic effects that break isospin symmetry may arise. The analog for the SUSY corrections is that for degenerate up- and down-squarks, the gluino loop contributions will cancel from the sum of vertex and external leg corrections, a consequence of \lq\lq super CVC"~\cite{kurylov}.  This degeneracy will be broken by the difference in up- and down-quark mass contributions to the squark masses as well as by any mixing between left- and right-handed squarks. Since the latter is typically taken to be proportional to quark Yukawa couplings, the breakdown of super CVC will lead to a negligible sensitivity to the gluino mass.

\subsection{Present Constraints
\label{constrain}}

Collider searches for superpartners have placed lower bounds on many of the masses relevant for radiative corrections computed here. Prior to the operation of the LHC, results from superpartner searches at LEP and the Tevatron yielded the following lower bounds~\cite{data_group}:

\beqn
&& m_{\tilde{\chi}^{0}_{1}} > 46~{\rm GeV},~m_{\tilde{\chi}^{0}_{2}} > 62.4~{\rm GeV},~m_{\tilde{\chi}^{0}_{3}} > 99.9~{\rm GeV},~m_{\tilde{\chi}^{0}_{4}} > 116~{\rm GeV}~,\nonumber \\
&& m_{\tilde{\chi}_{1}}^{\pm},~m_{\tilde{\chi}_{2}}^{\pm} > 94~{\rm GeV}~, \nonumber \\
&& m_{\tilde{e}} > 107~{\rm GeV},~m_{\tilde{\mu}} > 94~{\rm GeV},~m_{\tilde{\tau}} > 81.9~{\rm GeV}~, \nonumber \\
&& m_{\tilde{q}} > 379~{\rm GeV},~m_{\tilde{b}} > 89~{\rm GeV},~m_{\tilde{t}} > 95.7~{\rm GeV},~m_{\tilde{g}} > 308~{\rm GeV}~.
\label{mass_constr}
\eeqn

In the case of first and second generation squarks and gluinos, more stringent lower bounds have recently been reported by the ATLAS and CMS collaborations. Namely, squark masses are constrained to be above roughly $1000$ GeV~\cite{lhc_bound}. The exact limits on the squark masses of course depend on assumptions about the MSSM, and these are summarized in Ref.~\cite{lepton_photon}. However, for the purposes of this paper, we will simply take $m_{\tilde{q}} > 1000~{\rm GeV}$ when illustrating the general implications of present LHC search results.

In general, electroweak precision observables (EWPOs) imply additional indirect constraints on the MSSM parameter space due to superpartner contributions at the one-loop level.\footnote{Here, we distinguish \lq\lq EWPO" as referring to these indirect constraints, and \lq\lq LEP" as referring to the direct search bounds.} To the extent that loop corrections to EWPOs are dominated by contributions to the electroweak gauge boson propagators, one may derive constraints using the oblique parameters. At present, a global analysis of EWPO leads to the allowed ranges 
\beq
S = -0.13 \pm 0.10,~T = -0.13 \pm 0.11,~U = 0.20 \pm 0.12 ~,
\label{STU_constr}
\eeq
where correlations between the errors are described below. The corresponding  $\chi^2$-fit function is
\beq
\chi^2 = (\sigma^2)^{-1}_{ij}(S_i - \bar{S}_i)(S_j - \bar{S}_j)~,
\label{chi_sqr}
\eeq
where the indices $i$ and $j$ are summed from $1$ to $3$, with $i=1,2,3$ corresponding, respectively, to $S$, $T$, and $U$. 
The quantities $\bar{S}_i$ are the corresponding central values and are listed in Eq.~(\ref{STU_constr}), while
the matrix $\sigma^2$ is given as
\beq
(\sigma^2)_{ij} = \rho_{ij} \sigma_i \sigma_j~,
\label{sigma_sqr}
\eeq
where the $\sigma_i$ are the errors in the oblique parameters appearing in Eq.~(\ref{STU_constr}) and the correlation  matrix $\rho$ is
\begin{equation*}
\rho = \left(
\begin{array}{ccc}
1 & 0.866 & -0.588 \\
0.866 & 1 & -0.392 \\
-0.588 & -0.392 & 1
\end{array} \right).
\label{rho}
\end{equation*}

Setting $\chi^2 < 7.815$ in Eq.(\ref{chi_sqr}) defines a 95\% confidence level allowed region for a three parameter fit.\footnote{We note that in many theories of new physics, the oblique parameter $U$ automatically is very small (see, {\it e.g.}, Ref.~\cite{data_group}). Therefore, one might presume that it would be appropriate to perform a $\chi^2$-fit for two oblique parameters instead of three. However, it turns out that the oblique parameter $U$ is {\it not} small for the scans over MSSM parameters performed in this paper. This is demonstrated in Figs.~\ref{fig7} and~\ref{fig11}.}

For each choice of the MSSM parameters that we use to compute the corrections to the low-energy CC observables, we also compute the contributions to the oblique parameters. We discard any parameter set that falls outside the allowed region for the latter. Of course, a more complete treatment of present EWPO constraints would require computing non-oblique corrections to each observable, and performing a new global fit, and retaining only those parameters that satisfy an appropriate criterion for a goodness of the fit (see, {\em e.g.}, Ref.~\cite{Erler:1998ur}). In the present case, we find that implementation of the oblique parameter constraints does not lead to a significant restriction of the MSSM parameter space that is most relevant to the low-energy observables. Consequently, we anticipate that a more comprehensive EWPO analysis is unlikely to yield significant additional constraints. Nonetheless, our expectations should be checked by a more comprehensive, model-independent EWPO analysis -- a task that goes beyond the scope of the present study. 

\section{Parameter Scans and Numerical Analysis
\label{scan}}

In order to evaluate the possible magnitude of the corrections~(\ref{eq:betarcsusy}) and~(\ref{eq:pircsusy}), we have scanned over the relevant MSSM parameters, taking into account the aforementioned constraints and then computing the resultant CC radiative corrections. The relevant parameters include:
gaugino masses $M_1$, $M_2$, $M_3$; the supersymmetric Higgs-Higgsino mass parameter $\mu$; the ratio of up- and down-type Higgs vacuum expectation values, $\tan \beta$; and the slepton and squark mass matrices  $m^{2}_{L}$, $m^{2}_{R}$, $m^{2}_{Q}$, $m^{2}_{U}$ and $m^{2}_{D}$. (We take the triscalar couplings $A_{\ell}$, $A_u$ and $A_d$ to vanish.) To avoid unacceptably large flavor changing neutral currents, we have taken the mass matrices to be flavor diagonal. In what follows, we discuss results of two types of variations of MSSM parameters. In the first type, we vary a single MSSM parameter while keeping the other parameters fixed. In the second type, a random number generator is used to select parameters distributed uniformly over  ranges. Parameter ranges are listed in Table~\ref{table2}. In all figures, left-right sfermion mass mixings vanish.

\begin{figure}[htb]
   \epsfxsize 3.00 truein \epsfbox {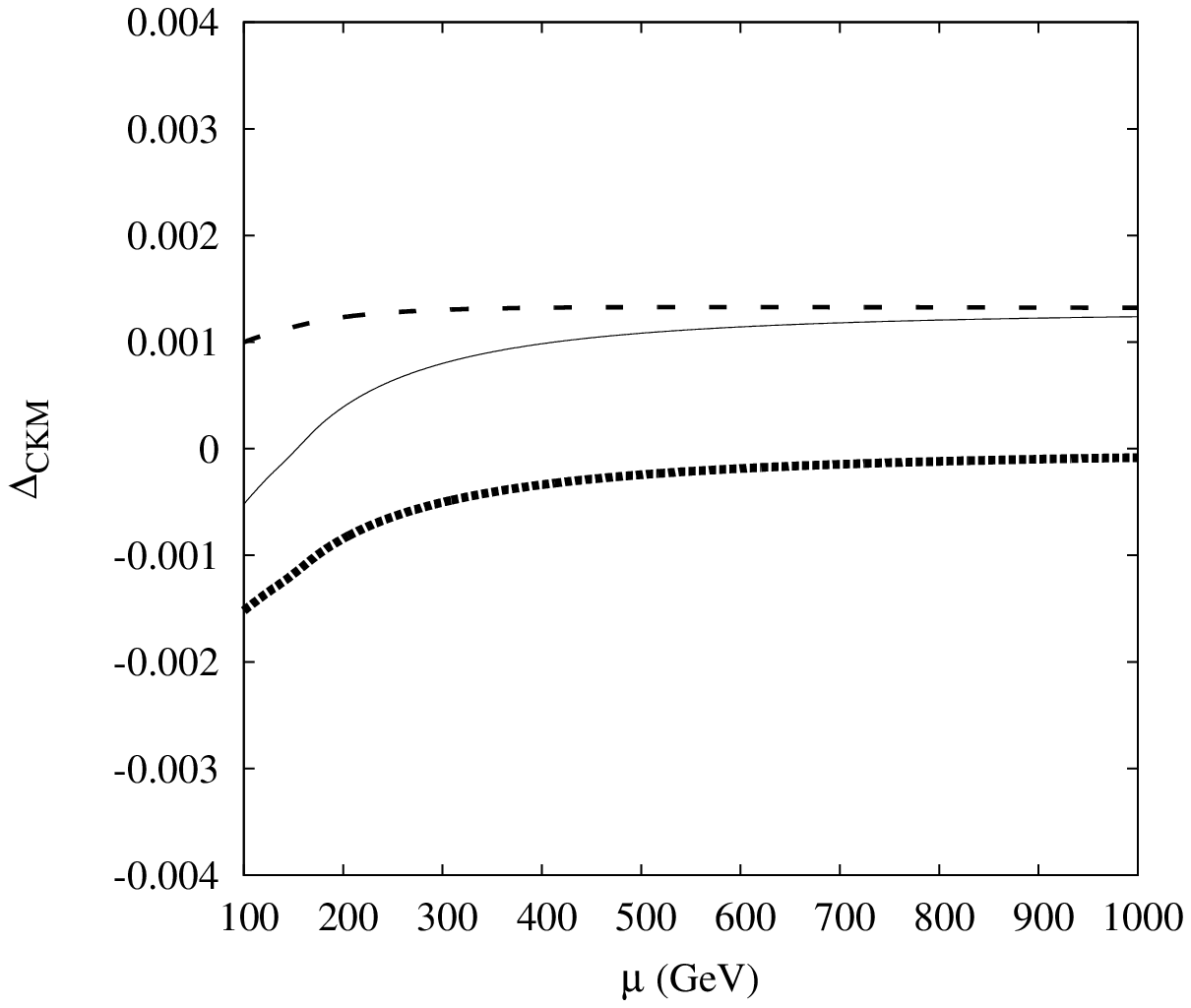}
\caption{$\Delta_{CKM}$ vs. $\mu$. The total correction is given by the light solid line. The vertex and external leg contribution is given by the heavy line. The box graph contribution is given by the dashed line.}
\label{fig1}
\end{figure}

\begin{figure}[htb]
\subfloat[$\Delta_{CKM}$ and the contributions to this quantity. The total correction is given by the light solid line. The vertex and external leg contribution is given by the heavy line. The box graph contribution is given by the dashed line.]{
   \epsfxsize 2.85 truein \epsfbox {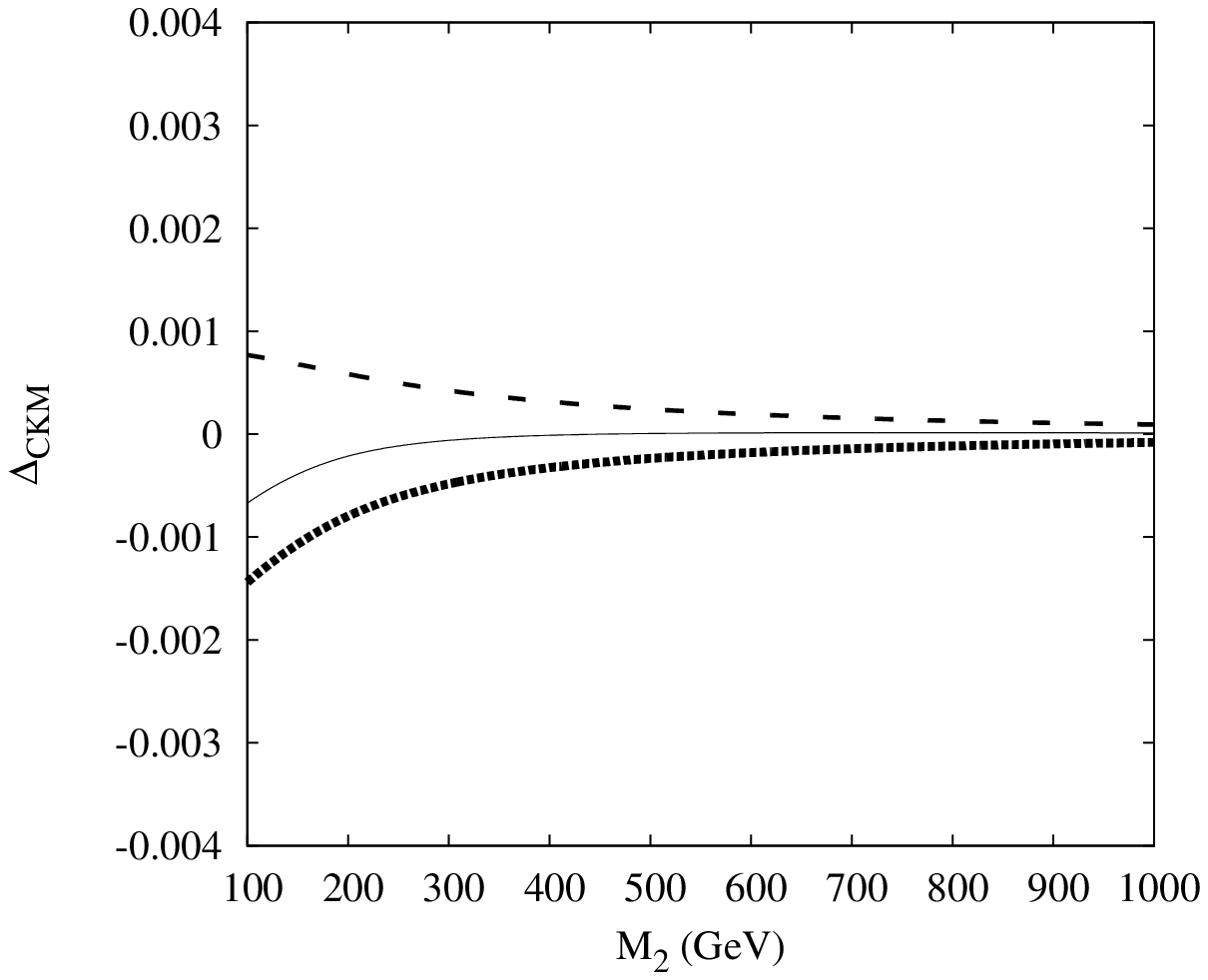}
\label{fig2a}}
\subfloat[The ratio of the box graph correction to the vertex and external leg correction.]{
   \epsfxsize 2.85 truein \epsfbox {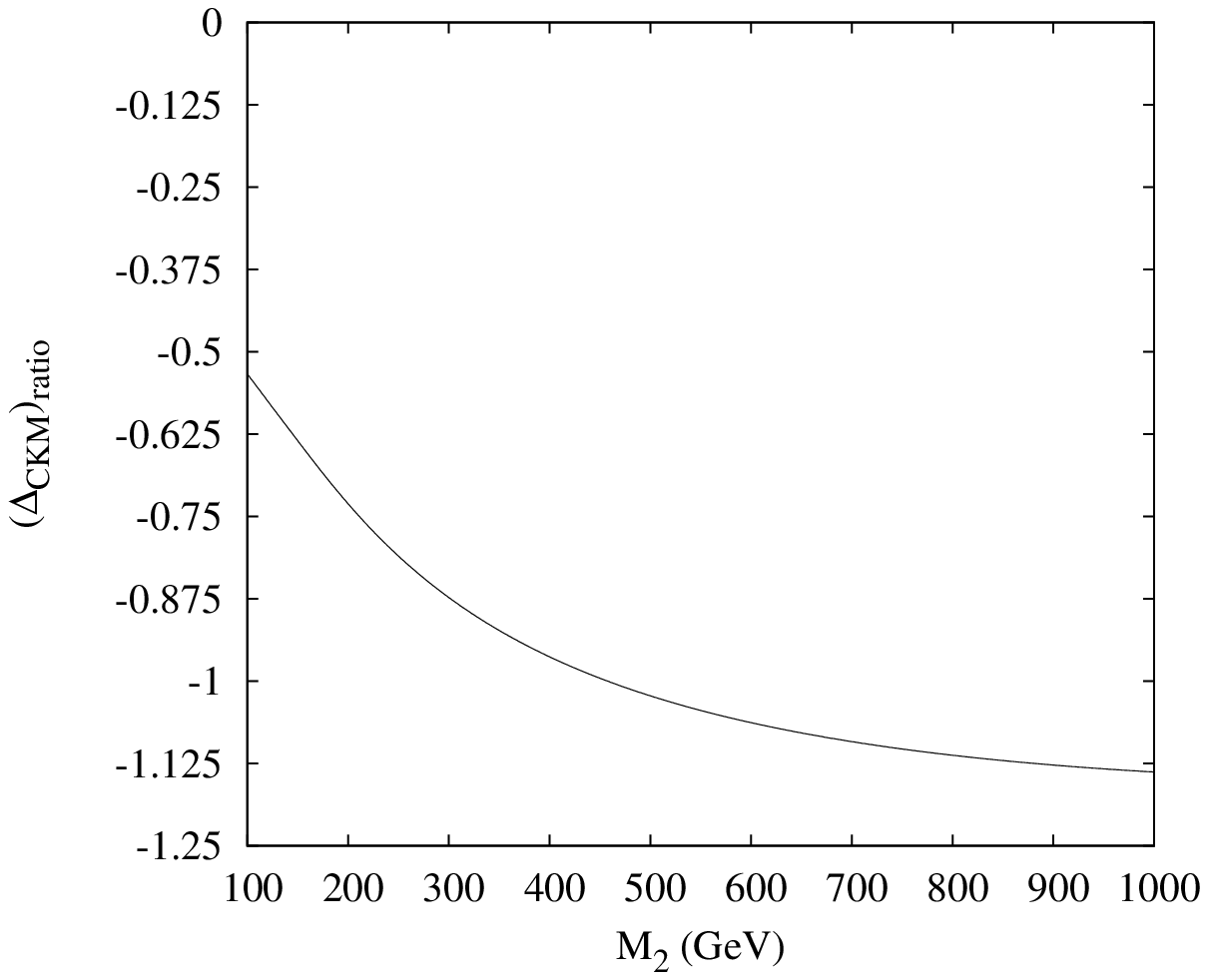}
\label{fig2b}}
\newline
\subfloat[The ratio of the box graph correction to the vertex and external leg correction.]{
   \epsfxsize 2.85 truein \epsfbox {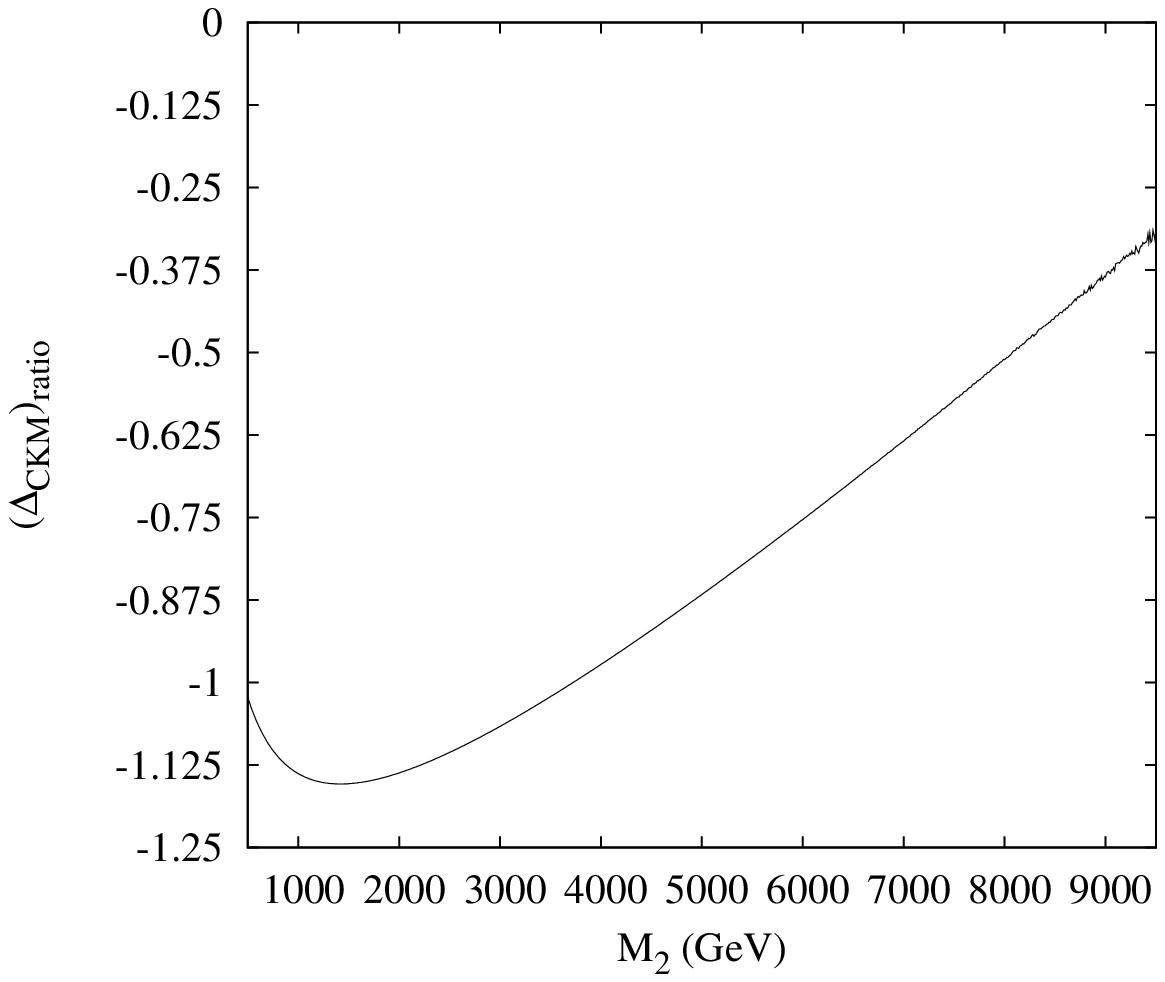}
\label{fig2c}}
\caption{Corrections as functions of $M_2$.}
\label{fig2}
\end{figure}

\begin{figure}[htb]
   \epsfxsize 3.00 truein \epsfbox {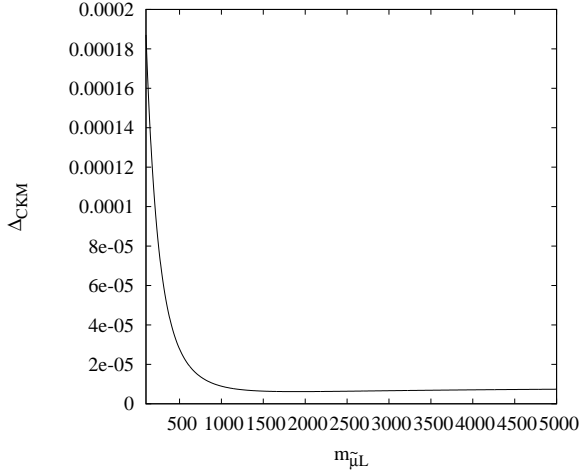}
\caption{$\Delta_{CKM}$ as a function of $m_{\tilde{\mu}L}$.}
\label{fig3}
\end{figure}

\begin{figure}[htb]
   \epsfxsize 3.00 truein \epsfbox {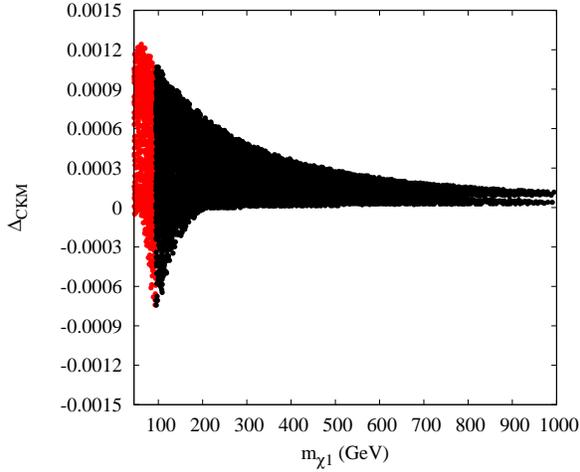}
\caption{$\Delta_{CKM}$ vs. $m_{\chi 1}$. Points inside electroweak bounds are black. Points outside electroweak bounds are red.}
\label{fig4}
\end{figure}

\begin{figure}[htbp]
\subfloat[$\Delta_{CKM}$ vs. $M_1$.]{
   \epsfxsize 2.85 truein \epsfbox {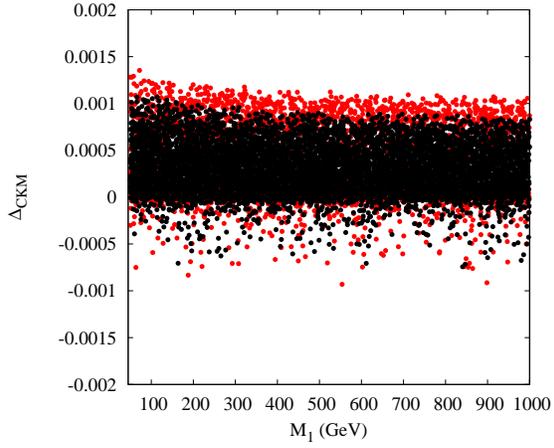}
\label{fig5a}}
\subfloat[$\Delta_{CKM}$ vs. $M_2$.]{
  \epsfxsize 2.85 truein \epsfbox {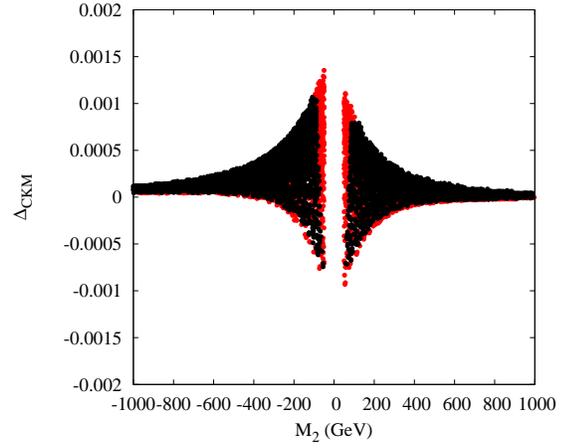}
\label{fig5b}}
\newline
\subfloat[$\Delta_{CKM}$ vs. $\mu$.]{
   \epsfxsize 2.85 truein \epsfbox {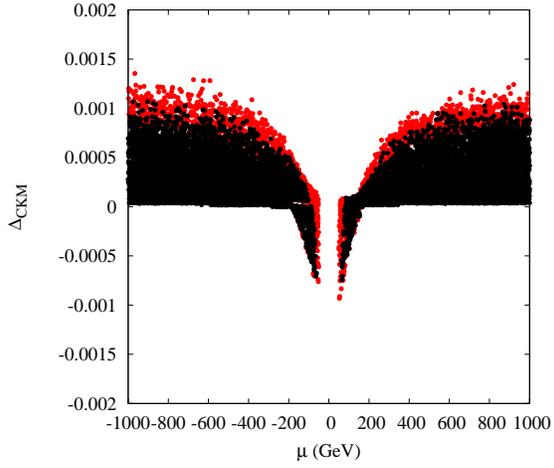}
\label{fig5c}}
\caption{Corrections to $\Delta_{CKM}$, as functions of $M_1$, $M_2$ and $\mu$.}
\label{fig5}
\end{figure}

\begin{figure}[htbp]
\subfloat[The Vertex and External Leg Contribution to $\Delta_{CKM}$ vs. $\mu$.]{
   \epsfxsize 2.85 truein \epsfbox {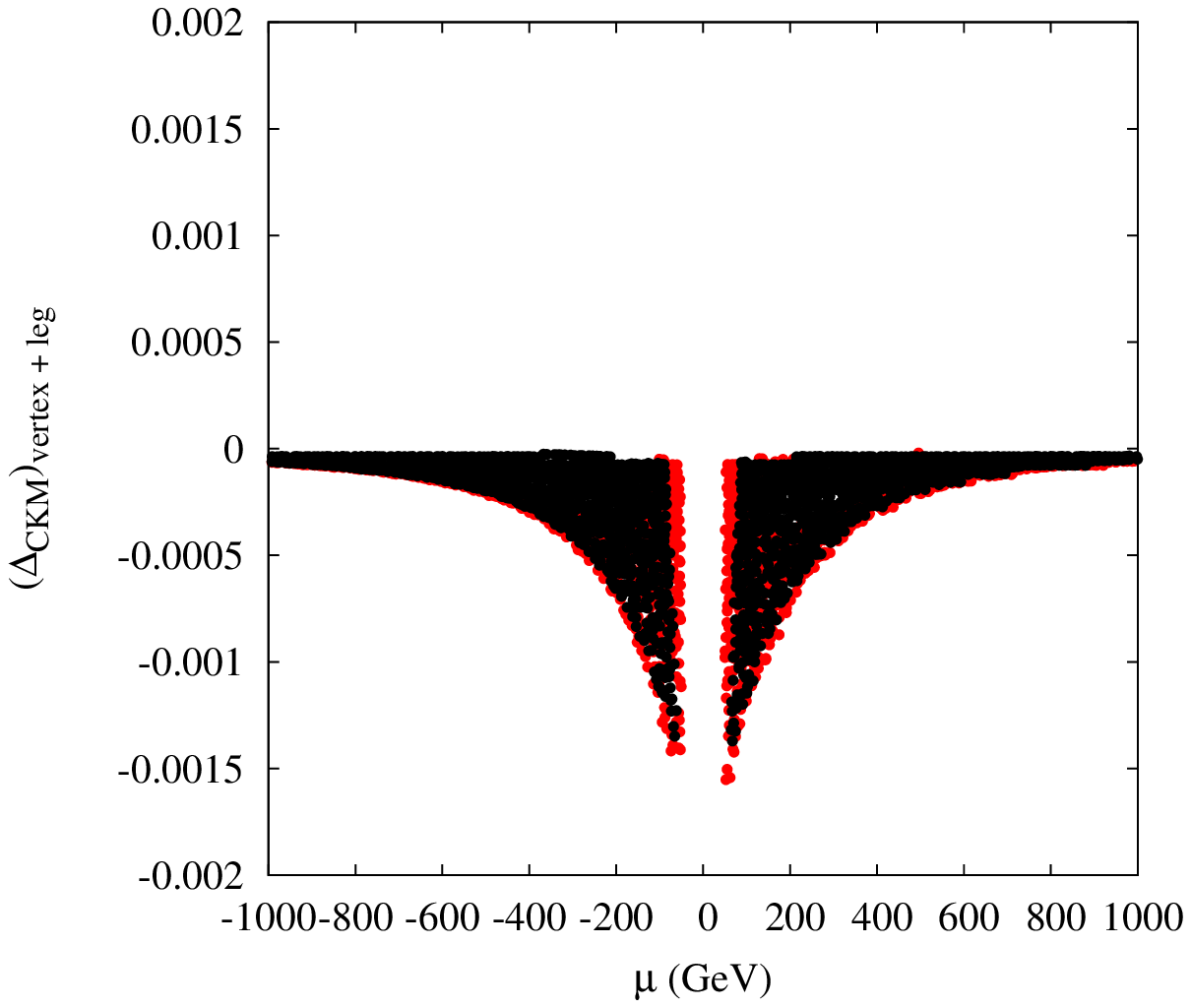}
\label{fig6a}}
\subfloat[The Box Graph Contribution to $\Delta_{CKM}$ vs. $\mu$.]{
  \epsfxsize 2.85 truein \epsfbox {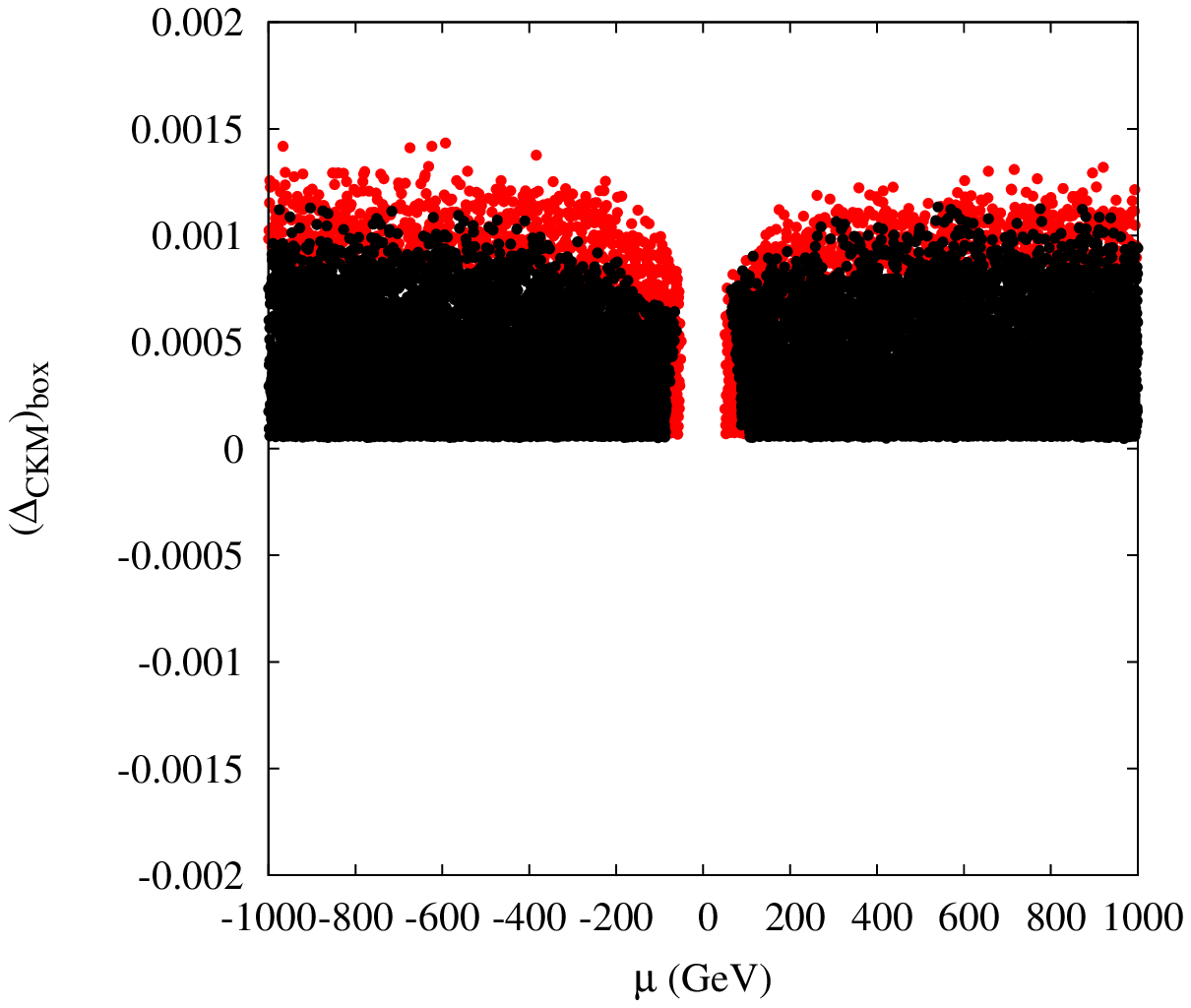}
\label{fig6b}}
\caption{Contributions to $\Delta_{CKM}$ as functions of $\mu$.}
\label{fig6}
\end{figure}

\begin{figure}[htbp]
\subfloat[$U$ vs. $S$.]{
   \epsfxsize 2.85 truein \epsfbox {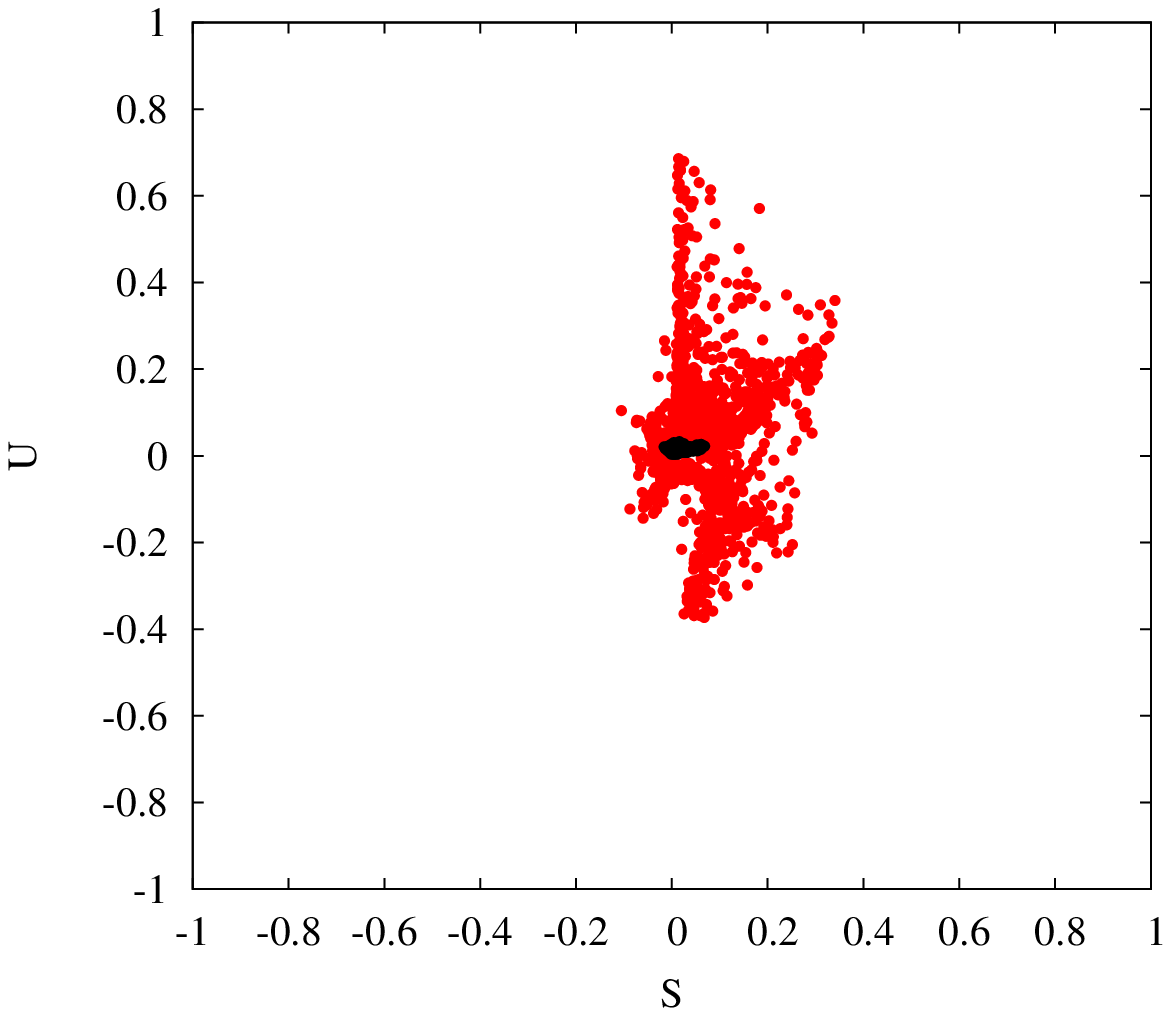}
\label{fig7a}}
\subfloat[$U$ vs. $T$.]{
  \epsfxsize 2.85 truein \epsfbox {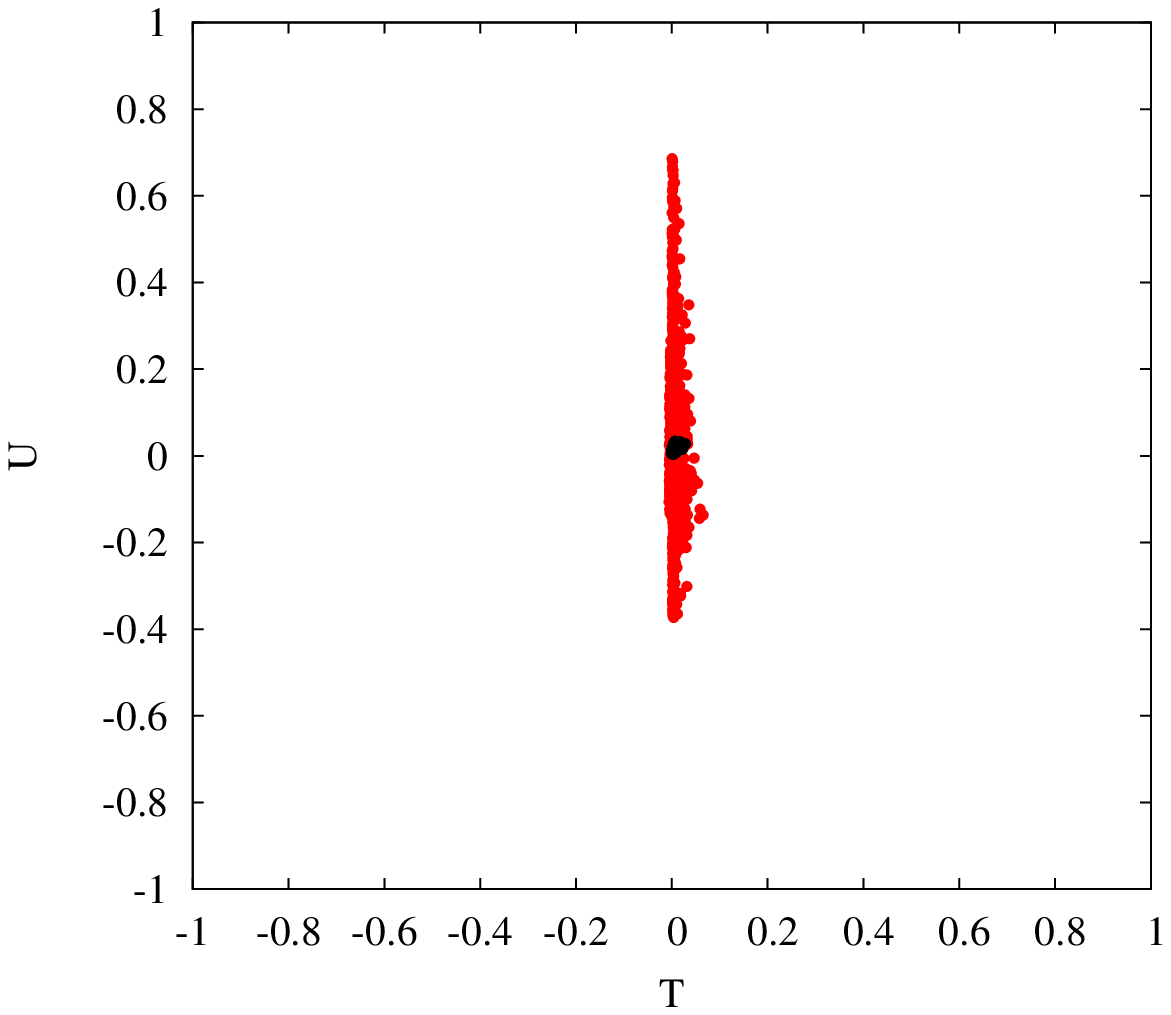}
\label{fig7b}}
\newline
\subfloat[$T$ vs. $S$.]{
  \epsfxsize 2.85 truein \epsfbox {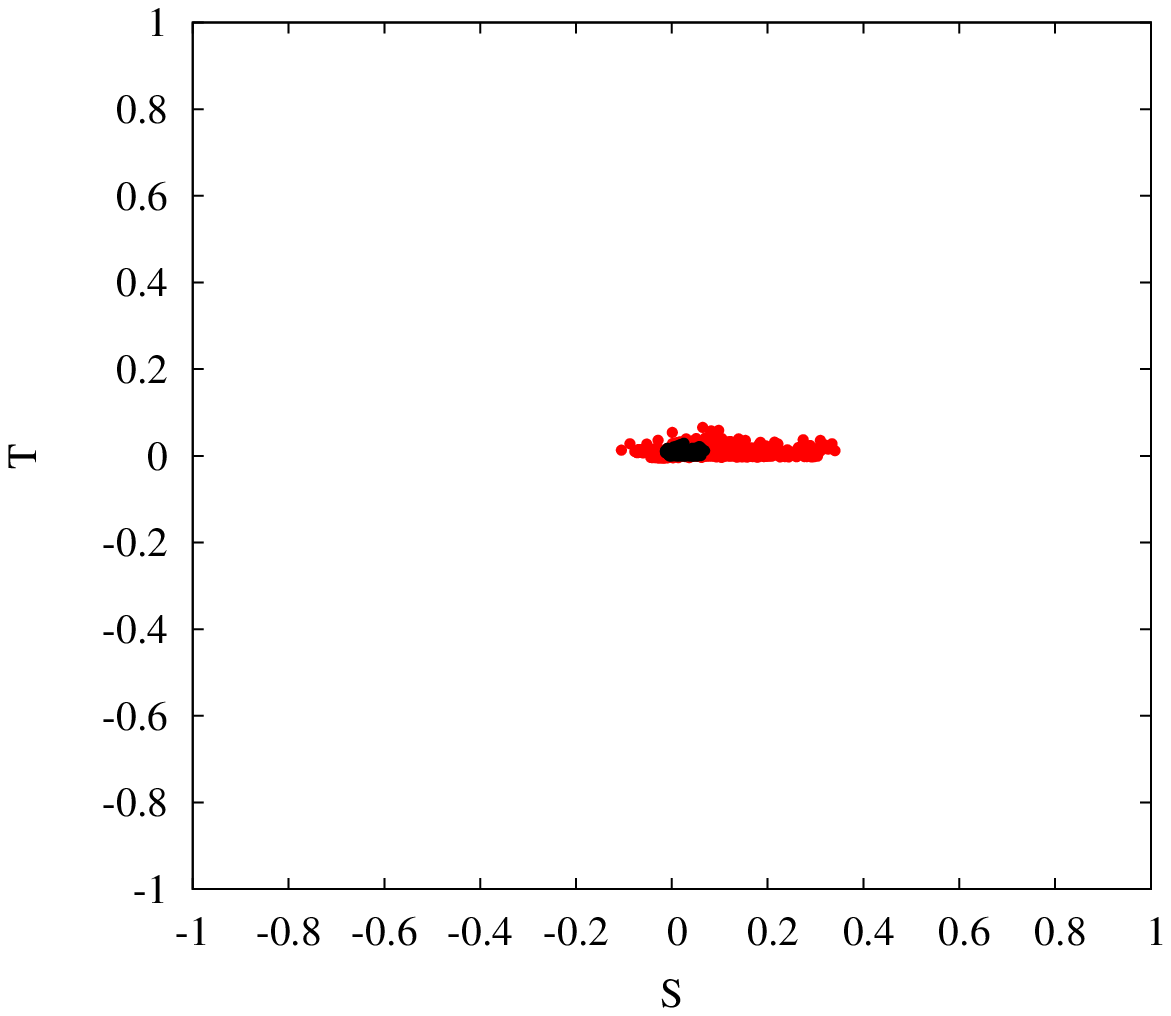}
\label{fig7c}}
\caption{Scatter plots of the oblique parameters corresponding to Figs.~\ref{fig4}, \ref{fig5} and~\ref{fig6}.}
\label{fig7}
\end{figure}

\begin{figure}[htb]
\subfloat[$\Delta_{e/\mu}$ vs. $\mu$.]{
   \epsfxsize 2.85 truein \epsfbox {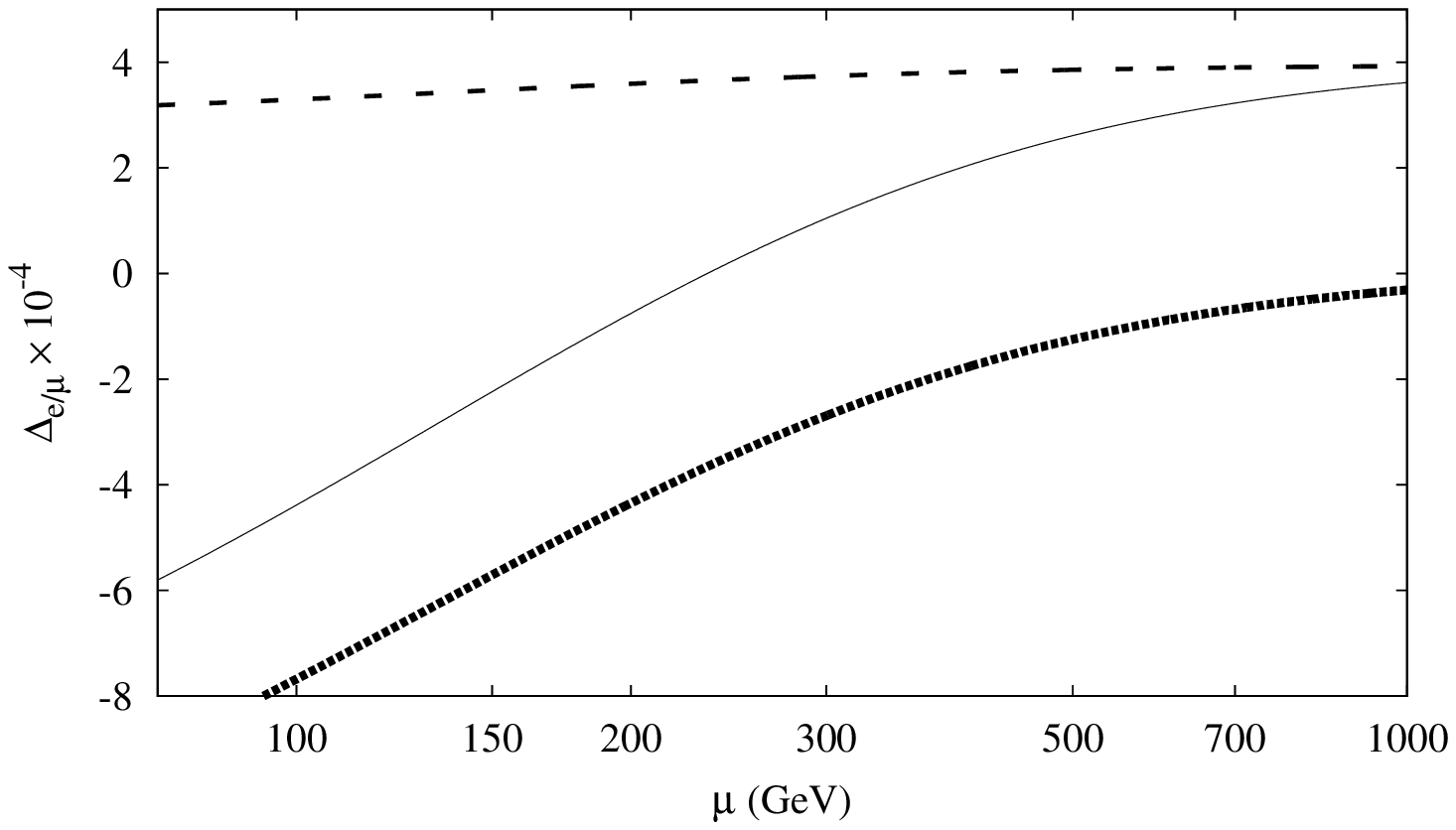}
\label{fig8a}}
\subfloat[$\Delta_{e/\mu}$ vs. $M_2$.]{
   \epsfxsize 2.85 truein \epsfbox {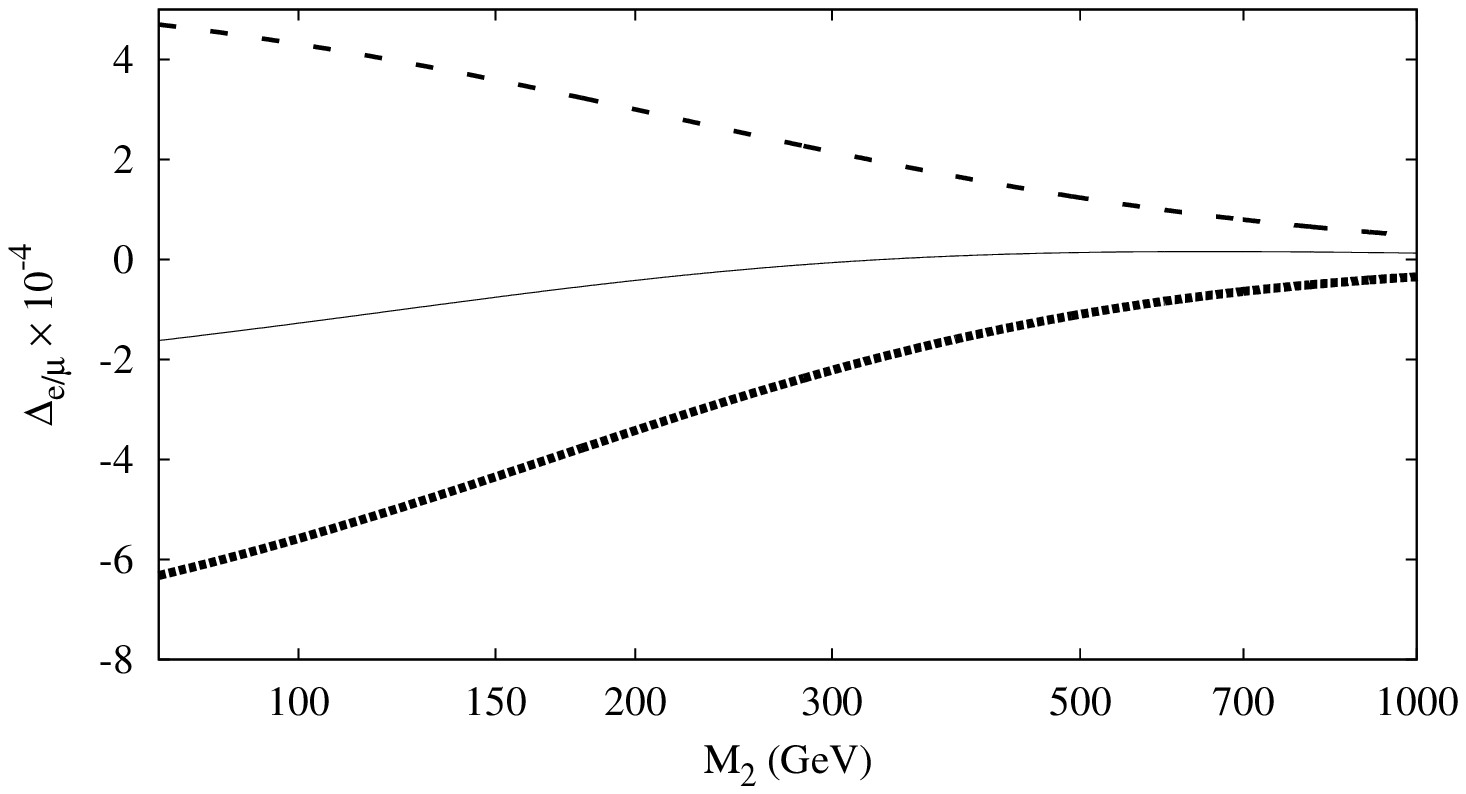}
\label{fig8b}}
\caption{$\Delta_{e/\mu}$ vs. $\mu$ and $M_2$. The total correction is given by the light solid line. The vertex and external leg contribution is given by the heavy line. The box graph contribution is given by the dashed line.}
\label{fig8}
\end{figure}

\begin{figure}[htb]
\subfloat[$|\Delta_{e/\mu}|$ vs. $m_{\chi 1}$.]{
   \epsfxsize 2.85 truein \epsfbox {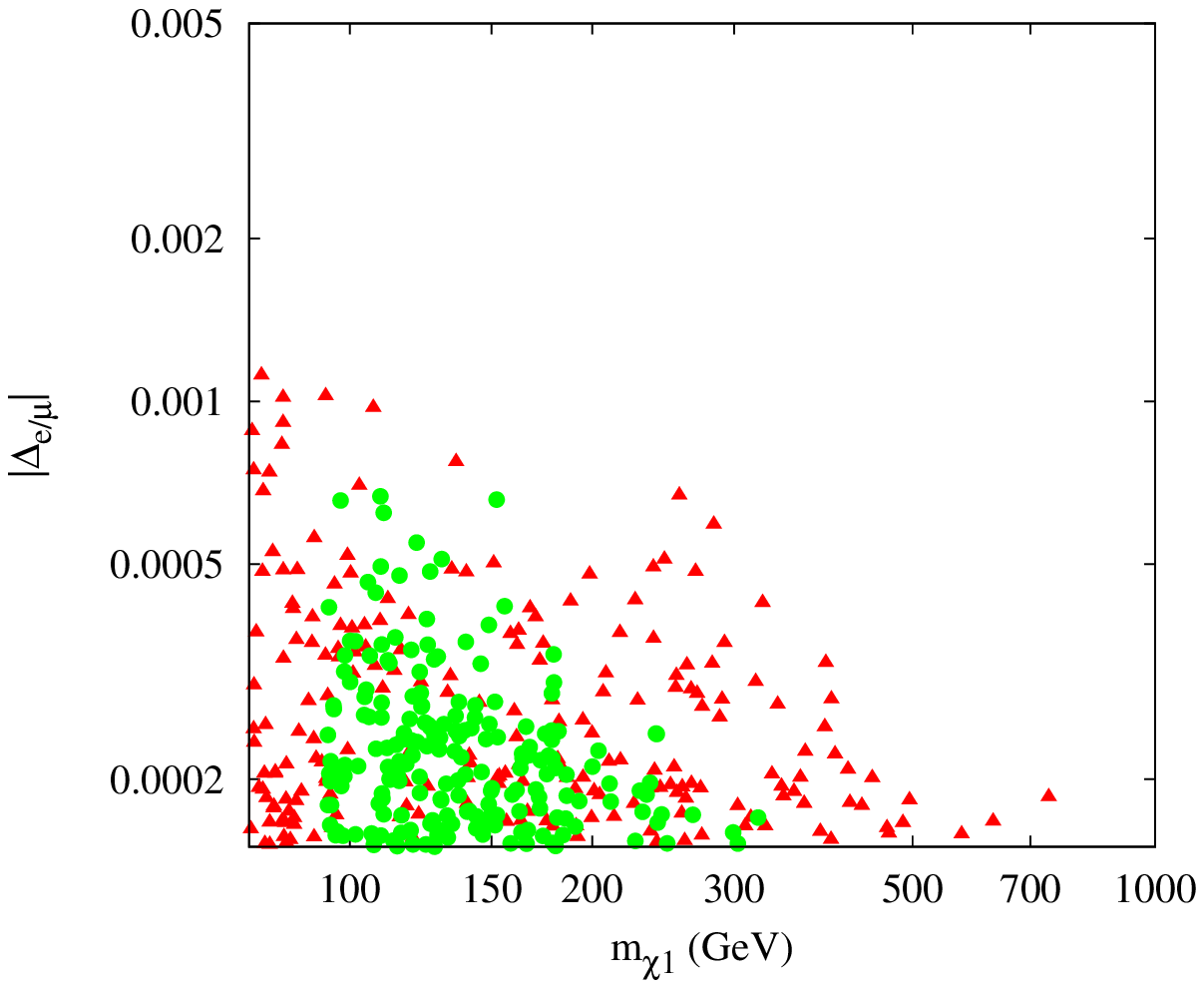}
\label{fig9a}}
\subfloat[$|\Delta_{e/\mu}|$ vs. $Min(m_{\tilde{e}_L},m_{\tilde{\mu}_L})$.]{
   \epsfxsize 2.85 truein \epsfbox {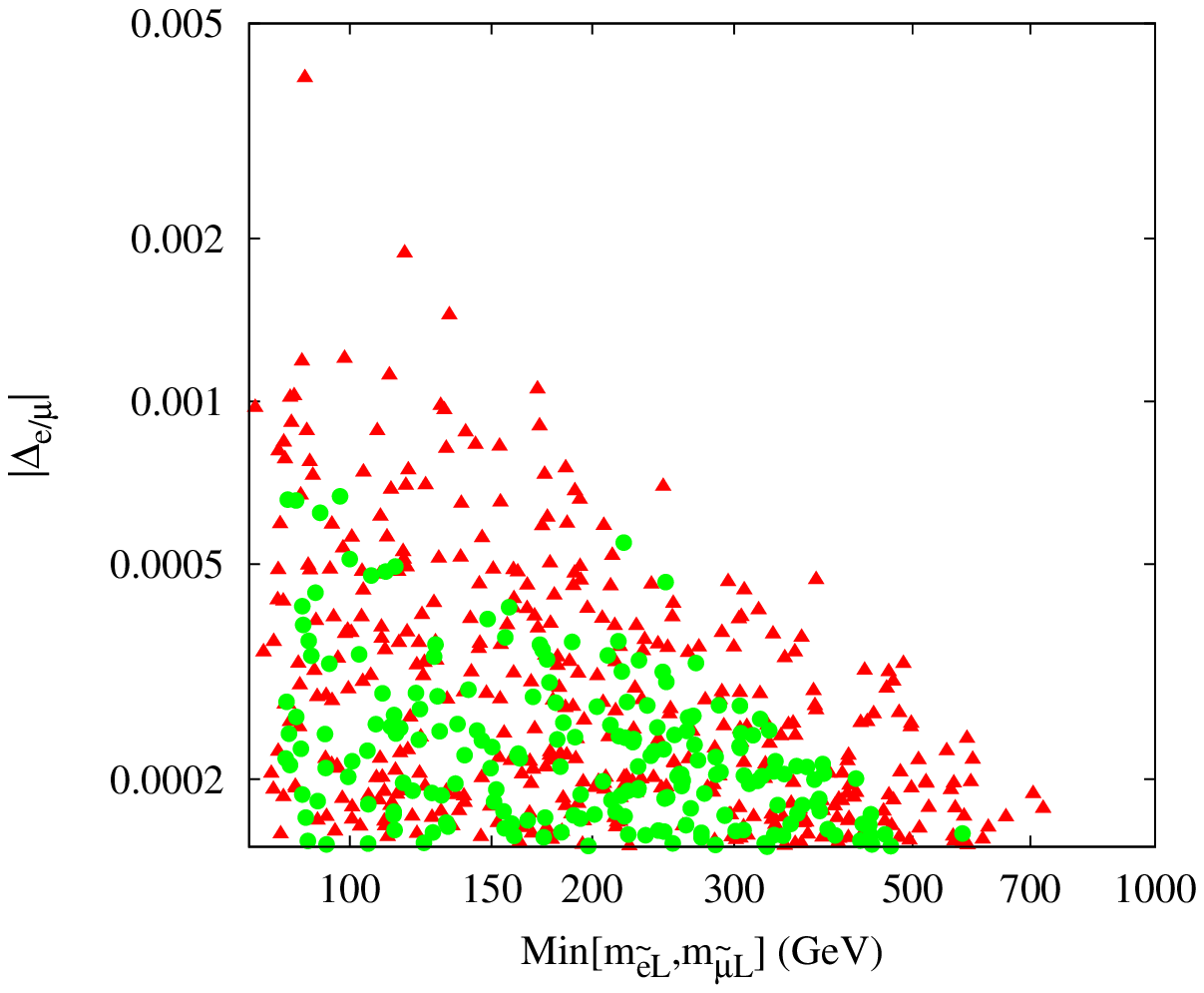}
\label{fig9b}}
\newline
\subfloat[$|\Delta_{e/\mu}|$ vs. $m_{\tilde{e}}/m_{\tilde{\mu}}$.]{
   \epsfxsize 2.85 truein \epsfbox {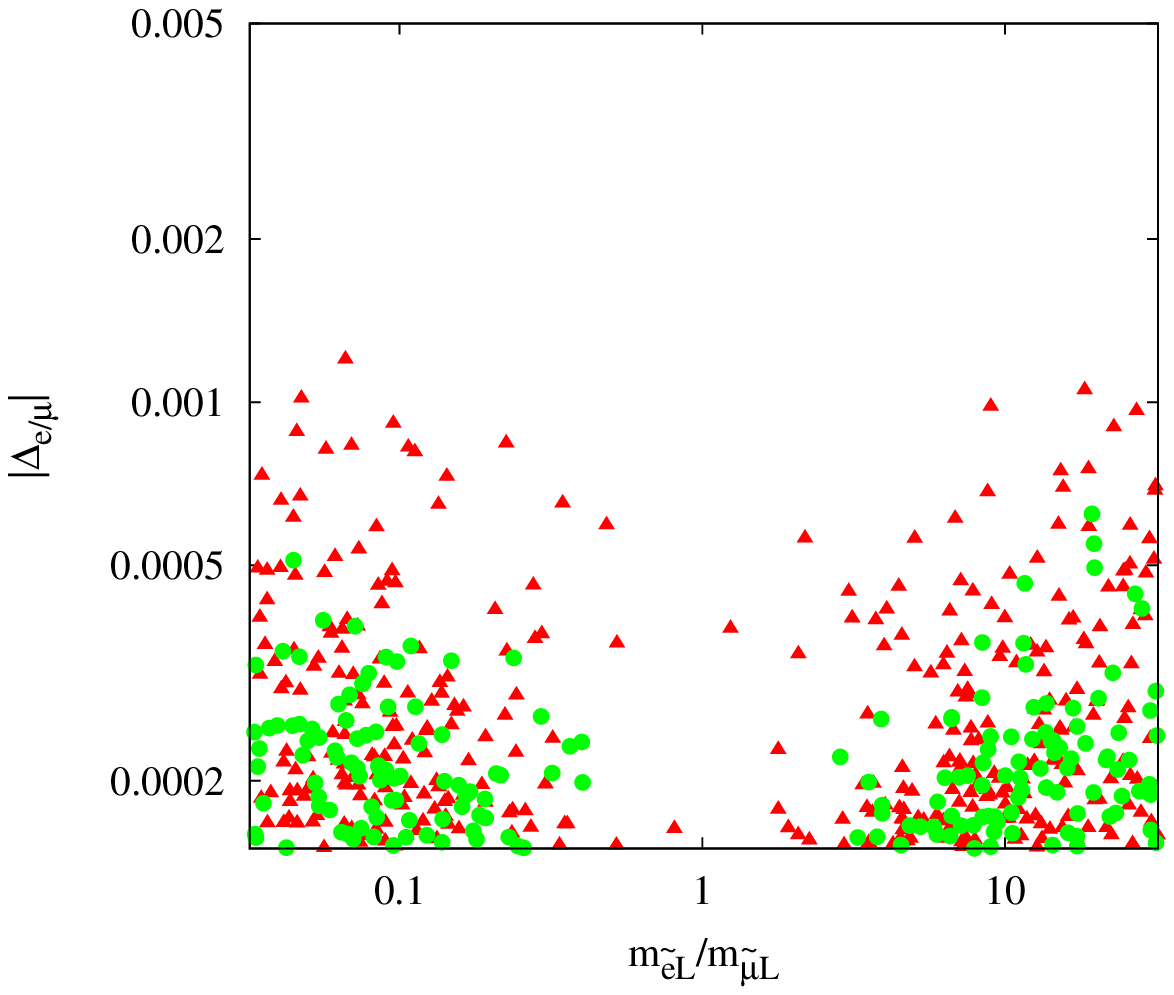}
\label{fig9c}}
\caption{Scatter plots of $|\Delta_{e/\mu}|$. Points inside electroweak bounds are given by green circles. Points outside electroweak bounds are given by red triangles.}
\label{fig9}
\end{figure}

\begin{figure}[htb]
\subfloat[$\Delta_{e/\mu}$ vs. $m_{\chi_1}$.]{
   \epsfxsize 2.85 truein \epsfbox {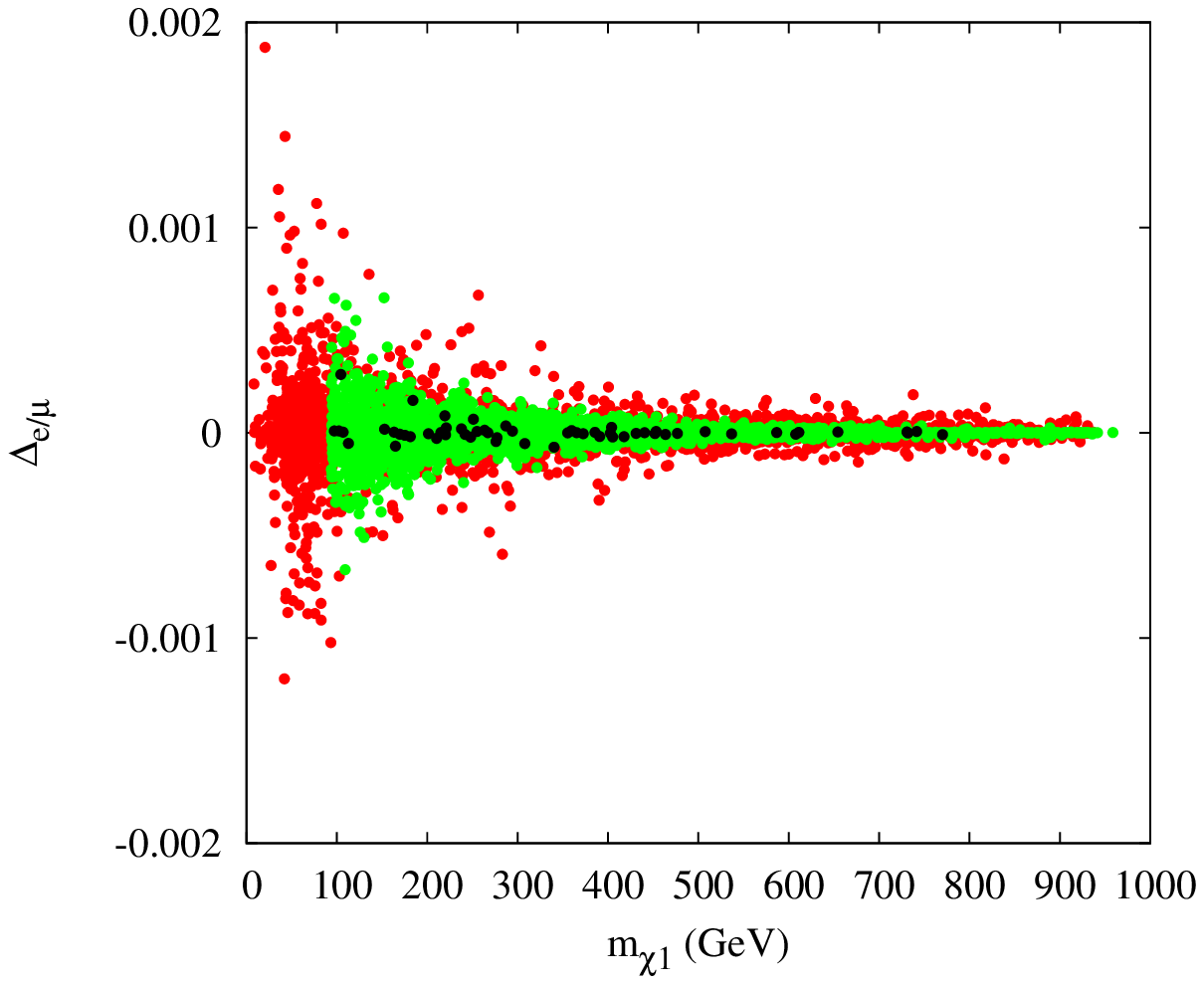}
\label{fig10a}}
\subfloat[$\Delta_{CKM}$ vs. $m_{\chi_1}$.]{
  \epsfxsize 2.85 truein \epsfbox {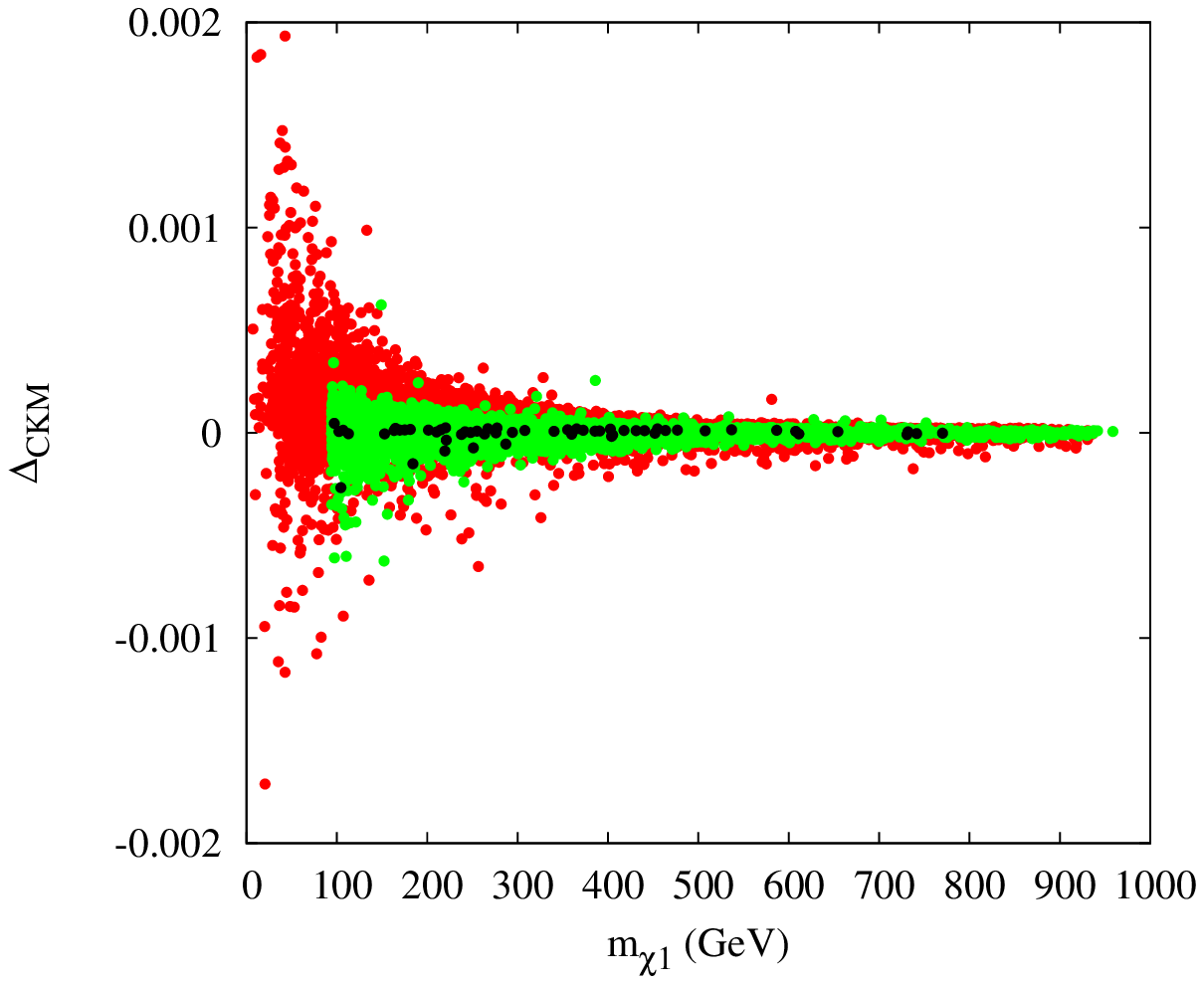}
\label{fig10b}}
\newline
\subfloat[$\Delta_{CKM}$ vs. $\Delta_{e/\mu}$.]{
   \epsfxsize 2.85 truein \epsfbox {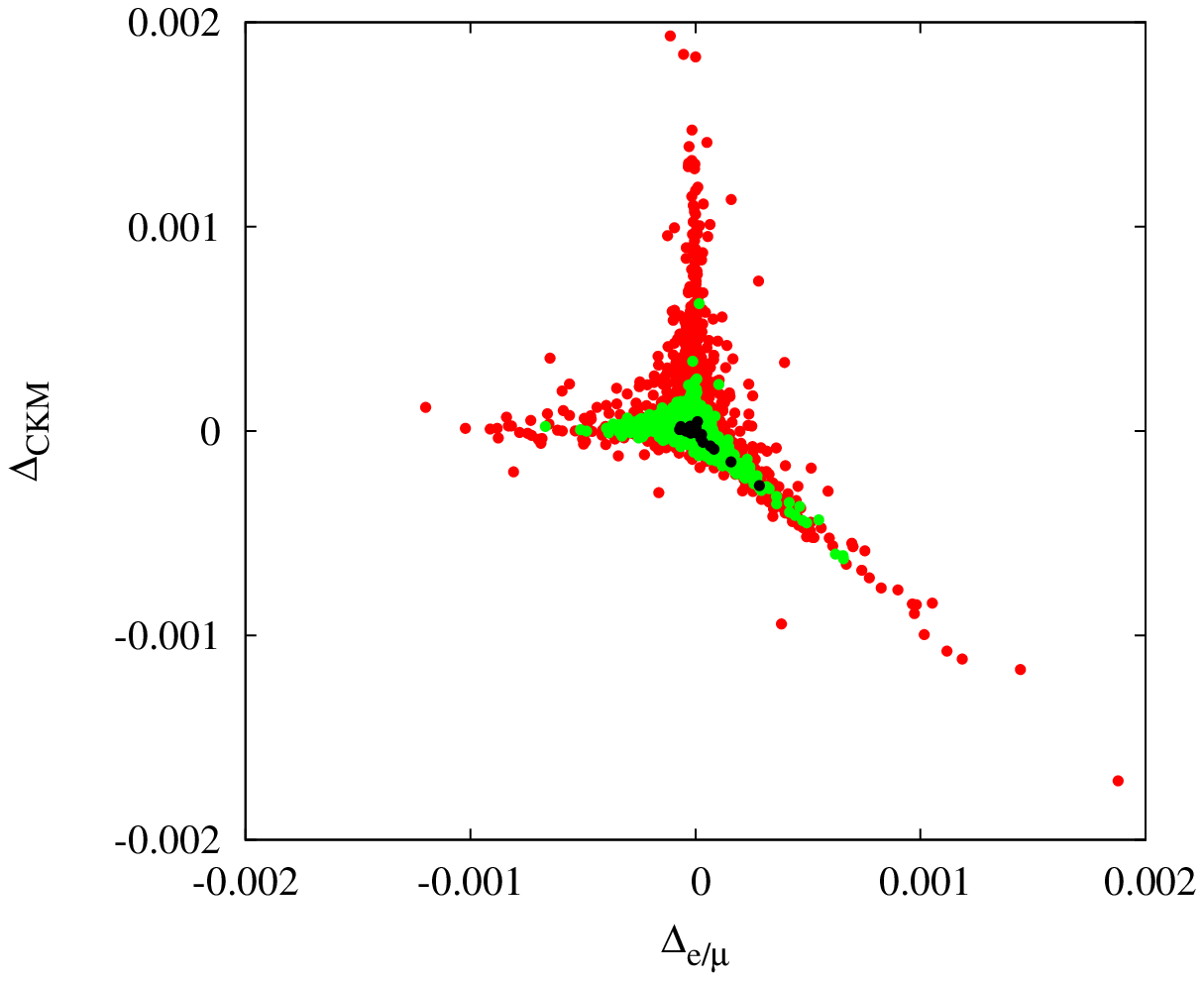}
\label{fig10c}}
\subfloat[$\Delta_{CKM}$ vs. $\Delta_{e/\mu}$.]{
   \epsfxsize 2.85 truein \epsfbox {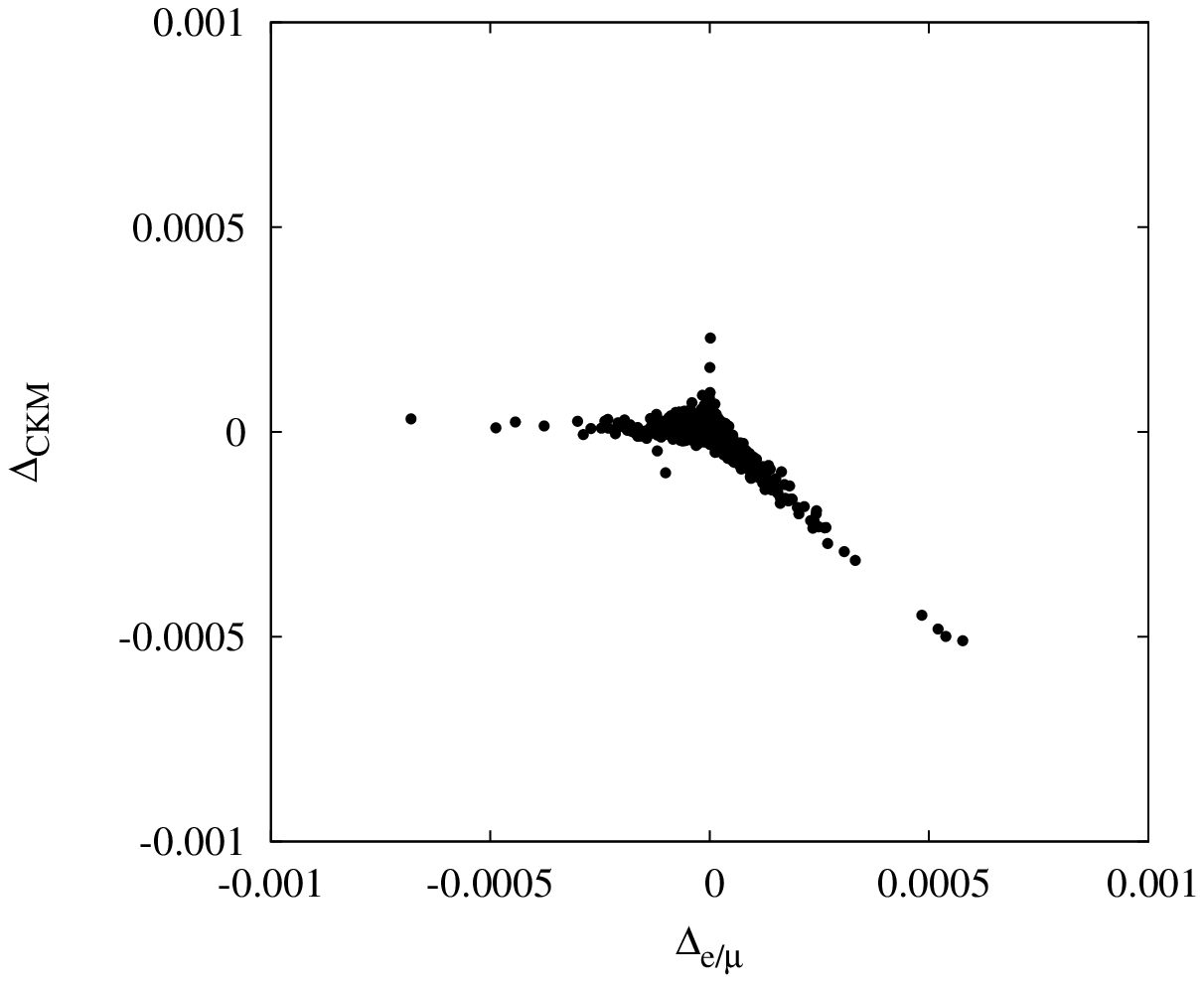}
\label{fig10d}}
\caption{Corrections to $\Delta_{e/\mu}$ and $\Delta_{CKM}$. Points inside electroweak bounds (including LHC bounds) are black. Points inside electroweak bounds (excluding LHC bounds) are green. Points outside electroweak bounds are red. Note that Figs.~\ref{fig10a}--\ref{fig10c} all are of the same points. Fig.~\ref{fig10d} shows a larger set of points within LHC bounds. Note also that, in contrast to Fig. \ref{fig4} the squark and slepton mass parameters are also varied in these plots. }
\label{fig10}
\end{figure}

\begin{figure}[htbp]
\subfloat[$U$ vs. $S$.]{
   \epsfxsize 2.85 truein \epsfbox {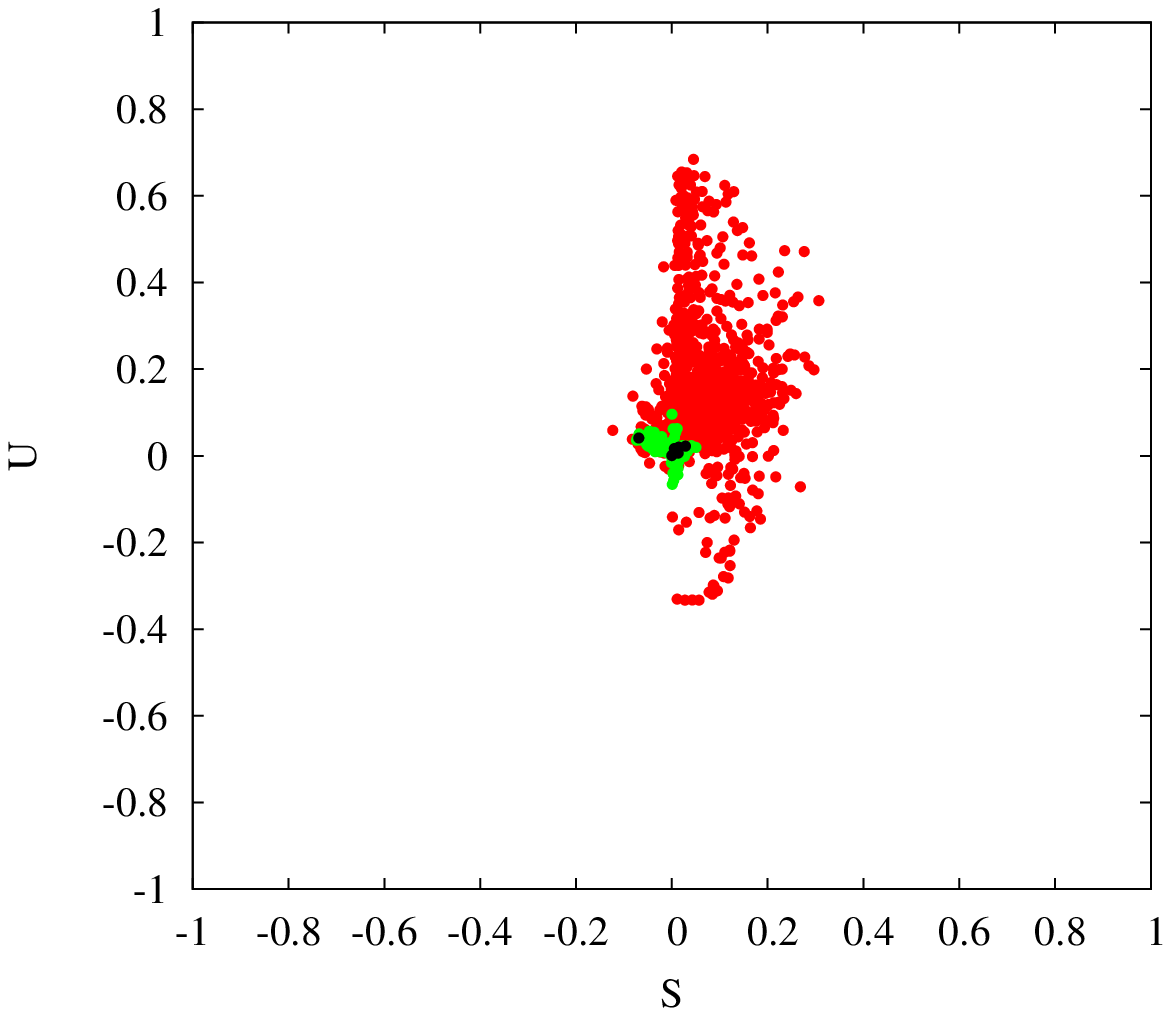}
\label{fig11a}}
\subfloat[$U$ vs. $T$.]{
  \epsfxsize 2.85 truein \epsfbox {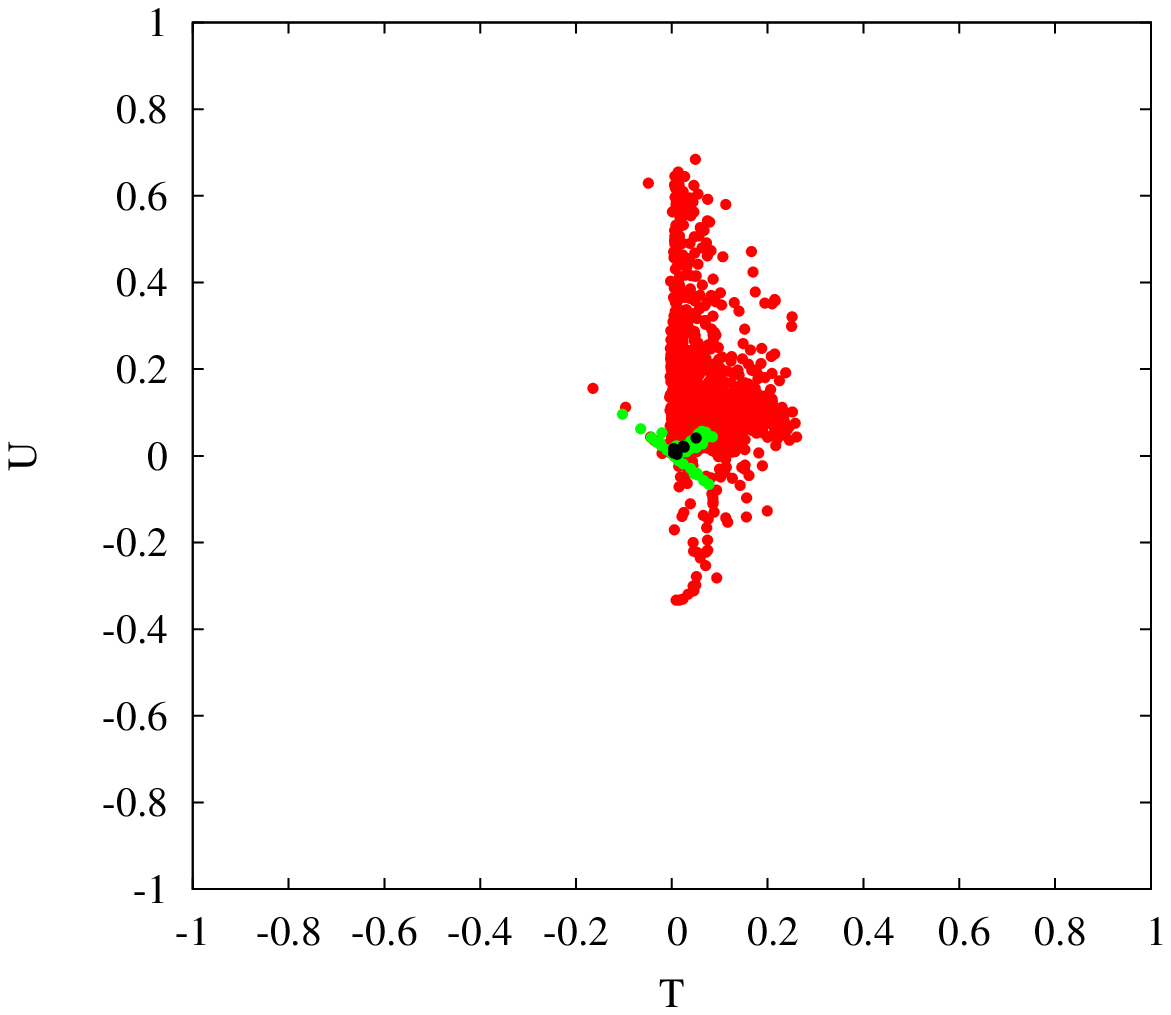}
\label{fig11b}}
\newline
\subfloat[$T$ vs. $S$.]{
  \epsfxsize 2.85 truein \epsfbox {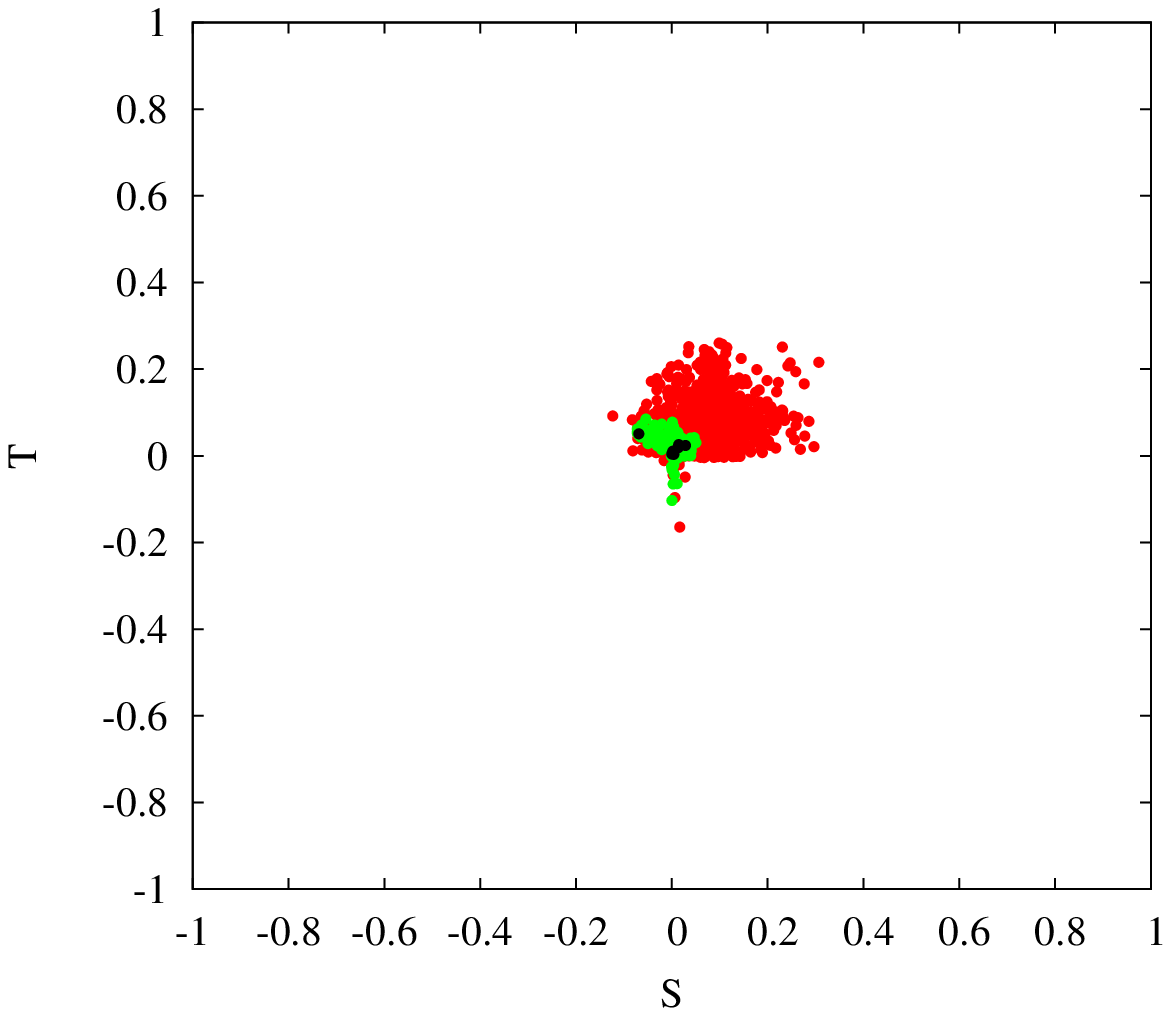}
\label{fig11c}}
\caption{Scatter plots of the oblique parameters corresponding to Figs.~\ref{fig9} and~\ref{fig10}.}
\label{fig11}
\end{figure}

\begin{figure}[htbp]
\subfloat[$\Delta_{CKM}$ vs. $m_{\tilde{\ell}_1}$.]{
   \epsfxsize 2.85 truein \epsfbox {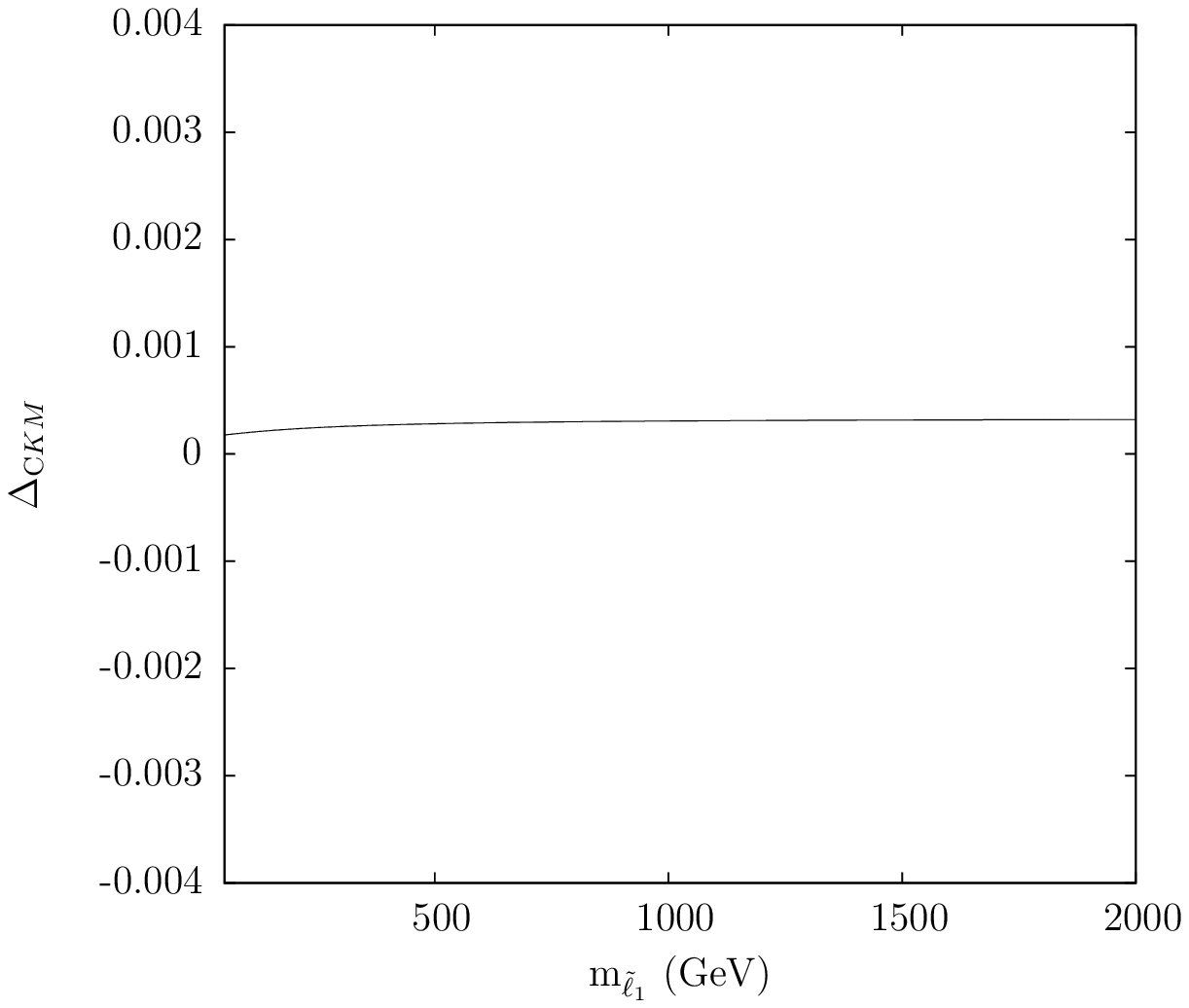}
\label{fig12a}}
\subfloat[$\Delta_{e/\mu}$ vs. $m_{\tilde{\ell}_1}$.]{
  \epsfxsize 2.85 truein \epsfbox {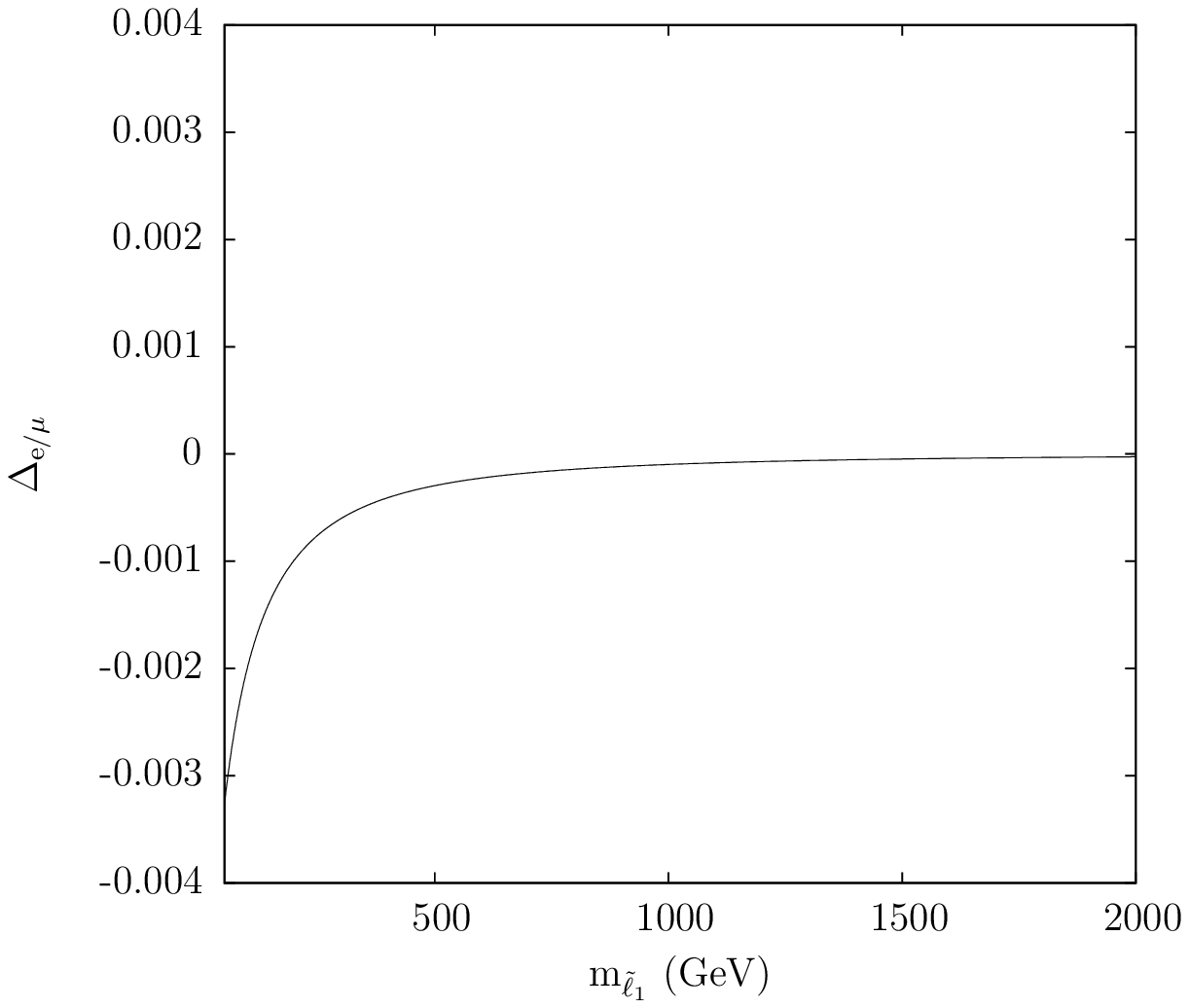}
\label{fig12b}}
\newline
\subfloat[$\Delta_{CKM}$ vs. $m_{\tilde{\ell}_2}$.]{
   \epsfxsize 2.85 truein \epsfbox {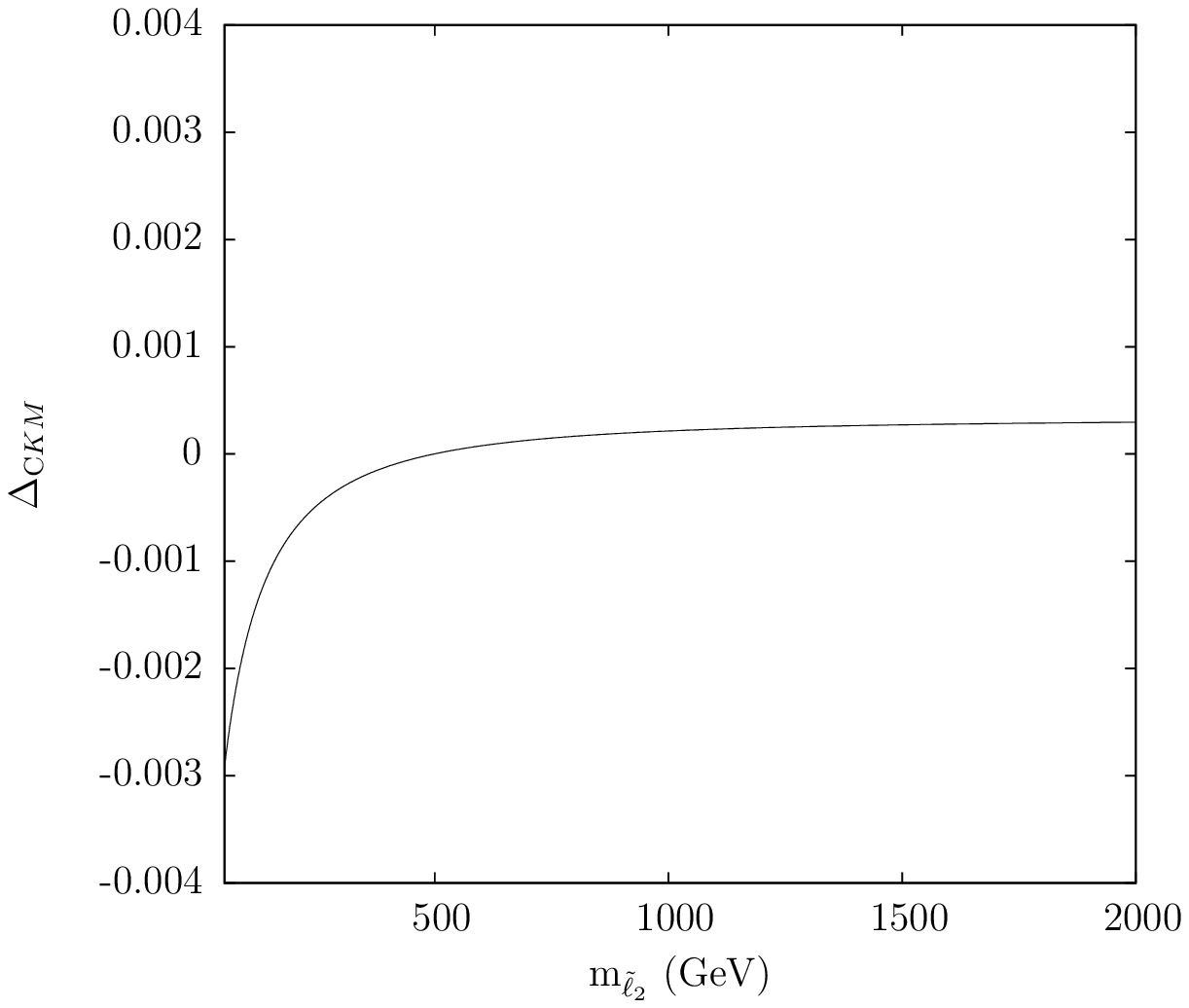}
\label{fig12c}}
\subfloat[$\Delta_{e/\mu}$ vs. $m_{\tilde{\ell}_2}$.]{
  \epsfxsize 2.85 truein \epsfbox {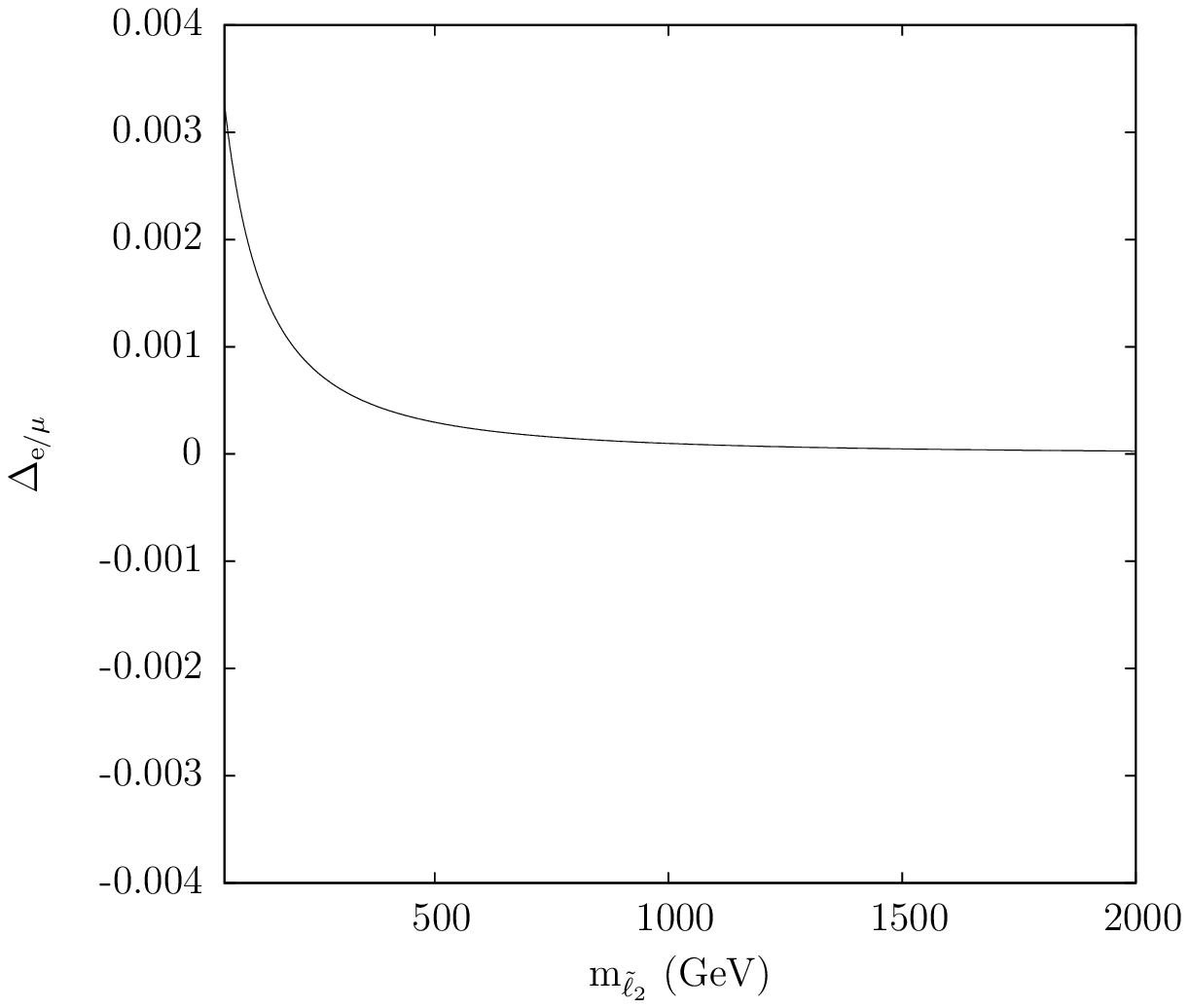}
\label{fig12d}}
\newline
\subfloat[$\Delta_{CKM}$ vs. $m_{\tilde{q}_1}$.]{
   \epsfxsize 2.85 truein \epsfbox {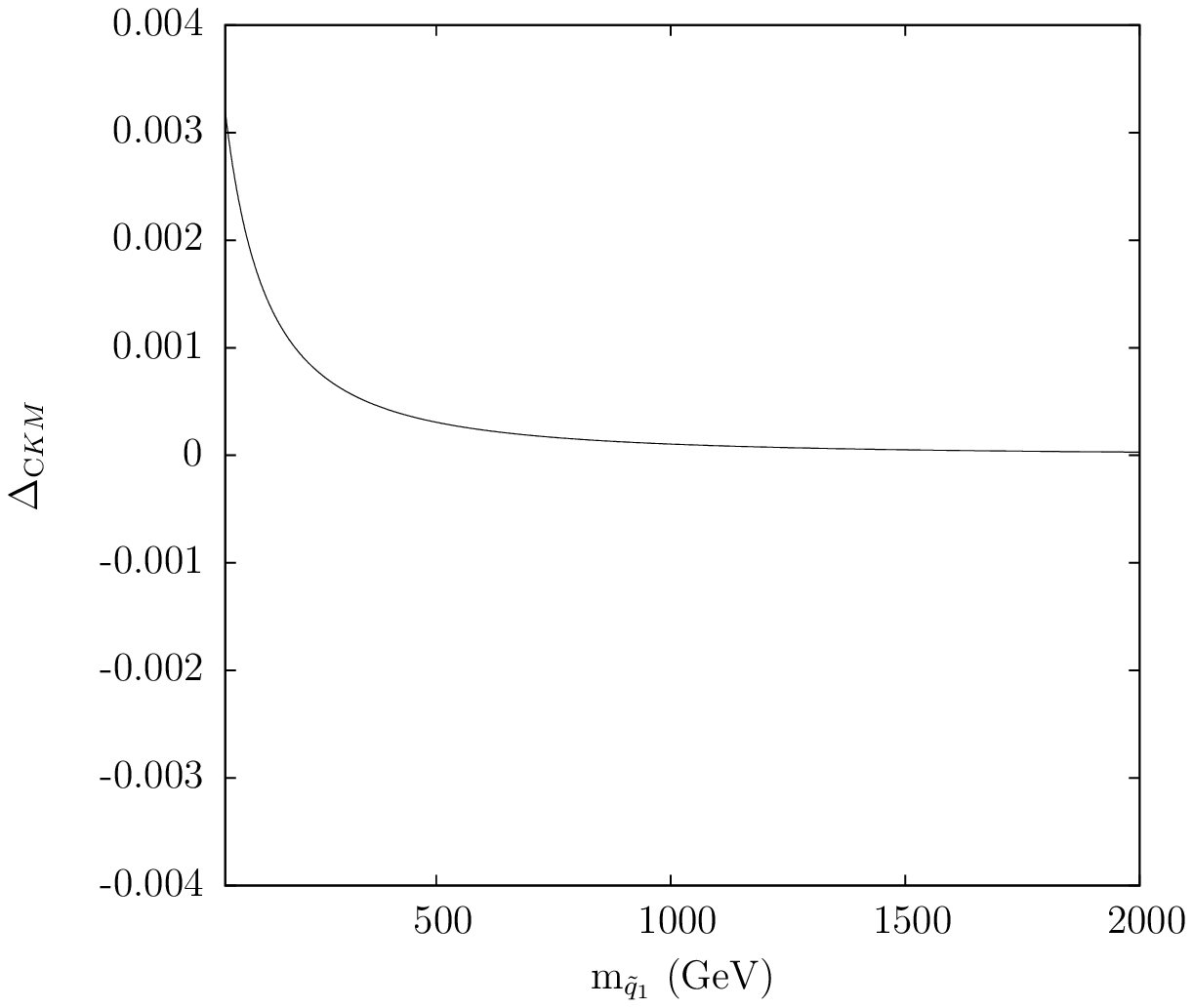}
\label{fig12e}}
\subfloat[$\Delta_{e/\mu}$ vs. $m_{\tilde{q}_1}$.]{
  \epsfxsize 2.85 truein \epsfbox {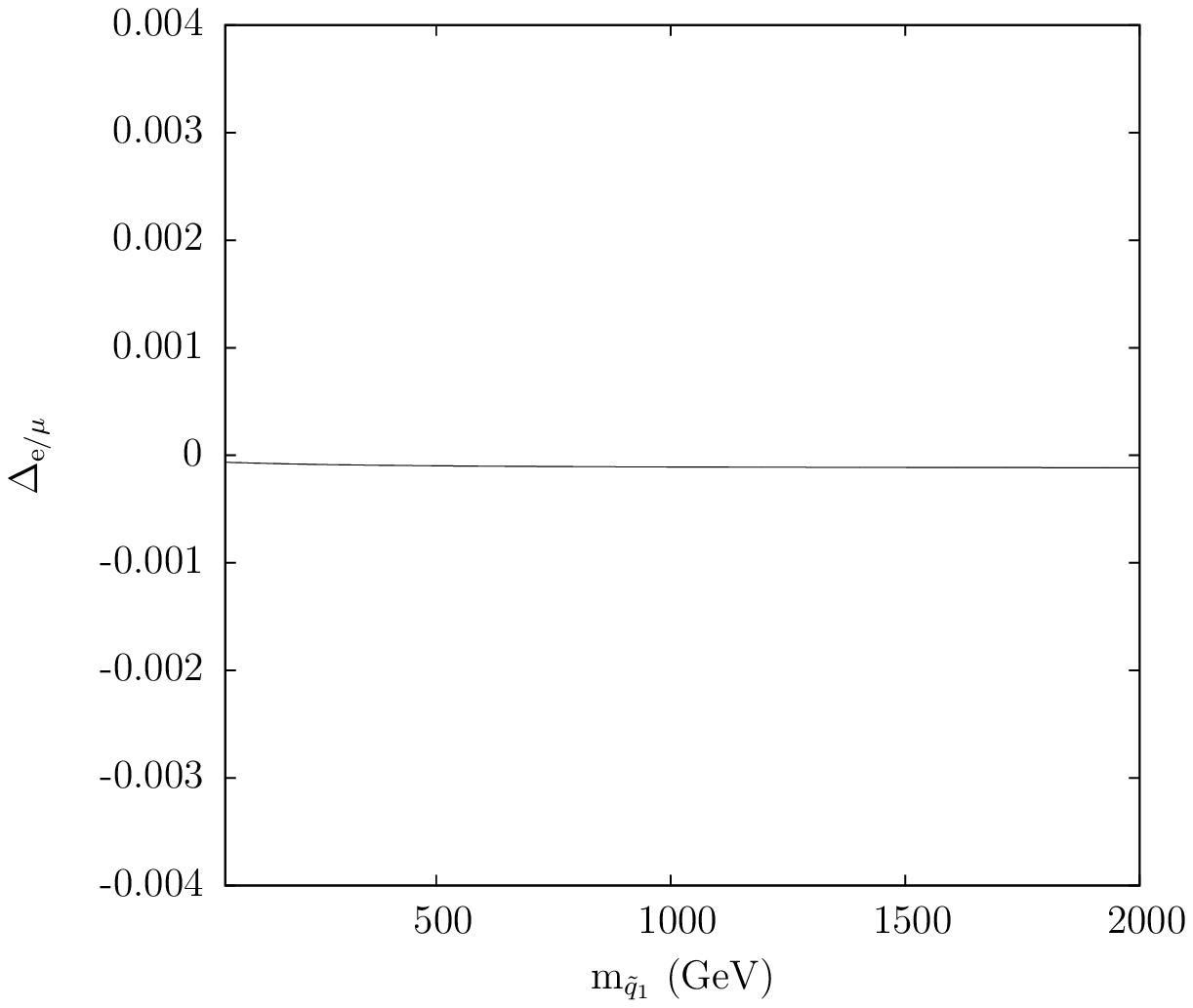}
\label{fig12f}}
\caption{Corrections to $\Delta_{e/\mu}$ and $\Delta_{CKM}$, as functions of sfermion masses.}
\label{fig12}
\end{figure}

The results of our numerical study are indicated by the plots in Figures \ref{fig1}-\ref{fig12}.

\begin{table}[hbp]
\caption{Parameter values for plots. All mass scales are in GeV.}
{ \scriptsize
\begin{tabular}{|c|c|c|c|c|c|c|c|c|}
\hline
Fig. & $\mu$ & $\tan \beta$ & $M_1$ & $M_2$ & $M_3$ & $m_{L11}$ & $m_{L22}$ & $m_{Q11}$ \\
\hline
\ref{fig1} & 100--1000 & 1 & 50 & 50 & 10000 & 110 & 110 & 10000 \\
\hline
\ref{fig2a} \& \ref{fig2b} & 50 & 1 & 50 & 100--1000 & 10000 & 110 & 110 & 10000 \\
\hline
\ref{fig2c} & 50 & 1 & 50 & 500--9500 & 10000 & 110 & 110 & 10000 \\
\hline
\ref{fig3} & 250 & 1 & 150 & 200 & 10000 & 150 & 100--5000 & 1000 \\
\hline
\ref{fig4} & $\pm$(50--1000) & 1 & 50--1000 & $\pm$(50--1000) & 10000 & 110 & 110 & 1000 \\
\hline
\ref{fig5} & $\pm$(50--1000) & 1 & 50--1000 & $\pm$(50--1000) & 10000 & 110 & 110 & 1000 \\
\hline
\ref{fig6} & $\pm$(50--1000) & 1 & 50--1000 & $\pm$(50--1000) & 10000 & 110 & 110 & 1000 \\
\hline
\ref{fig7} & $\pm$(50--1000) & 1 & 50--1000 & $\pm$(50--1000) & 10000 & 110 & 110 & 1000 \\
\hline
\ref{fig8a} & 75--1000 & 20 & 100 & 150 & 10000 & 100 & 500 & 200 \\
\hline
\ref{fig8b} & 200 & 20 & 100 & 75--1000 & 10000 & 100 & 500 & 200 \\
\hline
\ref{fig9} & $\pm$(45--1000) & 1--50 & 45--1000 & $\pm$(45--1000) & 10000 & 45--5000 & 45--5000 & 45--1000 \\
\hline
\ref{fig10} & $\pm$(45--1000) & 1--50 & 45--1000 & $\pm$(45--1000) & 10000 & 45--5000 & 45--5000 & 45--1000 \\
\hline
\ref{fig11} & $\pm$(45--1000) & 1--50 & 45--1000 & $\pm$(45--1000) & 10000 & 45--5000 & 45--5000 & 45--1000 \\
\hline
\ref{fig12a} \& \ref{fig12b} & 45 & 1 & 45 & 45 & 10000 & 50--2000 & 5000 & 500 \\
\hline
\ref{fig12c} \& \ref{fig12d} & 45 & 1 & 45 & 45 & 10000 & 5000 & 50--2000 & 500 \\
\hline
\ref{fig12e} \& \ref{fig12f} & 45 & 1 & 45 & 45 & 10000 & 1000 & 5000 & 50--2000 \\
\hline
\end{tabular}}
\label{table2}
\end{table}

We first observe that the corrections to $\Delta_\mathrm{CKM}$ are at most of order $10^{-3}$ in the region of relatively light superpartner masses.  On the basis of the size of electroweak couplings and the masses of superpartners, one might expect corrections to be larger.  However, significant cancellations between loop corrections occur, thereby reducing the magnitude of the total. These cancellations can be understood in part from the $\mu$-dependence shown in
Fig.~\ref{fig1}. For large $\mu$, and light $M_{1,2}$, the electroweak gauginos are nearly pure bino, wino, and Higgsino states -- corresponding to the spectrum which would occur in the limit of unbroken electroweak symmetry. In this limit, the vertex and external leg corrections for each $Wff$ interaction sum to a mass-independent constant: ${\hat\alpha}/4\pi\sin^2{\hat\theta}_W$ in the $\overline{\mathrm{DR}}$ scheme~\cite{sean}. This constant then cancels exactly in the difference $\Delta_\mathrm{vertex}+\Delta_\mathrm{leg} $. For this regime, the box graph contribution $\Delta_\mathrm{box}$ dominates, so that the dashed and light solid lines in Fig. \ref{fig1} nearly coincide.

As $\mu$ becomes lighter, the cancellation of vertex and external leg corrections no longer holds. Though $\Delta_\mathrm{vertex}+\Delta_\mathrm{leg} $ remains finite in this region, it introduces some dependence on the other MSSM parameters and for very light superpartners, can lead to corrections of order $10^{-3}$ (see the heavy line in Fig. \ref{fig1}). The box graph correction, in contrast, varies only gently with $\mu$. Moreover, its sign is opposite to that of $\Delta_\mathrm{vertex}+\Delta_\mathrm{leg} $. Consequently, in the light $\mu$ region, the latter cancels against $\Delta_\mathrm{box}$. For the particular parameter choice used in generating Fig. \ref{fig1}, the cancellation is quite strong, leading to a total correction that is much smaller in magnitude than that of either $\Delta_\mathrm{box}$ or $\Delta_\mathrm{vertex}+\Delta_\mathrm{leg} $ individually. At present, we have no physical explanation for this cancellation, but simply report it as a result. 

In Fig.~\ref{fig2a}, we show $\Delta_{CKM}$ as a function of $M_2$ for light $\mu$. The same cancellation as indicated above occurs for light $M_2$ and persists into the heavy $M_2$ region, where $\Delta_\mathrm{box}$ and $\Delta_\mathrm{vertex}+\Delta_\mathrm{leg} $ are becoming small individually. The cancellation becomes exact in the region of $M_2\sim 500$ GeV, as we observe from Fig. \ref{fig2b}. There, we give the ratio of $\Delta_\mathrm{box}/(\Delta_\mathrm{vertex}+\Delta_\mathrm{leg})$ as a function of $M_2$. The ratio becomes close to $-1$ for $M_2$ in the vicinity of 500 GeV. 

Since generically the finite corrections $\Delta r^{(V)}_{\beta}$ and $\Delta r_{\mu}$ tend to cancel each other in the difference $\Delta r^{(V)}_{\beta} - \Delta r_{\mu}$, the overall effect on $\Delta_\mathrm{CKM}$ will be maximized in special regions of parameter space where one contribution which otherwise cancels is relatively suppressed. One situation is the region of a large splitting between the first generation squark and second generation slepton masses.  Given the present LHC lower bounds on squark masses, such a situation is phenomenologically more viable for light ${\tilde\mu}$ and heavy ${\tilde u}$, ${\tilde d}$. In Fig.~\ref{fig3}, we show $\Delta_{CKM}$ as a function of $m_{\tilde{\mu}L}$ for heavy squark masses in the region of light $\mu$ and $M_2$.

The resulting possibility of an overall correction having a magnitude comparable or larger than the present sensitivity is indicated by the scatter plot in  in Fig.~\ref{fig4}. Here, we show  $\Delta_{CKM}$ as a function of the lightest chargino mass for heavy squarks and light sleptons, scanning over the remaining electroweak gaugino-Higgsino mass parameters, as indicated in Table \ref{table2}. We also illustrate the impact of imposing  indirect electroweak precision observable (EWPO) constraints and direct LEP bounds. The red points are outside these bounds. 
Points consistent with LHC data and all other constraints are black (note that we have set the squark mass parameter to 1 TeV for this plot). 
The impact from the oblique parameter constraints is negligible. The primary impact from including LEP constraints is to impose a lower bound on the lightest chargino mass. The weak dependence on oblique parameter constraints is expected, since the oblique parameters characterize the impact of superpartner loops on the electroweak gauge boson propagators and since the propagator corrections cancel from $\Delta_{CKM}$. Scatter plots for the oblique parameters are shown in Fig.~\ref{fig7}.

For $m_{\chi 1}$ close to the LEP lower bound, $\Delta_{CKM}$ can be as large as $10^{-3}$ and have either sign. We note that the maximum magnitude of $\Delta_{CKM}$ is slightly larger than the present combined experimental and theoretical error. Further reductions in these uncertainties would be needed in order for a first row CKM unitarity test to provide a significant probe of superpartner loop effects. The sign of $\Delta_{CKM}$ in Fig.~\ref{fig4} is sensitive to the value of the $\mu$-parameter, but not to $M_1$ or $M_2$. The magnitude of $\Delta_{CKM}$ in this figure is sensitive to $\mu$ and $M_2$, but not $M_1$. This is illustrated in Fig.~\ref{fig5}. An interesting feature of Fig.~\ref{fig5} is the sharp sign change of $\Delta_{CKM}$ as a function of $\mu$. This is due to competition between the vertex and external leg contributions to $\Delta_{CKM}$ and the box graph contribution. As shown in Fig.~\ref{fig6}, the vertex and external leg contributions are negative, whereas the box graph contribution is positive. The vertex and external leg contribution approaches zero in the limit of large $|\mu|$, due to the effects of electroweak symmetry breaking becoming negligible in this limit.

Ref.~\cite{kurylov} investigated superpartner corrections in special regions of parameter space.  The authors of that work showed that when $\tilde{u}$ and $\tilde{d}$ had the same masses and mixing angles, $M_{LR}^{2} \gg |M_{L}^{2} - M_{R}^{2}|$, $\kappa_{\tilde{f}} = M_{\tilde{f}_2}/M_{\tilde{f}_1}$ was large,
\beq
\left[\Delta r^{(V)}_{\beta} - \Delta r_{\mu}\right]^{\mathrm{SUSY}} \sim \alpha_{\rm EM} \frac{(c_{W}^{2} - s_{W}^{2})}{32\pi s_{W}^{2} c_{W}^{2}} \ln \left(\kappa_{\tilde{q}^2}/\kappa_{\tilde{\mu}}^4 \right)~,
\label{kur_lim1}
\eeq
where $c_W$ is the cosine of the weak mixing angle and $s_W$ is the sine. Under the assumed limits on MSSM parameters, box graphs can be neglected. 
Ref.~\cite{kurylov} also considered the limit of large sfermion masses and $M_{LR}^{2} \approx 0$. In this case,
\beq
\left[\Delta r^{(V)}_{\beta} - \Delta r_{\mu}\right]^{\mathrm{SUSY}}  \sim \frac{\alpha_{\rm EM}}{2\pi} \cos (2\beta) \left[ \frac{M_{Z}^{2}}{3 m_{\tilde{q}}^2}\ln \left(\frac{m_{\tilde{q}}^{2}}{<M_{\tilde{\chi}}^{2}>} \right) - \frac{M_{Z}^{2}}{m_{\tilde{\mu}}^{2}}\ln \left(\frac{m_{\tilde{\mu}}^{2}}{<M_{\tilde{\chi}}^{2}>} \right) \right]~,
\label{kur_lim2}
\eeq
where $<M_{\tilde{\chi}}^{2}>$ is the squared-mass scale for the charginos and neutralinos. As a cross check on our analysis, we have numerically verified Eqs.~(\ref{kur_lim1}) and~(\ref{kur_lim2})  using our scans in the limits assumed in Ref.~\cite{kurylov}.

Turning now to $R_{e/\mu}$, we first compare our computation with the results of Ref.~\cite{sean}. For completeness, we show our results in Figs.~\ref{fig8a}--\ref{fig9c}, which agree with the corresponding graphs in Ref.~\cite{sean}. In Fig.~\ref{fig9}, the absolute value of $\Delta_{e/\mu}$ is plotted, in accordance with Ref.~\cite{sean}. We also plot the SUSY correction without the absolute value in Fig.~\ref{fig10a}. For comparison, we show the $\beta$-decay correction for the same set of mass constraints in Fig.~\ref{fig10b}. The correlation between the $\beta$-decay correction and the pion decay correction is shown in Fig.~\ref{fig10}. We note that both Figs.~\ref{fig4} and~\ref{fig10b} are scatter plots of $\Delta_{CKM}$ vs. $m_{\chi 1}$. However, as noted in Table~\ref{table2}, the bounds on the MSSM parameters are different. In particular, $\tan \beta$ and the sfermion masses are fixed in Fig.~\ref{fig4} but are not in Fig.~\ref{fig10b}.

In Fig.~\ref{fig9}, the absolute value of $\Delta_{e/\mu}$ is plotted, in accordance with Ref.~\cite{sean}. The corresponding corrections obtained without using absolute values are indicated in Figs.~\ref{fig10a} and  \ref{fig10b}. As expected, $\Delta_{e/\mu}$ goes to zero for degenerate first and second generation sleptons, corresponding to \lq\lq slepton universality". Away from this regime, the magnitude of $\Delta_{e/\mu}$ can be as large as the expected experimental uncertainty of the present experiments and as much as five times larger than the theoretical SM uncertainty when the lightest chargino and lighter of the two slepton generations is sufficiently light. Moreover, for ${\tilde e}_L$ lighter (heavier) than ${\tilde \mu_L}$, the sign of $\Delta_{e/\mu}$ is negative (positive). Thus, a reduction in the experimental error to a level commensurate with the theoretical SM uncertainty would allow for a significant probe of the MSSM parameter space, including the nature of the first and second generation slepton spectrum. Fig.~\ref{fig11} shows scatter plots of oblique parameters corresponding to Figs.~\ref{fig9} and~\ref{fig10}.

In Figs.~\ref{fig10c} and~\ref{fig10d}, we give the correlation between the corrections $\Delta_{CKM}$ and $\Delta_{e/\mu}$. (Fig.~\ref{fig10c} shows points inside and outside LHC bounds, whereas Fig.~\ref{fig10d} only shows points within LHC bounds. The purpose of Fig.~\ref{fig10d} is to show a larger number of points, since only a small proportion of points in Fig.~\ref{fig10c} are within LHC bounds.) We observe the existence of three branches in which loop corrections are enhanced: (a) $\Delta_{e/\mu}$ is negatively enhanced, but $\Delta_{CKM}$ receives no enhancement; (b) both are enhanced, but $\Delta_{e/\mu}$ is positive and $\Delta_{CKM}$ is negative; (c) $\Delta_{CKM}$ is positively enhanced, but $\Delta_{e/\mu}$ receives no enhancement; imposition of LHC constraints cuts into this branch. As we discuss below, these features are associated with dependencies of the corrections on sfermion masses, as illustrated in Fig.~\ref{fig12}.

First, Figs.~\ref{fig12a} and~\ref{fig12b} show the dependence on the first generation slepton mass for heavy first generation squarks and second generation sleptons. In the limit of a small first generation slepton mass but a large second generation slepton mass, $\Delta_{e/\mu}$ is enhanced and negative, but $\Delta r^{(V)}_{\beta} - \Delta r_{\mu}$ is not. This is due to the fact that $\Delta_{e/\mu}$ depends on the difference between the first and second generation slepton masses, whereas  $\Delta r^{(V)}_{\beta} - \Delta r_{\mu}$ is relatively insensitive to the first generation slepton mass. Thus, branch (a) corresponds to this sfermion spectrum. 

Figs.~\ref{fig12c} and~\ref{fig12d} then show the dependence on the second generation slepton mass for heavy first generation squarks and sleptons. In the limit of a small second generation slepton mass but large first generation slepton and squark masses, $\Delta_{e/\mu}$ and $\Delta r^{(V)}_{\beta} - \Delta r_{\mu}$ both are positively enhanced, corresponding to branch (b). In the case of $\Delta_{e/\mu}$, this is again due to the dependence on the difference between first and second generation slepton masses. Moreover, as discussed above, the quantity $\Delta r^{(V)}_{\beta} - \Delta r_{\mu}$ depends on the difference between the first generation squark and the second generation slepton masses and thus can become relatively large when one or the other of these sfermion masses is relatively light. 

Finally, Figs.~\ref{fig12e} and~\ref{fig12f} show the dependence on the first generation squark mass for heavy sleptons. In the limit of a small first generation squark mass and a large second generation slepton mass, $\Delta r^{(V)}_{\beta} - \Delta r_{\mu}$ is negatively enhanced, but $\Delta_{e/\mu}$ receives no enhancement. For this spectrum, one thus obtains branch (c). As stated previously, $\Delta r^{(V)}_{\beta} - \Delta r_{\mu}$ is sensitive to the difference between the first generation squark and the second generation slepton masses. However, dependence on squark masses cancel in $\Delta_{e/\mu}$. We also note that the sign of $\Delta r^{(V)}_{\beta} - \Delta r_{\mu}$ indicates whether the first generation squarks or second generation sleptons are heavier.

New results from the LHC raised the limit on the squark mass. As seen in Figs.~\ref{fig10c} and~\ref{fig10d}, LHC constraints contract branch~(c). This is because this branch corresponds to the limit of a low squark mass.

\section{Discussion and Conclusions
\label{conclude}}

Precision tests of CKM unitarity and lepton universality provide powerful indirect probes of BSM physics.  Here, we have uncovered a novel correlation between MSSM corrections to first row CKM unitarity and  the ratio $R_{e/\mu}$ of $\pi_{\ell 2}$-decays that would be indicative of interesting details for the first and second generation sfermion spectrum.\footnote{As noted in Sect.~\ref{beta_decay}, we computed {\it apparent} violations of first row CKM unitarity, due to the measured value of $|V_{ud}|$ possibly deviating from the value appearing in the MSSM Lagrangian. (Standard Model assumptions go into the measurement of $|V_{ud}|$.) We did not compute corrected values of CKM matrix elements in an effective Lagrangian or $\beta$-functions for CKM parameters. For a study of the latter, see {\it e.g.}, Ref.~\cite{true_ckm}.} If superpartners are discovered at the LHC, these correlations could provide a particularly interesting diagnostic tool in an effort to specify the underlying Lagrangian. For example, the observation of a significant and positive deviation of $\Delta_{CKM}$ would either be inconsistent with the present LHC bounds on first generation squarks -- thereby pointing to the presence of some other new physics -- or would suggest a spectrum for gluinos and first generation squarks that evades these bounds. On the other hand, agreement of $\Delta_{CKM}$ with the SM but a significant deviation for $\Delta_{e/\mu}$ would indicate heavy first generation squarks, consistent with the early LHC constraints, but relatively lighter and non-degenerate first or second generation sleptons, with the sign of the effect indicating the mass hierarchy. 

More generally, the study of these charged current observables introduces considerable simplifications of the MSSM parameter space analysis, since superpartner contributions to the gauge boson propagators cancel from the relevant ratios -- thereby weakening the impact of indirect constraints from electroweak precision data -- as do some classes of vertex plus external leg corrections. In this sense, the present study provides a concrete illustration of the unique potential for insight that comparisons of low-energy CC processes may provide in the LHC era.

As our analysis indicates, application of this diagnostic tool to the R-parity conserving MSSM requires a precision of a few times $10^{-4}$, at least when a subset of the superpartners are relatively light. The sensitivities of present tests of first row CKM unitarity and lepton universality with $\pi_{\ell 2}$-decays are just beginning to probe the relevant region of MSSM parameter space. Importantly, however, the prospects for more sensitive probes are promising, particularly in light of reduced hadronic uncertainties in the SM predictions for $R_{e/\mu}$. In the case of first row CKM unitarity, similar reductions in the hadronic uncertainties associated with SM contributions to $\Delta r^{(V)}_{\beta} - \Delta r_{\mu}$ and with the determination of $V_{us}$, along with commensurate reductions in the experimental uncertainty in the determination of $V_{ud}$ from nuclear and neutron $\beta$-decay, would be desirable. 

\vskip 0.25in

\noindent{\bf Acknowledgments} The authors thank S. Su and S. Tulin for numerous discussions and numerical cross checks. This work was supported in part under U.S. Department of Energy Contract DE-FG02-08ER41531 (SB and MJRM), the Wisconsin Alumni Research Foundation, and CONACYT (M\'exico) Project 82291--F (JE).

\newpage

\appendix

\section{Loop Graphs
\label{individual}}

In this appendix, we give explicit expressions for one-loop graphs involving Standard Model superpartners, which contribute to $\beta$-decay corrections. Specifically, we give results for external leg, vertex and box graphs which contribute to $\Delta r^{(V)}_{\beta}$ and $\Delta r_{\mu}$. Extensive discussions of the corrections entering $\Delta_{e/\mu}$ appear in Ref.~\cite{sean} so we do not reproduce them here. 

\subsection{External Leg Corrections
\label{leg}}

\begin{figure}[htb]
   \epsfxsize 3.00 truein \epsfbox {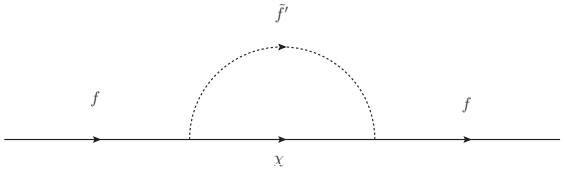}
\caption{A one-loop correction to the graph for an external fermion leg. The internal scalar ${\tilde{f}}^{\prime}$ is a sfermion. The internal $\chi$-field is a chargino, a neutralino or a gluino, depending on the specific diagram. (We shall use similar notations for subsequent loop diagram figures.) The fermion-sfermion interactions are defined according to Fig.~\ref{fig14}.}
\label{fig13}
\end{figure}

\begin{figure}[htb]
   \epsfxsize 3.00 truein \epsfbox {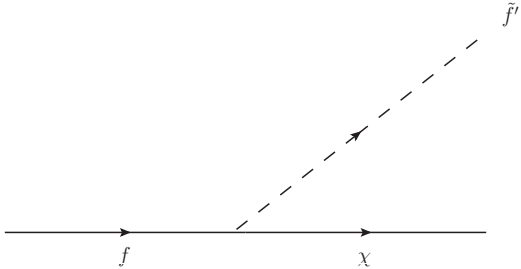}
\caption{The tree-level interaction between $f$, $\chi$ and ${\tilde{f}}^{\prime}$. The value of this vertex interaction is $ig(g_L P_L + g_R P_R)$.}
\label{fig14}
\end{figure}

A generic external leg contribution for either $\beta$-decay or muon decay is shown in Fig.~\ref{fig13}. The general result for an electroweak correction given by Fig.~\ref{fig13} is
\beq
\Delta_{\rm diagram}^{(\ref{fig13})} = \frac{g^2}{(4\pi)^2} \left( \frac{1}{2} \right) C \left[ \frac{\alpha_{\epsilon}}{2} - \int_{0}^{1}dx x \log((x m_{{\tilde{f}}^{\prime}}^{2} + (1 - x)m_{\chi}^{2})/\mu^2) \right]~,
\label{leg_diagram}
\eeq
where $m_{{\tilde{f}}^{\prime}}$ and $m_{\chi}$ are the masses of the ${\tilde{f}}^{\prime}$-field and the $\tilde{\chi}$-field respectively in Fig.~\ref{fig13}. A similar result holds for a supersymmetric QCD correction, except with the weak coupling $g$ replaced with the strong coupling $g_s$. The superscript on $\Delta_{\rm leg}^{\rm (diagram)}$ indicates that the quantity is the contribution to the external leg correction from a single diagram, in contrast to the total external leg correction $\Delta_{\rm leg}$ given in Eq.~(\ref{Delta_contr}). The variable $x$ is the Feynman parameter. The coefficient of $1/2$ in front of the Feynman parameter integral is due to the fact that the square root of the fermion field strength renormalization appears in a coupling correction. The quantity $\mu$ is the 't Hooft scale. The quantity $C$ is
\beq
C = -\left( |g_L|^2 + |g_R|^2 \right)~,
\label{C_def}
\eeq
where $g_L$ and $g_R$ are couplings defined in Fig.~\ref{fig14}. This computation was performed in dimensional reduction. The term $\alpha_{\epsilon}$ is given as
\beq
\alpha_{\epsilon} = \frac{1}{\epsilon} - \gamma + \log(4\pi) + \mathcal{O}(\epsilon)~,
\label{alpha_eps}
\eeq
where $\epsilon$ is the dimensional reduction parameter defining the loop diagram in $d = 4 - 2\epsilon$ dimensions, and $\gamma$ is the Euler-Mascheroni constant. 
Both $\alpha_{\epsilon}$ and $\mu$ cancel in $\Delta r^{(V)}_{\beta} - \Delta r_{\mu}$.

\begin{table}[h]
\caption{The values of $C$ from Eq.~(\ref{leg_diagram}), for the external leg graphs in $\beta$-decay. Note that the weak coupling $g$ is replaced by the strong coupling $g_s$ for the diagrams involving the gluino $\Lambda_G$.}
\begin{tabular}{|c|c|c|c|}
\hline
$$ & $$ & $$ & $$ \\
$f$ & ${\tilde{f}}^{\prime}$ & $\chi$ & $C$ \\
$$ & $$ & $$ & $$ \\
\hline
$$ & $$ & $$ & $$ \\
$u$ & ${\tilde{u}}_i$ & $\chi^{0}_{j}$ & $-\frac{1}{2}|Z_{U}^{1i}|^2 (N_{j2} + \tan \theta_W N_{j1}/3)^2$ \\
$$ & $$ & $$ & $$ \\
\hline
$$ & $$ & $$ & $$ \\
$u$ & ${\tilde{d}}_i$ & $\chi_j$ & $-|U_{j1}|^2 |Z_{D}^{1i}|^2$ \\
$$ & $$ & $$ & $$ \\
\hline
$$ & $$ & $$ & $$ \\
$u$ & ${\tilde{u}}_i$ & $\Lambda_G$ & $-\frac{8}{3}|Z_{U}^{1i}|^2$ \\
$$ & $$ & $$ & $$ \\
\hline
$$ & $$ & $$ & $$ \\
$d$ & ${\tilde{u}}_i$ & $\chi^{C}_{j}$ & $-|V_{j1}|^2 |Z_{U}^{1i}|^2$ \\
$$ & $$ & $$ & $$ \\
\hline
$$ & $$ & $$ & $$ \\
$d$ & ${\tilde{d}}_i$ & $\chi^{0}_{j}$ & $-\frac{1}{2}|Z_{D}^{1i}|^2 (N_{j2} - \tan \theta_W N_{j1}/3)^2$ \\
$$ & $$ & $$ & $$ \\
\hline
$$ & $$ & $$ & $$ \\
$d$ & ${\tilde{d}}_i$ & $\Lambda_G$ & $-\frac{8}{3}|Z_{D}^{1i}|^2$ \\
$$ & $$ & $$ & $$ \\
\hline
\end{tabular}
\label{table3}
\end{table}

\begin{table}[h]
\caption{The values of $C$ from Eq.~(\ref{leg_diagram}), for the external leg graphs in muon decay.}
\begin{tabular}{|c|c|c|c|}
\hline
$$ & $$ & $$ & $$ \\
$f$ & ${\tilde{f}}^{\prime}$ & $\chi$ & $C$ \\
$$ & $$ & $$ & $$ \\
\hline
$$ & $$ & $$ & $$ \\
$\mu$ & ${\tilde{\ell}}_i$ & $\chi^{0}_{j}$ & $-\frac{1}{2}|Z_{L}^{2i}|^2 (N_{j2} + \tan \theta_W N_{j1})^2$ \\
$$ & $$ & $$ & $$ \\
\hline
$$ & $$ & $$ & $$ \\
$\mu$ & ${\tilde{\nu}}_J$ & $\chi^{C}_{j}$ & $-|V_{j1}|^2 |Z_{\nu}^{2J}|^2$ \\
$$ & $$ & $$ & $$ \\
\hline
$$ & $$ & $$ & $$ \\
$\nu_{\mu}$ & ${\tilde{\ell}}_i$ & $\chi_j$ & $-|U_{j1}|^2 |Z_{L}^{2i}|^2$ \\
$$ & $$ & $$ & $$ \\
\hline
$$ & $$ & $$ & $$ \\
$\nu_{\mu}$ & ${\tilde{\nu}}_J$ & $\chi^{0}_{j}$ & $-\frac{1}{2}|Z_{\nu}^{2J}|^2 (N_{j2} - \tan \theta_W N_{j1})^2$ \\
$$ & $$ & $$ & $$ \\
\hline
\end{tabular}
\label{table4}
\end{table}

Tables~\ref{table3} and~\ref{table4} list the values of the coefficient $C$ from Eq.~(\ref{leg_diagram}) for external leg contributions to $\beta$-decay and muon decay. According to the convention in Ref.~\cite{rosiek}, $Z_L$ is the selectron mixing matrix, $Z_{\nu}$ is the sneutrino mixing matrix, $Z_U$ is the up squark mixing matrix and $Z_D$ is the down squark mixing matrix. According to the convention in Ref.~\cite{haber_kane}, $N$ is the neutralino mixing matrix, and $U$ and $V$ are the chargino mixing matrices. We note that unlike the other mixing matrices, absolute values do not appear on $N$ in Tables~\ref{table3} and~\ref{table4}. This is because $N$ is an orthogonal matrix whose components are real.

\subsection{Vertex Corrections
\label{vertex}}

\begin{figure}[h]
\subfloat[]{
   \epsfxsize 2.50 truein \epsfbox {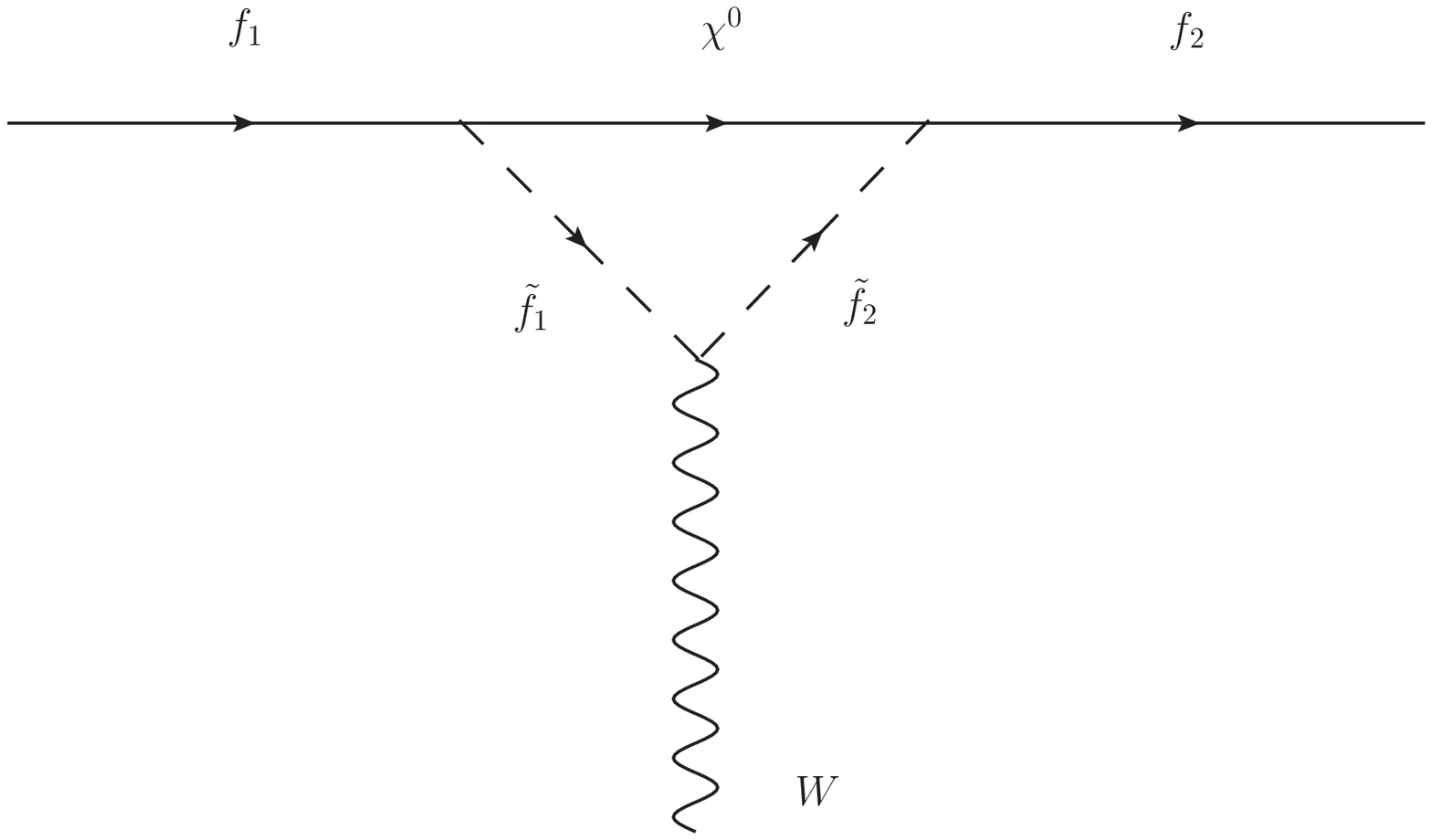}
\label{fig15a}
    }
\subfloat[]{
   \epsfxsize 2.50 truein \epsfbox {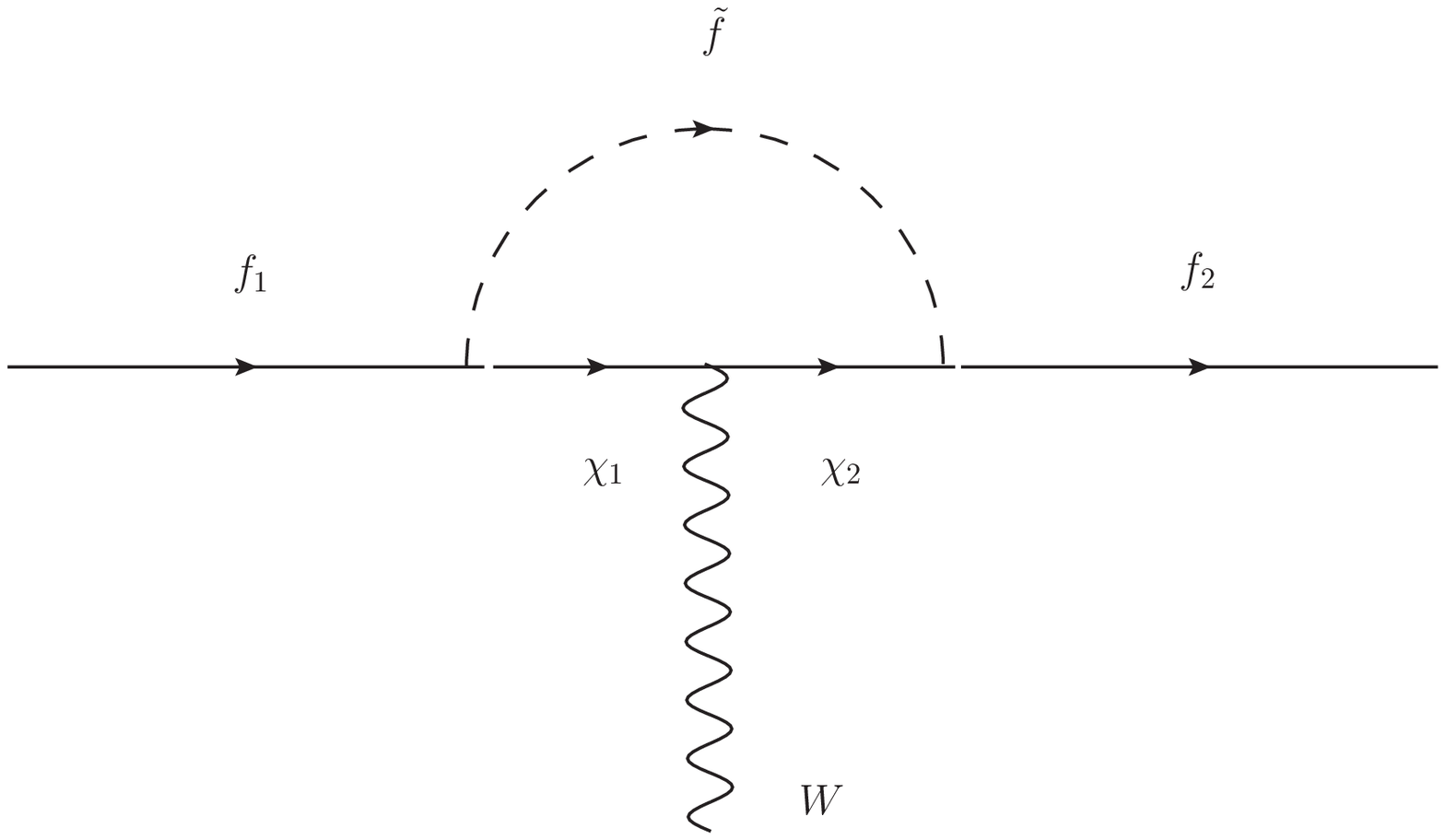}
\label{fig15b}
    }
\caption{Vertex corrections involving superpartners. The internal fermion in Fig.~\ref{fig15a} is a neutralino or a gluino. The $W$-$\tilde{f}_1$-$\tilde{f}_2$ interaction in Fig.~\ref{fig15a} is defined as $-igG_{LL}(q_1 + q_2)^{\mu}$, where $q_1$ is the momentum of $\tilde{f}_1$ and $q_2$ is the momentum of $\tilde{f}_2$. The $W$-$\chi_1$-$\chi_2$ interaction in Fig.~\ref{fig15b} is given as $ig\gamma^{\mu}(a + b\gamma^5)$. The $\tilde{f}$-$f_1$-$\chi_1$ interaction is given as $-ig(g^{1}_{L} P_L + g^{1}_{R} P_R)$, and the $\tilde{f}$-$f_2$-$\chi_2$ interaction is $-ig(g^{2*}_{L}P_R + g^{2*}_{R}P_L)$. The fermion-sfermion interactions in both Figs.~\ref{fig15a} and~\ref{fig15b} are defined according to Fig.~\ref{fig14}.}
\label{fig15}
\end{figure}

\begin{table}[h]
\caption{The values of $H$ from Eq.~(\ref{vert_left}), for the vertex graphs of the type shown in Fig.~\ref{fig15a}, for $\beta$-decay. Note that the weak coupling $g$ is replaced with the strong coupling $g_s$ for the graph involving the gluino $\Lambda_G$.}
\begin{tabular}{|c|c|c|c|c|c|}
\hline
$$ & $$ & $$ & $$ & $$ & $$ \\
$f_1$ & $f_2$ & ${\tilde{f}}^{\prime}_{1}$ & ${\tilde{f}}^{\prime}_{2}$ & $\chi^0$ & $H$ \\
$$ & $$ & $$ & $$ & $$ & $$ \\
\hline
$$ & $$ & $$ & $$ & $$ & $$ \\
$d$ & $u$ & $\tilde{d}_i$ & $\tilde{u}_{i^{\prime}}$ & $\chi^{0}_{j}$ & $-\frac{1}{2\sqrt{2}}Z_{D}^{ki*}Z_{D}^{1i}Z_{U}^{ki^{\prime}*}Z_{U}^{1i^{\prime}}$ \\
$$ & $$ & $$ & $$ & $$ & $(N_{j2} + \tan \theta_W N_{j1}/3)(-N_{j2} + \tan \theta_W N_{j1}/3)$ \\
$$ & $$ & $$ & $$ & $$ & $$ \\
\hline
$$ & $$ & $$ & $$ & $$ & $$ \\
$d$ & $u$ & $\tilde{d}_i$ & $\tilde{u}_j$ & $\Lambda_G$ & $-\frac{8}{3\sqrt{2}}Z_{U}^{*kj}Z_{U}^{1j}Z_{D}^{*ki}Z_{D}^{1i}$ \\
$$ & $$ & $$ & $$ & $$ & $$ \\
\hline
\end{tabular}
\label{table5}
\end{table}

\begin{table}[h]
\caption{The value of $H$ from Eq.~(\ref{vert_left}), for the vertex graph of the type shown in Fig.~\ref{fig15a}, for muon decay.}
\begin{tabular}{|c|c|c|c|c|c|}
\hline
$$ & $$ & $$ & $$ & $$ & $$ \\
$f_1$ & $f_2$ & ${\tilde{f}}^{\prime}_{1}$ & ${\tilde{f}}^{\prime}_{2}$ & $\chi^0$ & $H$ \\
$$ & $$ & $$ & $$ & $$ & $$ \\
\hline
$$ & $$ & $$ & $$ & $$ & $$ \\
$\mu$ & $\nu_{\mu}$ & $\tilde{\ell}_i$ & $\tilde{\nu}_J$ & $\chi^{0}_{j}$ & $\frac{1}{2\sqrt{2}}Z_{\nu}^{kJ*}Z_{\nu}^{2J}Z_{L}^{ki*}Z_{L}^{2i}(N_{j2} - \tan \theta_W N_{j1})(N_{j2} + \tan \theta_W N_{j1})$ \\
$$ & $$ & $$ & $$ & $$ & $$ \\
\hline
\end{tabular}
\label{table6}
\end{table}

\begin{table}[h]
\caption{The values of $\Lambda_1$ from Eq.~(\ref{vert_right}), for the vertex graph shown in Fig.~\ref{fig15b}, for $\beta$-decay.}
\begin{tabular}{|c|c|c|c|c|c|}
\hline
$$ & $$ & $$ & $$ & $$ & $$ \\
$f_1$ & $f_2$ & $\tilde{f}$ & $\chi_1$ & $\chi_2$ & $\Lambda_1$ \\
$$ & $$ & $$ & $$ & $$ & $$ \\
\hline
$$ & $$ & $$ & $$ & $$ & $$ \\
$d$ & $u$ & $\tilde{u}_i$ & $\chi_{j}^{-}$ & $\chi_{j^{\prime}}^{0}$ & $\frac{1}{2\sqrt{2}}V_{j1}^{*}|Z_{U}^{1i}|^2 (N_{j^{\prime}2} + \tan \theta_W N_{j^{\prime}1}/3)$ \\
$$ & $$ & $$ & $$ & $$ & $(-N_{j^{\prime}2}V_{j1} + N_{j^{\prime}4}V_{j2}/\sqrt{2} - N_{j^{\prime}2}U_{j1} - N_{j^{\prime}3}U_{j2}/\sqrt{2}$ \\
$$ & $$ & $$ & $$ & $$ & $+ N_{j^{\prime}2}U_{j1} + N_{j^{\prime}3}U_{j2}/\sqrt{2} - N_{j^{\prime}2}V_{j1} + N_{j^{\prime}4}V_{j2}/\sqrt{2})$ \\
$$ & $$ & $$ & $$ & $$ & $$ \\
\hline
$$ & $$ & $$ & $$ & $$ & $$ \\
$d$ & $u$ & $\tilde{d}_i$ & $\chi^{0}_{j}$ & $\chi_{j^{\prime}}$ & $\frac{1}{2\sqrt{2}}U_{j^{\prime}1}|Z_{D}^{1i}|^2(-N_{j2} + \tan \theta_W N_{j1}/3)$ \\
$$ & $$ & $$ & $$ & $$ & $(N_{j 2}V^{*}_{j^{\prime} 1} - N_{j 4}V^{*}_{j^{\prime} 2}/\sqrt{2} + N_{j 2}U^{*}_{j^{\prime} 1} + N_{j 3}U^{*}_{j^{\prime} 2}/\sqrt{2}$ \\
$$ & $$ & $$ & $$ & $$ & $+ N_{j 2}U^{*}_{j^{\prime} 1} + N_{j 3}U^{*}_{j^{\prime} 2}/\sqrt{2} - N_{j 2}V^{*}_{j^{\prime} 1} + N_{j 4}V^{*}_{j^{\prime} 2}/\sqrt{2})$ \\
$$ & $$ & $$ & $$ & $$ & $$ \\
\hline
\end{tabular}
\label{table7}
\end{table}

\begin{table}[h]
\caption{The values of $\Lambda_4$ from Eq.~(\ref{vert_right}), for the vertex graph shown in Fig.~\ref{fig15b}, for $\beta$-decay.}
\begin{tabular}{|c|c|c|c|c|c|}
\hline
$$ & $$ & $$ & $$ & $$ & $$ \\
$f_1$ & $f_2$ & $\tilde{f}$ & $\chi_1$ & $\chi_2$ & $\Lambda_4$ \\
$$ & $$ & $$ & $$ & $$ & $$ \\
\hline
$$ & $$ & $$ & $$ & $$ & $$ \\
$d$ & $u$ & $\tilde{u}_i$ & $\chi_{j}^{-}$ & $\chi_{j^{\prime}}^{0}$ & $\frac{1}{2\sqrt{2}}V_{j1}^{*}|Z_{U}^{1i}|^2 (N_{j^{\prime}2} + \tan \theta_W N_{j^{\prime}1}/3)$ \\
$$ & $$ & $$ & $$ & $$ & $(-N_{j^{\prime}2}V_{j1} + N_{j^{\prime}4}V_{j2}/\sqrt{2} - N_{j^{\prime}2}U_{j1} - N_{j^{\prime}3}U_{j2}/\sqrt{2}$ \\
$$ & $$ & $$ & $$ & $$ & $- N_{j^{\prime}2}U_{j1} - N_{j^{\prime}3}U_{j2}/\sqrt{2} + N_{j^{\prime}2}V_{j1} - N_{j^{\prime}4}V_{j2}/\sqrt{2})$ \\
$$ & $$ & $$ & $$ & $$ & $$ \\
\hline
$$ & $$ & $$ & $$ & $$ & $$ \\
$d$ & $u$ & $\tilde{d}_i$ & $\chi^{0}_{j}$ & $\chi_{j^{\prime}}$ & $\frac{1}{2\sqrt{2}}U_{j^{\prime}1}|Z_{D}^{1i}|^2(-N_{j2} + \tan \theta_W N_{j1}/3)$ \\
$$ & $$ & $$ & $$ & $$ & $(N_{j2}V^{*}_{j^{\prime}1} - N_{j4}V^{*}_{j^{\prime}2}/\sqrt{2} + N_{j2}U^{*}_{j^{\prime}1} + N_{j3}U^{*}_{j^{\prime}2}/\sqrt{2}$ \\
$$ & $$ & $$ & $$ & $$ & $-N_{j2}U^{*}_{j^{\prime}1} - N_{j3}U^{*}_{j^{\prime}2}/\sqrt{2} + N_{j2}V^{*}_{j^{\prime}1} - N_{j4}V^{*}_{j^{\prime}2}/\sqrt{2})$ \\
$$ & $$ & $$ & $$ & $$ & $$ \\
\hline
\end{tabular}
\label{table8}
\end{table}

\begin{table}[h]
\caption{The values of $\Lambda_1$ from Eq.~(\ref{vert_right}), for the vertex graph shown in Fig.~\ref{fig15b}, for muon decay.}
\begin{tabular}{|c|c|c|c|c|c|}
\hline
$$ & $$ & $$ & $$ & $$ & $$ \\
$f_1$ & $f_2$ & $\tilde{f}$ & $\chi_1$ & $\chi_2$ & $\Lambda_1$ \\
$$ & $$ & $$ & $$ & $$ & $$ \\
\hline
$$ & $$ & $$ & $$ & $$ & $$ \\
$\mu$ & $\nu_{\mu}$ & $\tilde{\nu}_J$ & $\chi^{-}_{j}$ & $\chi^{0}_{j^{\prime}}$ & $\frac{1}{2\sqrt{2}}V_{j1}^{*}|Z_{\nu}^{2J}|^2 (N_{j^{\prime}2} - \tan \theta_W N_{j^{\prime}1})$ \\
$$ & $$ & $$ & $$ & $$ & $(-N_{j^{\prime}2}V_{j1} + N_{j^{\prime}4}V_{j2}/\sqrt{2} - N_{j^{\prime}2}U_{j1} - N_{j^{\prime}3}U_{j2}/\sqrt{2}$ \\
$$ & $$ & $$ & $$ & $$ & $+ N_{j^{\prime}2}U_{j1} + N_{j^{\prime}3}U_{j2}/\sqrt{2} - N_{j^{\prime}2}V_{j1} + N_{j^{\prime}4}V_{j2}/\sqrt{2})$ \\
$$ & $$ & $$ & $$ & $$ & $$ \\
\hline
$$ & $$ & $$ & $$ & $$ & $$ \\
$\mu$ & $\nu_{\mu}$ & $\tilde{\ell}_i$ & $\chi^{0}_{j}$ & $\chi_{j^{\prime}}$ & $-\frac{1}{2\sqrt{2}}U_{j^{\prime}1}|Z_{L}^{2i}|^2 (N_{j2} + \tan \theta_W N_{j1})$ \\
$$ & $$ & $$ & $$ & $$ & $(N_{j2}V^{*}_{j^{\prime}1} - N_{j4}V^{*}_{j^{\prime}2}/\sqrt{2} + N_{j2}U^{*}_{j^{\prime}1} + N_{j3}U^{*}_{j^{\prime}2}/\sqrt{2}$ \\
$$ & $$ & $$ & $$ & $$ & $+ N_{j2}U^{*}_{j^{\prime}1} + N_{j3}U^{*}_{j^{\prime}2}/\sqrt{2} - N_{j2}V^{*}_{j^{\prime}1} + N_{j4}V^{*}_{j^{\prime}2}/\sqrt{2})$ \\
$$ & $$ & $$ & $$ & $$ & $$ \\
\hline
\end{tabular}
\label{table9}
\end{table}

\begin{table}[h]
\caption{The values of $\Lambda_4$ from Eq.~(\ref{vert_right}), for the vertex graph shown in Fig.~\ref{fig15b}, for muon decay.}
\begin{tabular}{|c|c|c|c|c|c|}
\hline
$$ & $$ & $$ & $$ & $$ & $$ \\
$f_1$ & $f_2$ & $\tilde{f}$ & $\chi_1$ & $\chi_2$ & $\Lambda_4$ \\
$$ & $$ & $$ & $$ & $$ & $$ \\
\hline
$$ & $$ & $$ & $$ & $$ & $$ \\
$\mu$ & $\nu_{\mu}$ & $\tilde{\nu}_J$ & $\chi^{-}_{j}$ & $\chi^{0}_{j^{\prime}}$ & $\frac{1}{2\sqrt{2}}V_{j1}^{*}|Z_{\nu}^{2J}|^2 (N_{j^{\prime}2} - \tan \theta_W N_{j^{\prime}1})$ \\
$$ & $$ & $$ & $$ & $$ & $(-N_{j^{\prime}2}V_{j1} + N_{j^{\prime}4}V_{j2}/\sqrt{2} - N_{j^{\prime}2}U_{j1} - N_{j^{\prime}3}U_{j2}/\sqrt{2}$ \\
$$ & $$ & $$ & $$ & $$ & $- N_{j^{\prime}2}U_{j1} - N_{j^{\prime}3}U_{j2}/\sqrt{2} + N_{j^{\prime}2}V_{j1} - N_{j^{\prime}4}V_{j2}/\sqrt{2})$ \\
$$ & $$ & $$ & $$ & $$ & $$ \\
\hline
$$ & $$ & $$ & $$ & $$ & $$ \\
$\mu$ & $\nu_{\mu}$ & $\tilde{\ell}_i$ & $\chi^{0}_{j}$ & $\chi_{j^{\prime}}$ & $-\frac{1}{2\sqrt{2}}U_{j^{\prime}1}|Z_{L}^{2i}|^2 (N_{j2} + \tan \theta_W N_{j1})$ \\
$$ & $$ & $$ & $$ & $$ & $(N_{j2}V^{*}_{j^{\prime}1} - N_{j4}V^{*}_{j^{\prime}2}/\sqrt{2} + N_{j2}U^{*}_{j^{\prime}1} + N_{j3}U^{*}_{j^{\prime}2}/\sqrt{2}$ \\
$$ & $$ & $$ & $$ & $$ & $- N_{j2}U^{*}_{j^{\prime}1} - N_{j3}U^{*}_{j^{\prime}2}/\sqrt{2} + N_{j2}V^{*}_{j^{\prime}1} - N_{j4}V^{*}_{j^{\prime}2}/\sqrt{2})$ \\
$$ & $$ & $$ & $$ & $$ & $$ \\
\hline
\end{tabular}
\label{table10}
\end{table}

Generic vertex graph corrections are shown in Fig.~\ref{fig15}. The  correction due to the vertex graph in Fig.~\ref{fig15a} has the general form,
\beqn
\Delta_{\rm vertex}^{\rm (15a)} & = & -\frac{g^2}{(4\pi)^2}\sqrt{2}H \Bigg\{ \frac{\alpha_{\epsilon}}{2} \nonumber \\
&& - \int_{0}^{1}du \int_{0}^{1}d \omega \omega \log \left[ \left( \omega \bar{m}_{\tilde{f}}^{2} + (1 - \omega)m_{\chi}^2 + \Delta m_{\tilde{f}}^{2} \omega (1 - 2u)/2 \right)/\mu^2 \right] \Bigg\}~, \nonumber \\
\label{vert_left}
\eeqn
where
\beq
\bar{m}_{\tilde{f}}^{2} = \frac{m_{\tilde{f}_{1}}^{2} + m_{\tilde{f}_{2}}^{2}}{2}~,
\label{bar_m_f}
\eeq
\beq
\Delta m_{\tilde{f}}^{2} = m_{\tilde{f}_{2}}^{2} - m_{\tilde{f}_{1}}^{2}~,
\label{Delt_m_f}
\eeq
$m_{\tilde{f}_1}$ is the mass of the $\tilde{f}_1$-particle in Fig.~\ref{fig15a}, $m_{\tilde{f}_2}$ is the mass of the $\tilde{f}_2$-particle, and $m_{\chi}$ is the mass of the $\chi$-particle. The values of $H$ for different vertex graphs are given in Tables~\ref{table5} and~\ref{table6}.

The  correction due to the vertex graph in Fig.~\ref{fig15b} has the form,
\beqn
\Delta_{\rm vertex}^{\rm (15b)} & = & -\frac{g^2}{(4\pi)^2}\sqrt{2}\Bigg[ \Lambda_1 \left( \frac{\alpha_{\epsilon}}{2} - \int_{0}^{1}du \int_{0}^{1}d \omega \omega \log (D/\mu^2) \right) \nonumber \\
&& + m_{\chi_1} m_{\chi_2} \Lambda_4 \int_{0}^{1}du \int_{0}^{1}d \omega \frac{\omega}{D} \Bigg]~,
\label{vert_right}
\eeqn
where
\beq
D = (1 - \omega)m_{\tilde{f}}^{2} + \omega \bar{m}_{\chi}^2 + \omega (1 - 2u) \Delta m_{\chi}^2 /2~,
\label{vert_right_D}
\eeq
where $m_{\tilde{f}}$ is the mass of the sfermion $\tilde{f}$ in the figure,
\beq
\bar{m}_{\chi}^{2} = \frac{m_{\chi_1}^{2} + m_{\chi_2}^{2}}{2}~,
\label{bar_m_chi}
\eeq
\beq
\Delta m_{\chi}^{2} = m_{\chi_2}^{2} - m_{\chi_1}^{2}~,
\label{Delt_m_chi}
\eeq
$m_{\chi_1}$ is the mass of the $\chi_1$-particle, and $m_{\chi_2}$ is the mass of the $\chi_2$-particle. The quantities $\Lambda_1$ and $\Lambda_4$ are given as
\beq
\Lambda_1 = g^{2*}_{L}g^{1}_{L}(a + b)
\label{Lambda1}
\eeq
and
\beq
\Lambda_4 = g^{2*}_{L}g^{1}_{L}(a - b)~,
\label{Lambda4}
\eeq
where $a$ and $b$ are interaction terms described in the caption of Fig.~\ref{fig15}. The coupling $g_L$ is defined in Fig.~\ref{fig14}. The quantities $g^{1}_{L}$ and $g^{2}_{L}$ represent $g_L$ for the $f_1$ and $f_2$ interactions in Fig.~\ref{fig15b}. The values of $\Lambda_1$ and $\Lambda_4$ for different graphs are given in Tables~\ref{table7}--\ref{table10}.

\subsection{Box Graph Corrections
\label{box}}

\begin{figure}[p]
\subfloat[]{
   \epsfxsize 2.85 truein \epsfbox {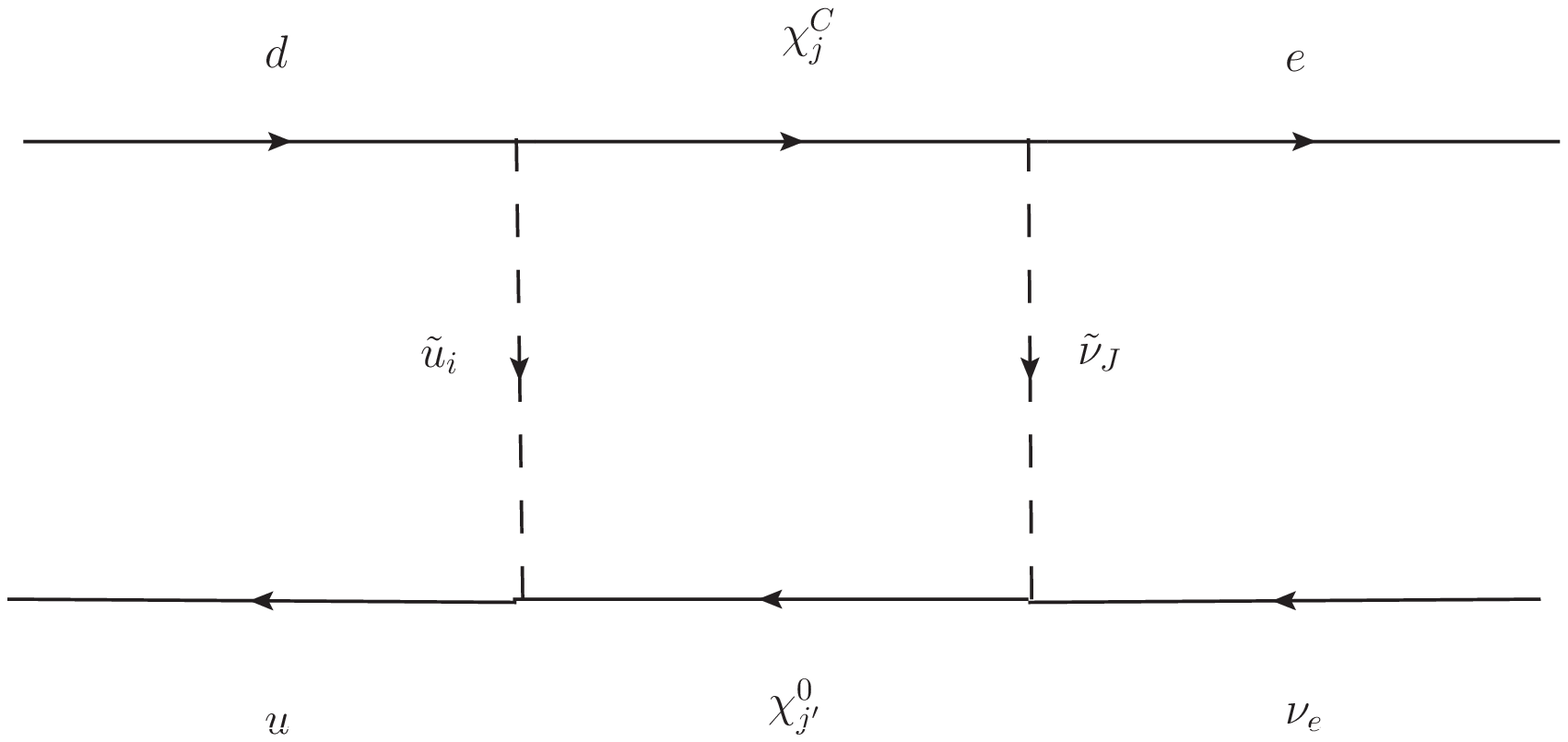}
\label{fig16a}
    }
\subfloat[]{
   \epsfxsize 2.85 truein \epsfbox {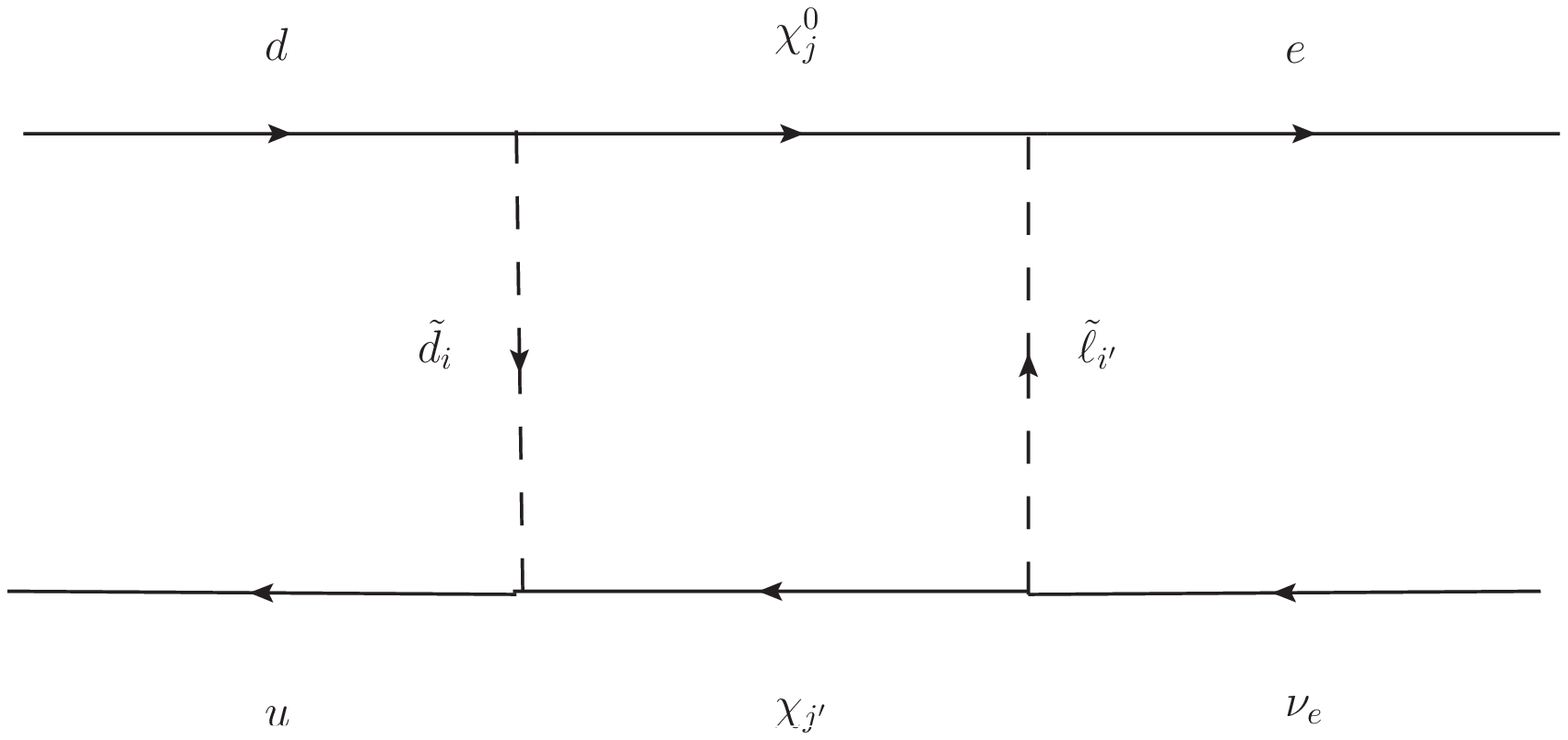}
\label{fig16b}
    }
\newline
\subfloat[]{
   \epsfxsize 2.85 truein \epsfbox {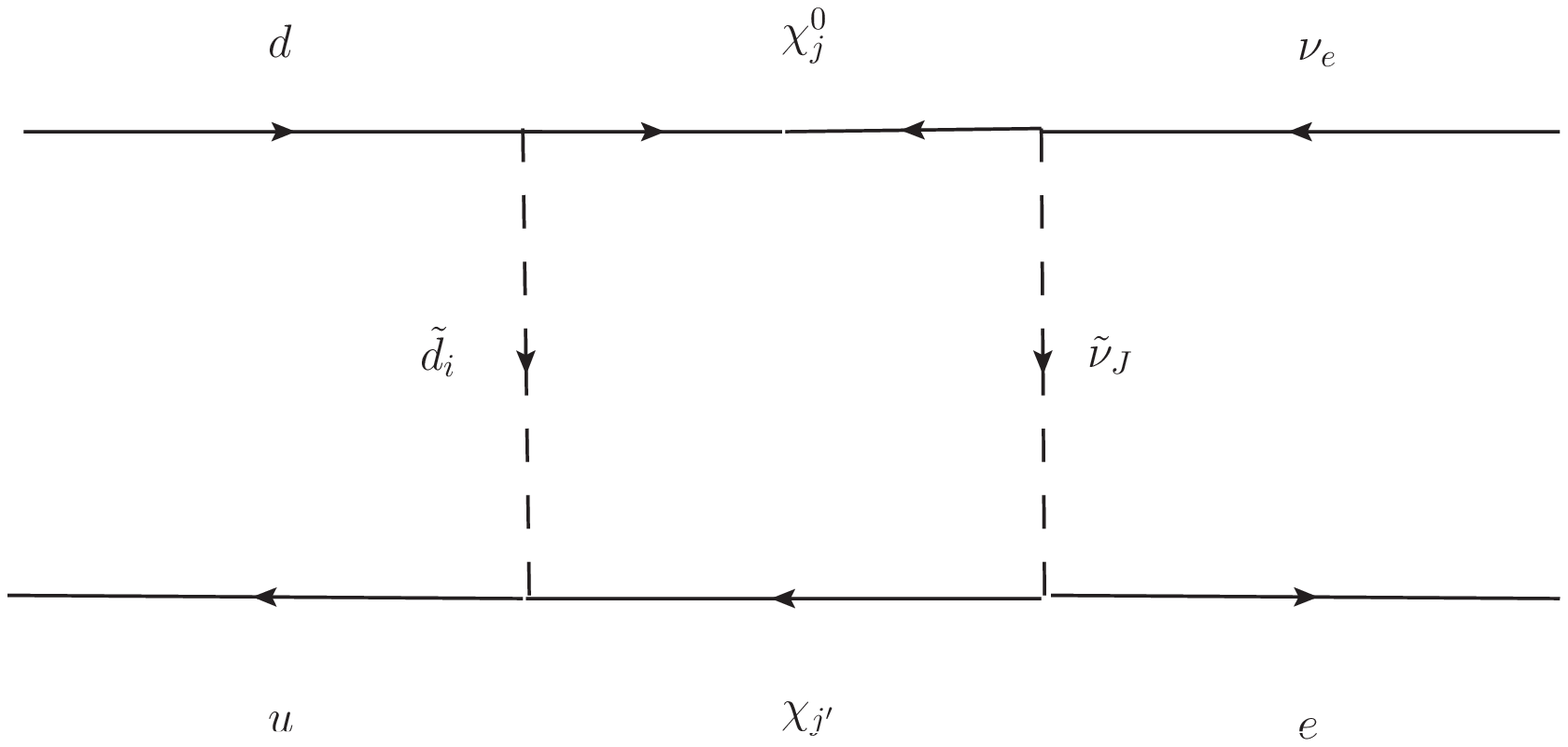}
\label{fig16c}
    }
\subfloat[]{
   \epsfxsize 2.85 truein \epsfbox {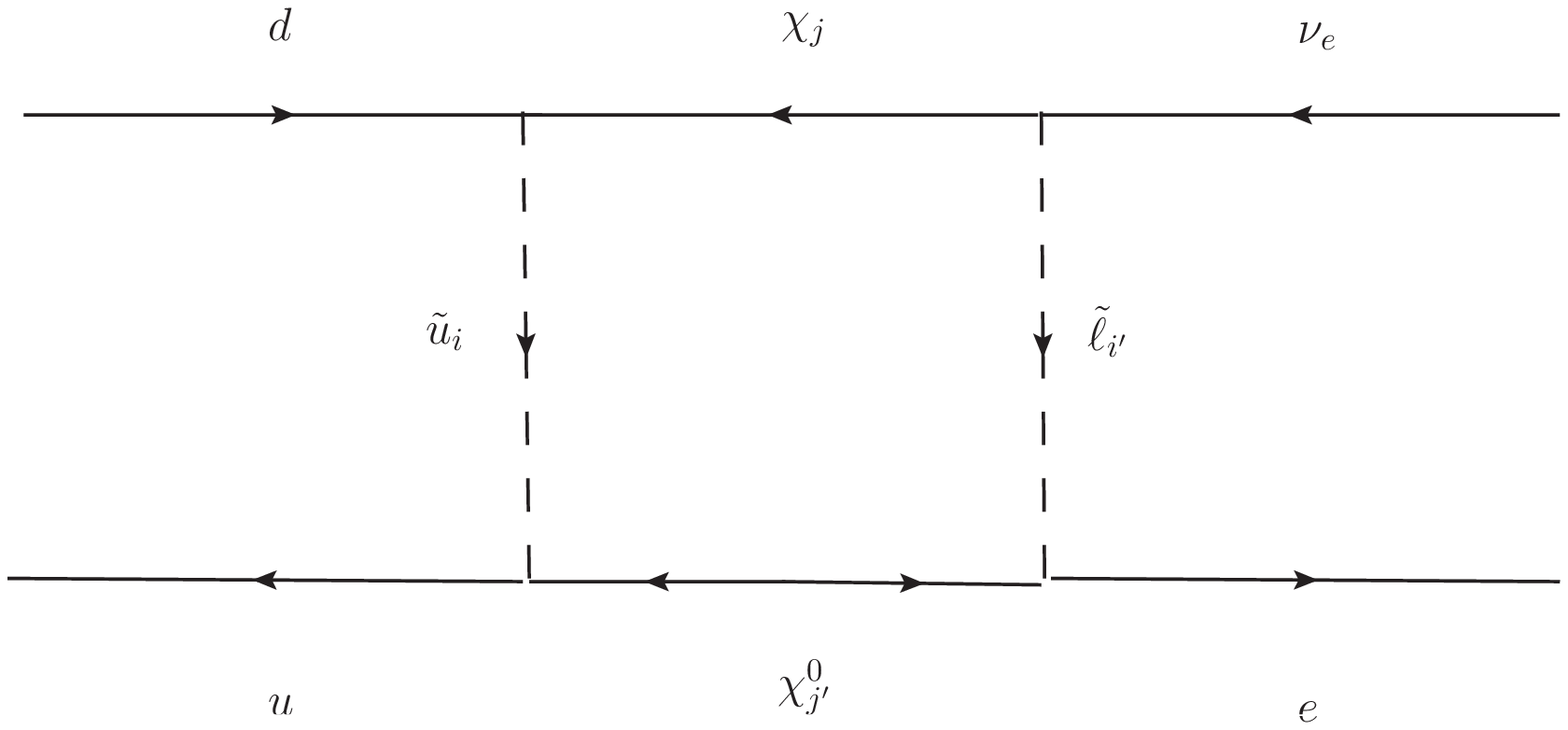}
\label{fig16d}
    }
\caption{Box graph contributions to $\Delta r^{(V)}_{\beta}$.}
\label{fig16}
\end{figure}

\begin{figure}[p]
\subfloat[]{
   \epsfxsize 2.85 truein \epsfbox {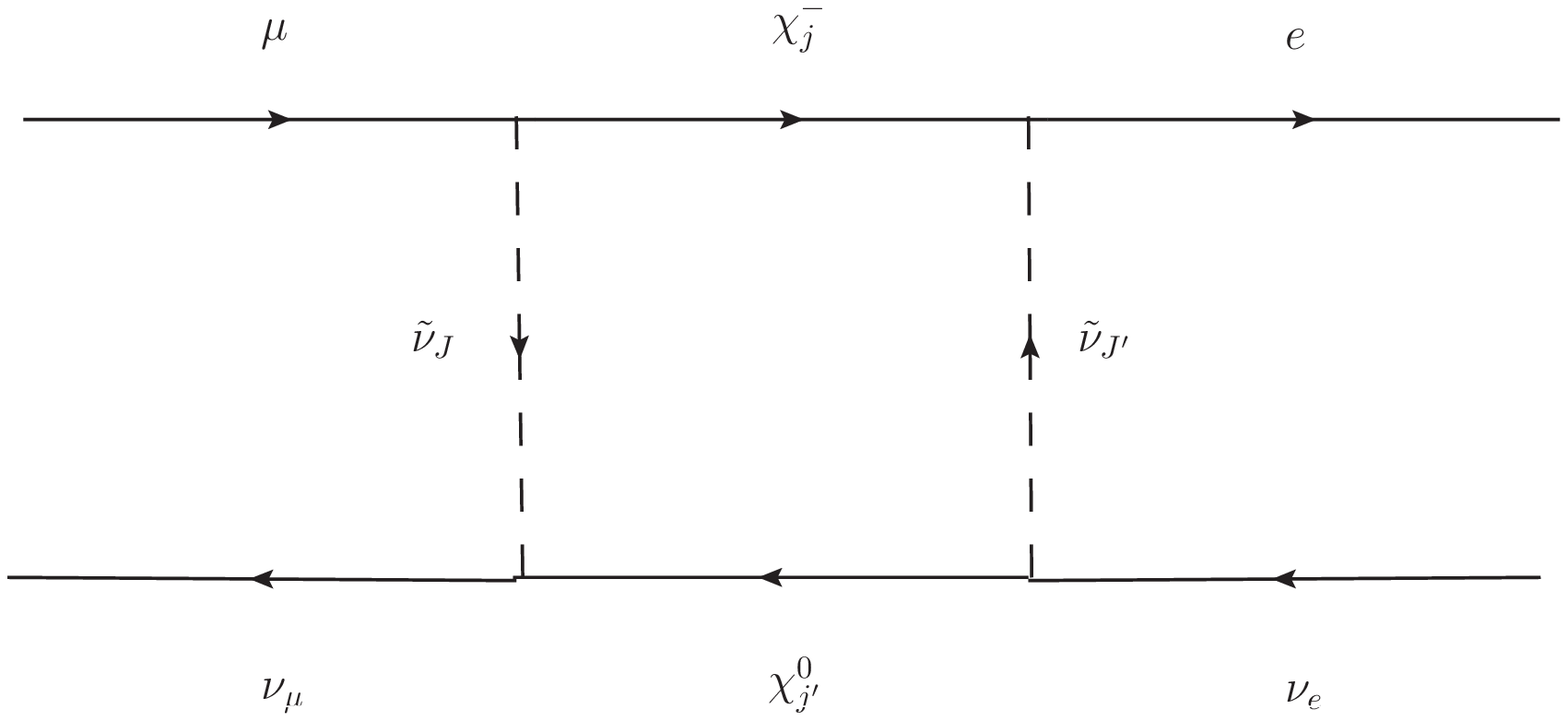}
\label{fig17a}
    }
\subfloat[]{
   \epsfxsize 2.85 truein \epsfbox {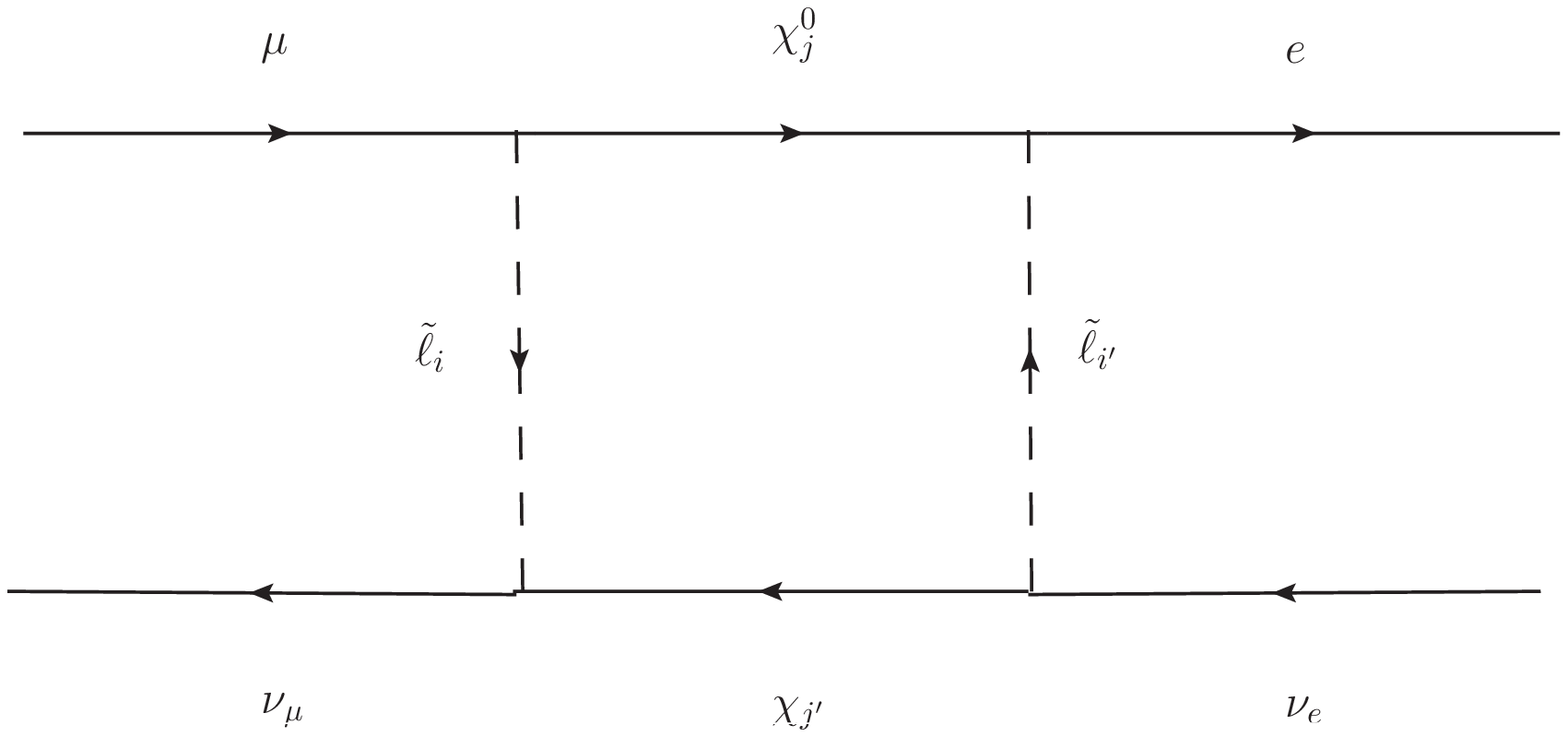}
\label{fig17b}
    }
\newline
\subfloat[]{
   \epsfxsize 2.85 truein \epsfbox {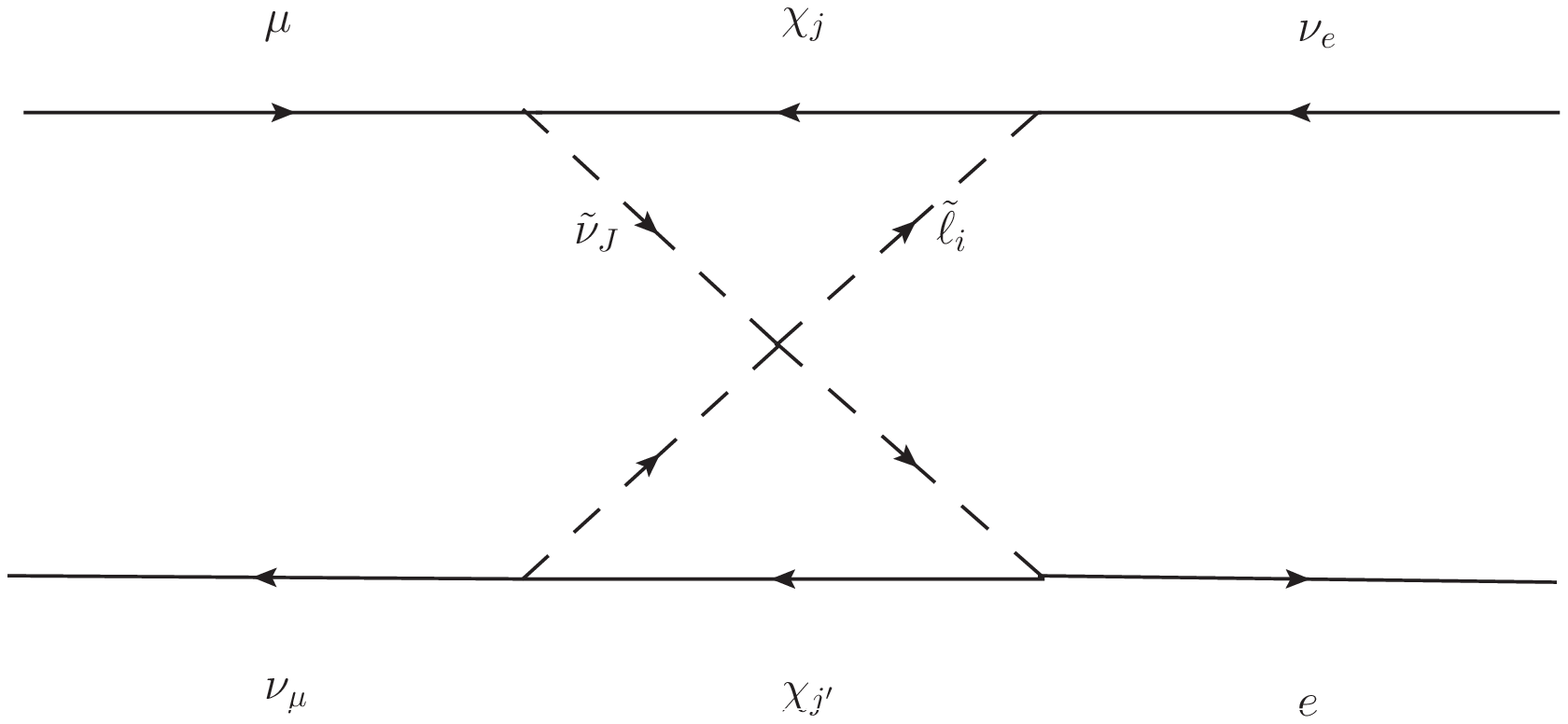}
\label{fig17c}
    }
\subfloat[]{
   \epsfxsize 2.85 truein \epsfbox {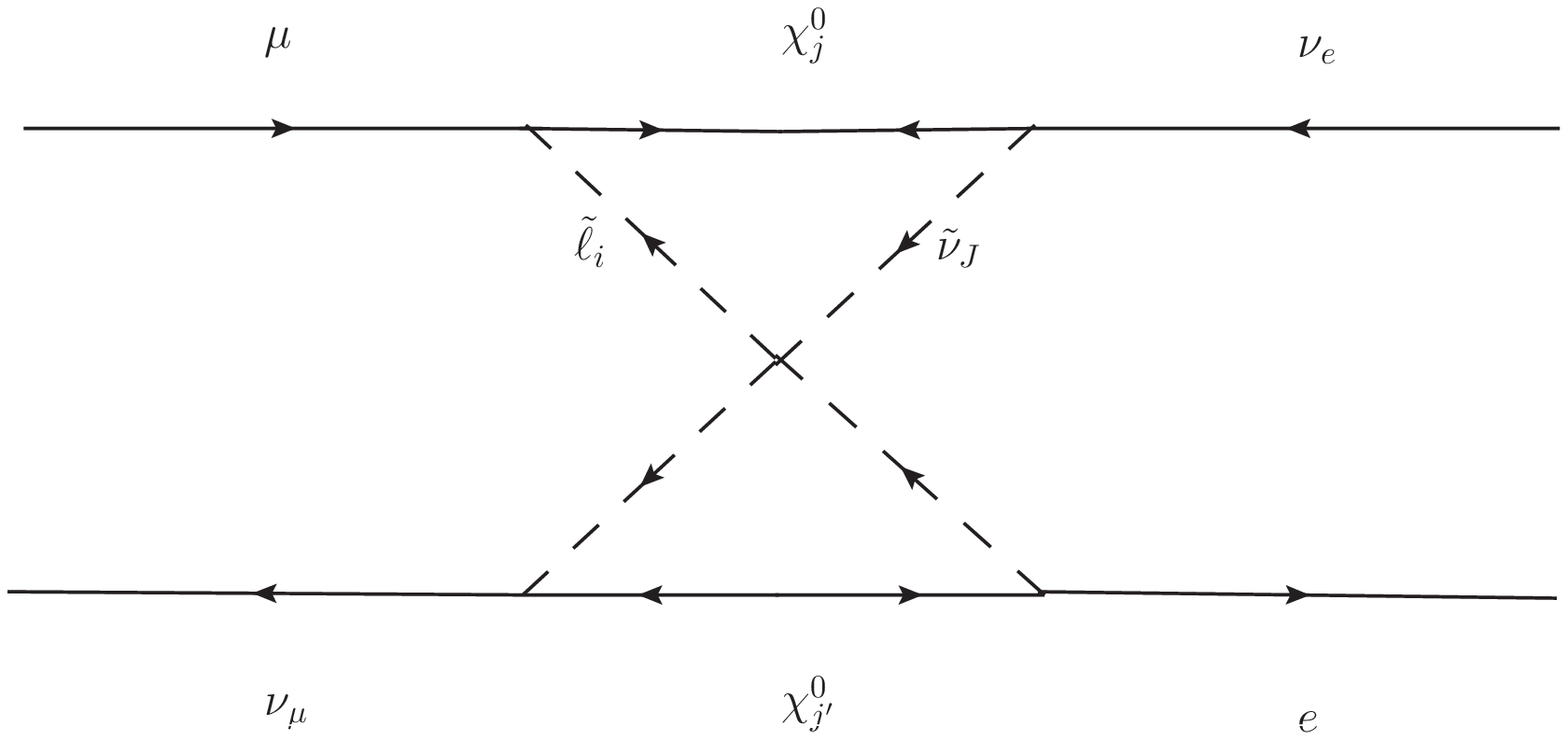}
\label{fig17d}
    }
\newline
\subfloat[]{
   \epsfxsize 2.85 truein \epsfbox {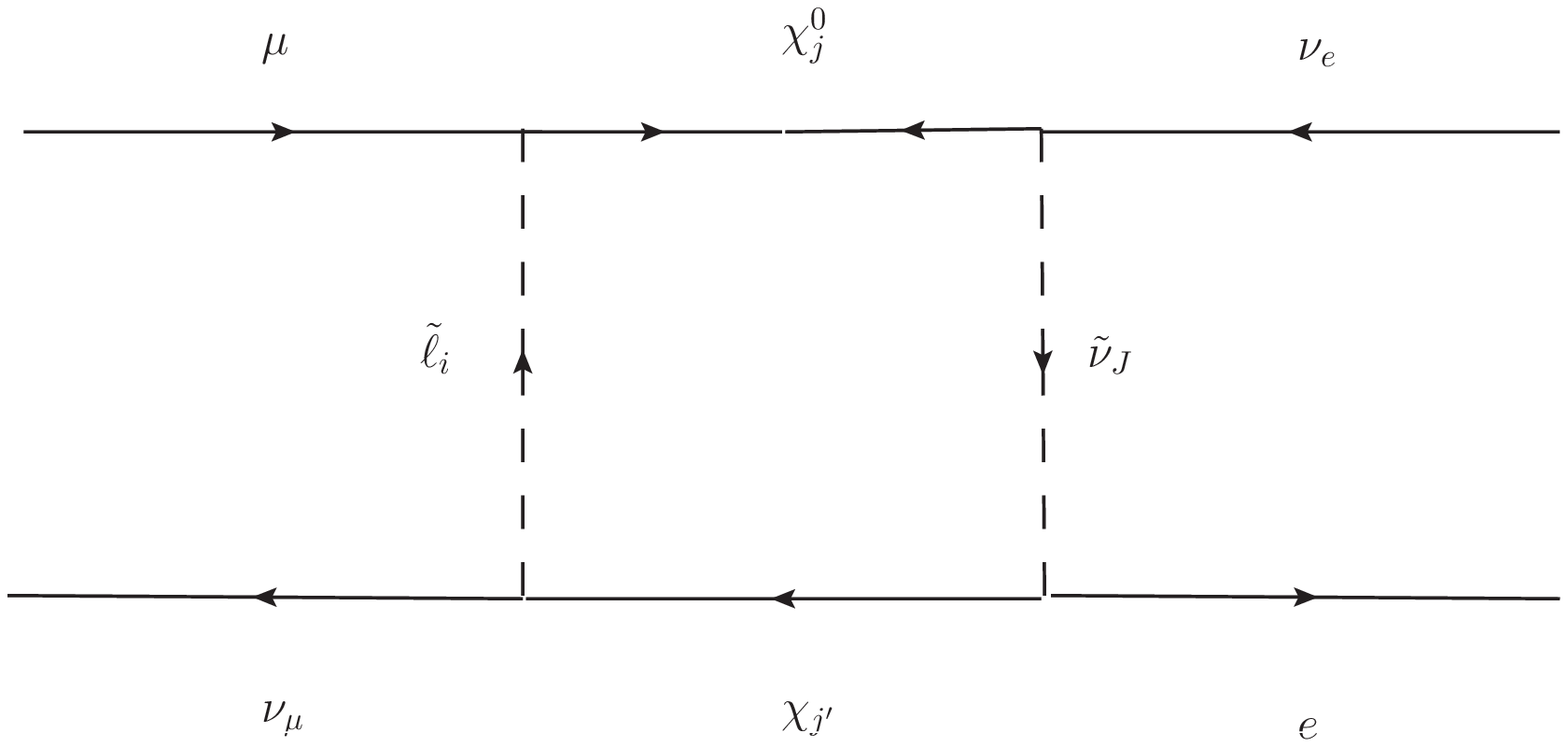}
\label{fig17e}
    }
\subfloat[]{
   \epsfxsize 2.85 truein \epsfbox {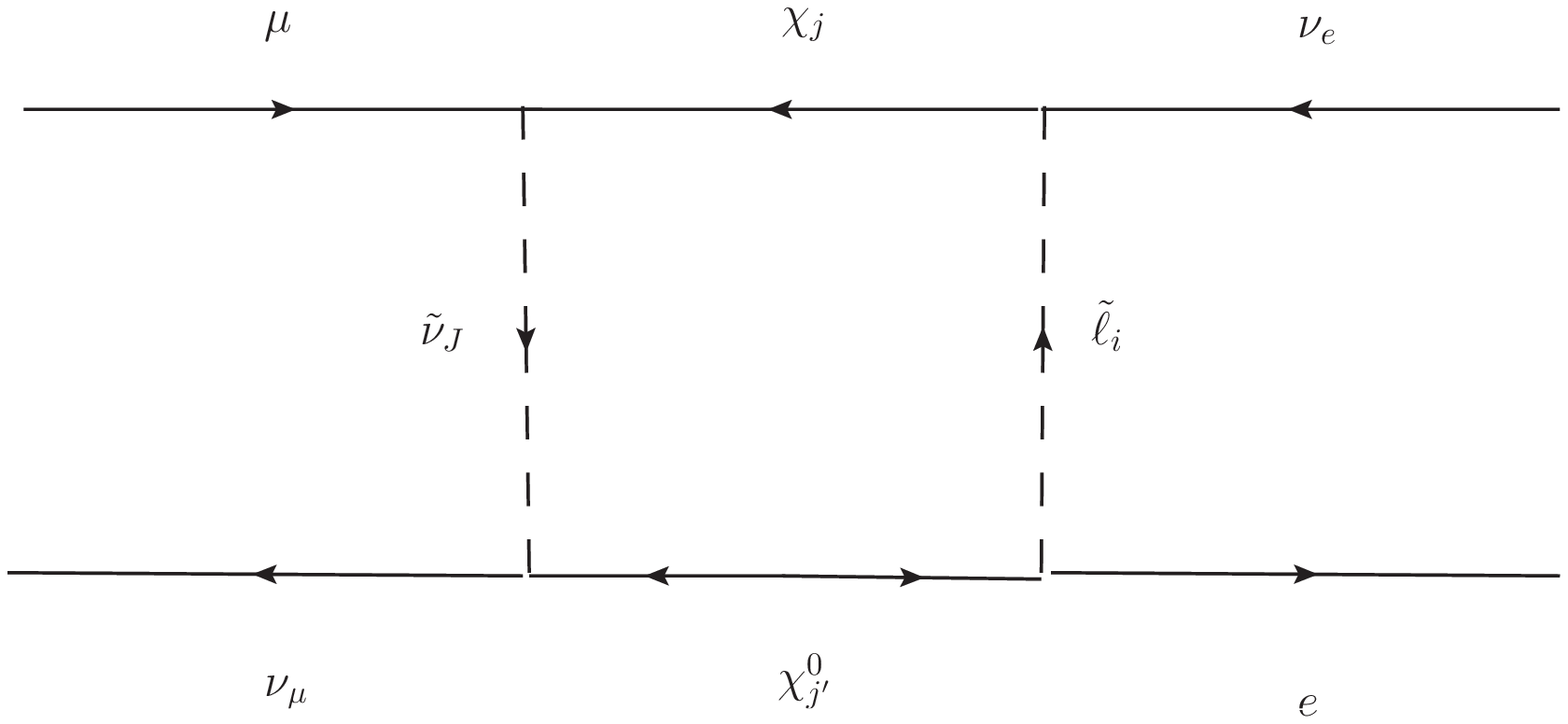}
\label{fig17f}
    }
\newline
\subfloat[]{
   \epsfxsize 2.85 truein \epsfbox {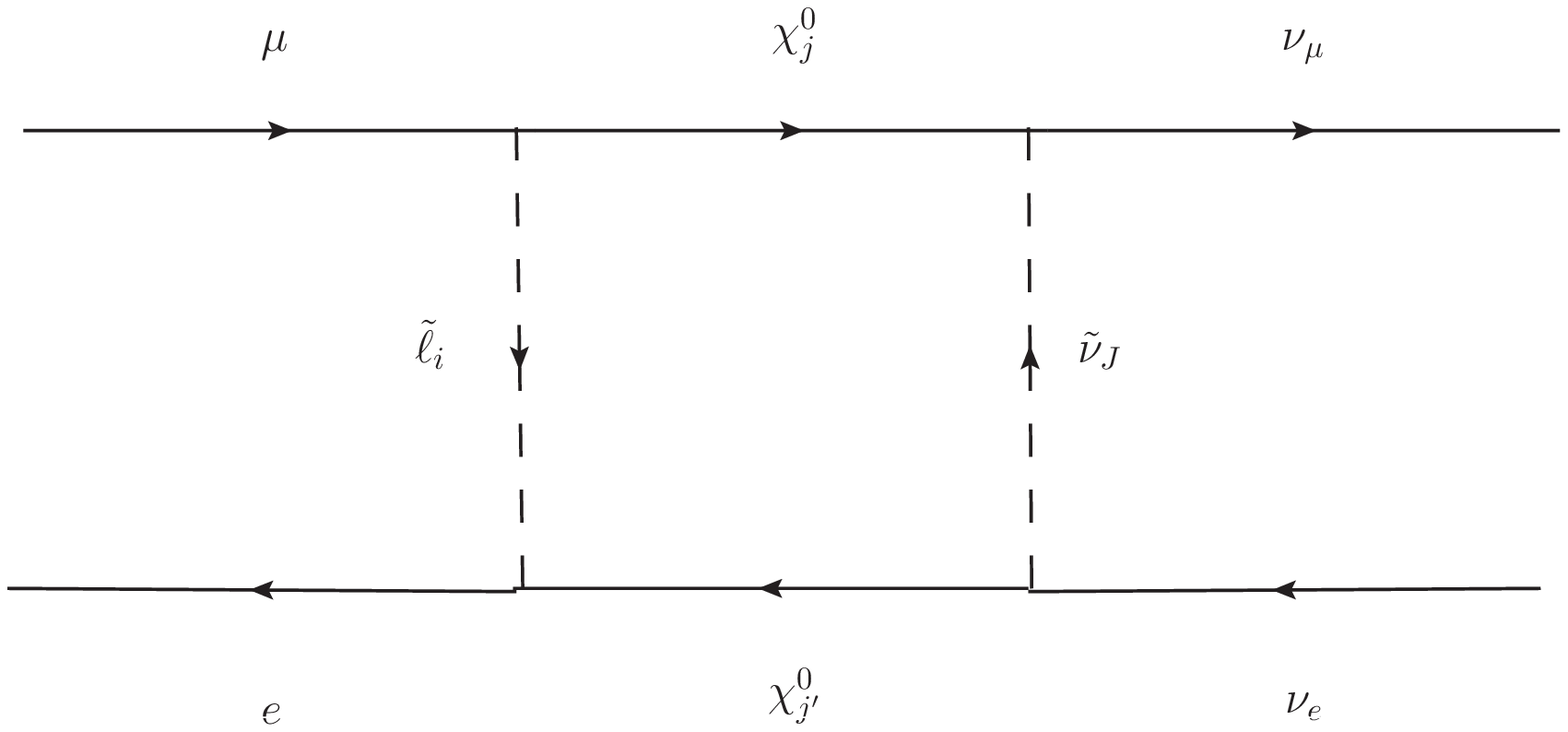}
\label{fig17g}
    }
\caption{Box graph contributions to $\Delta r_{\mu}$.}
\label{fig17}
\end{figure}

At one-loop order, there are box graph corrections to $\Delta r^{(V)}_{\beta}$ and $\Delta r_{\mu}$, as shown in Figs.~\ref{fig16} and~\ref{fig17}. The individual box graph contributions $\Delta_{\rm box}$  for either $\beta$-decay or muon decay take one of two forms:
\beqn
\Delta_{\rm box}^{\rm (diagram)} & = & \frac{g^2}{(4\pi)^2} Am_{W}^{2}m_{\chi}m_{\chi^{\prime}}\int_{0}^{1}dx\int_{0}^{1-x}dy\int_{0}^{1-x-y}dz \nonumber \\
&& \left[ xm_{\chi}^{2} + ym_{\chi^{\prime}}^{2} + zm_{\tilde{f}}^{2} + (1 - x - y - z)m_{\tilde{f}^{\prime}}^{2} \right]^{-2}
\label{box1}
\eeqn
or
\beqn
\Delta_{\rm box}^{\rm (diagram)} & = & \frac{g^2}{(4\pi)^2}Bm_{W}^{2}\int_{0}^{1}dx\int_{0}^{1-x}dy\int_{0}^{1-x-y}dz \nonumber \\
&& \left[ xm_{\chi}^{2} + ym_{\chi^{\prime}}^{2} + zm_{\tilde{f}}^{2} + (1 - x - y - z)m_{\tilde{f}^{\prime}}^{2} \right]^{-1}~,
\label{box2}
\eeqn
depending on the individual graph. Both equations describe box graphs for muon decay and for $\beta$-decay. The coefficients $A$ and $B$ are products of mixing matrices. The values of $A$ and $B$ are given in Tables~\ref{table11}--\ref{table14}.

\begin{table}[htbp]
\caption{The values of $A$ from Eq.~(\ref{box1}), for the $\beta$-decay box graphs in Fig.~\ref{fig16}.}
\begin{tabular}{|c|c|}
\hline
$$ & $$ \\
Fig. & $A$ \\
$$ & $$ \\
\hline
$$ & $$ \\
\ref{fig16c} & $-V^{*}_{j^{\prime}1}U_{j^{\prime}1}Z_{\nu}^{1J*}Z_{\nu}^{1J*}|Z_{D}^{1i}|^2 (N_{j2} - \tan \theta_W N_{j1})(-N_{j2} + \tan \theta_W N_{j1}/3)$ \\
$$ & $$ \\
\hline
$$ & $$ \\
\ref{fig16d} & $U^{*}_{j1}V^{*}_{j1}|Z_{L}^{1i^{\prime}}|^2 Z_{U}^{1i}Z_{U}^{1i}(N_{j^{\prime}2} + \tan \theta_W N_{j^{\prime}1})(N_{j^{\prime}2} + \tan \theta_W N_{j^{\prime}1}/3)$ \\
$$ & $$ \\
\hline
\end{tabular}
\label{table11}
\end{table}

\begin{table}[htbp]
\caption{The values of $B$ from Eq.~(\ref{box2}), for the $\beta$-decay box graphs in Fig.~\ref{fig16}.}
\begin{tabular}{|c|c|}
\hline
$$ & $$ \\
Fig. & $B$ \\
$$ & $$ \\
\hline
$$ & $$ \\
\ref{fig16a} & $|V_{j1}|^2 Z_{\nu}^{1J}Z_{U}^{1i*}Z_{U}^{1i}(N_{j^{\prime}2} + \tan \theta_W N_{j^{\prime}1}/3)Z_{\nu}^{1J*}(N_{j^{\prime}2} - \tan \theta_W N_{j^{\prime}1})$ \\
$$ & $$ \\
\hline
$$ & $$ \\
\ref{fig16b} & $-|U_{j^{\prime}1}|^2 Z_{L}^{1i^{\prime}}|Z_{D}^{1i}|^2 (N_{j2} + \tan \theta_W N_{j1})(-N_{j2} + \tan \theta_W N_{j1}/3)Z_{L}^{1i^{\prime}}$ \\
$$ & $$ \\
\hline
\end{tabular}
\label{table12}
\end{table}

\begin{table}[htbp]
\caption{The values of $A$ from Eq.~(\ref{box1}), for the muon decay box graphs in Fig.~\ref{fig17}.}
\begin{tabular}{|c|c|}
\hline
$$ & $$ \\
Fig. & $A$ \\
$$ & $$ \\
\hline
$$ & $$ \\
\ref{fig17c} & $-U^{*}_{j1}V^{*}_{j1}U_{j^{\prime}1}V_{j^{\prime}1}Z_{\nu}^{2J*}Z_{\nu}^{1J}Z_{L}^{1i}Z_{L}^{2i*}$ \\
$$ & $$ \\
\hline
$$ & $$ \\
\ref{fig17d} & $-Z_{\nu}^{1J*}Z_{\nu}^{2J}Z_{L}^{1i*}Z_{L}^{2i}(N_{j2} - \tan \theta_W N_{j1})(N_{j2} + \tan \theta_W N_{j1})$ \\
$$ & $(N_{j^{\prime}2} + \tan \theta_W N_{j^{\prime}1})(N_{j^{\prime}2} - \tan \theta_W N_{j^{\prime}1})$ \\
$$ & $$ \\
\hline
$$ & $$ \\
\ref{fig17e} & $U_{j^{\prime}1}V_{j^{\prime}1}|Z_{\nu}^{1J}|^2 |Z_{L}^{2i}|^2 (N_{j2} - \tan \theta_W N_{j1})(N_{j2} + \tan \theta_W N_{j1})$ \\
$$ & $$ \\
\hline
$$ & $$ \\
\ref{fig17f} & $U^{*}_{j1}V^{*}_{j1}|Z_{L}^{1i}|^2 |Z_{\nu}^{2J}|^2 (N_{j^{\prime}2} + \tan \theta_W N_{j^{\prime}1})(N_{j^{\prime}2} - \tan \theta_W N_{j^{\prime}1})$ \\
$$ & $$ \\
\hline
\end{tabular}
\label{table13}
\end{table}

\begin{table}[htbp]
\caption{The values of $B$ from Eq.~(\ref{box2}), for the muon decay box graphs in Fig.~\ref{fig17}.}
\begin{tabular}{|c|c|}
\hline
$$ & $$ \\
Fig. & $B$ \\
$$ & $$ \\
\hline
$$ & $$ \\
\ref{fig17a} & $|V_{j 1}|^2 |Z_{\nu}^{1 J^{\prime}}|^2 |Z_{\nu}^{2 J}|^2 (N_{j^{\prime} 2} - \tan \theta_W N_{j^{\prime} 1})(N_{j^{\prime} 2} - \tan \theta_W N_{j^{\prime} 1})$ \\
$$ & $$ \\
\hline
$$ & $$ \\
\ref{fig17b} & $|U_{j^{\prime}1}|^2 |Z_{L}^{2i}|^2 |Z_{L}^{1i^{\prime}}|^2 (N_{j2} + \tan \theta_W N_{j1})(N_{j2} + \tan \theta_W N_{j1})$ \\
$$ & $$ \\
\hline
$$ & $$ \\
\ref{fig17g} & $Z_{\nu}^{2J}Z_{\nu}^{1J*}Z_{L}^{2i}Z_{L}^{1i*}(N_{j2} - \tan \theta_W N_{j1})(N_{j2} + \tan \theta_W N_{j1})$ \\
$$ & $(N_{j^{\prime}2} - \tan \theta_W N_{j^{\prime}1})(N_{j^{\prime}2} + \tan \theta_W N_{j^{\prime}1})$ \\
$$ & $$ \\
\hline
\end{tabular}
\label{table14}
\end{table}

\newpage

\section{Cancellations of Divergences
\label{cancel}}

The Fermi constants for $\beta$-decay and muon decay both receive divergent corrections arising from vertex and external leg graphs (Box graphs are finite). In the limit of no electroweak symmetry breaking, the weak couplings mediating $\beta$-decay and muon decay are the same. When electroweak symmetry is broken, loop corrections generate a difference between the two. However, the difference is finite (\ie, divergences cancel between $\beta$-decay and muon decay). This is because divergences are due to effects in the UV, and the scale of electroweak symmetry breaking is finite.

In this appendix, we illustrate the cancellation of divergences. Specifically, we show that the divergent part of the difference $\Delta r^{(V)}_{\beta} - \Delta r_{\mu}$ is zero.

We define $\Delta^{\rm div}$ to be the divergent part of $\Delta r^{(V)}_{\beta} - \Delta r_{\mu}$. We may express $\Delta^{\rm div}$ as
\beq
\Delta^{\rm div} = \Delta^{\rm div}_{\rm leg} + \Delta^{\rm div}_{\rm vert(a)} + \Delta^{\rm div}_{\rm vert(b)}~,
\label{delt_expand}
\eeq
where $\Delta^{\rm div}_{\rm leg}$ is the contribution to $\Delta^{\rm div}$ from external legs, $\Delta^{\rm div}_{\rm vert(a)}$ is the contribution from vertex corrections of the type shown in Fig.~\ref{fig15a}, and $\Delta^{\rm div}_{\rm vert(b)}$ is the contribution from vertex corrections of the type shown in Fig.~\ref{fig15b}. According to Eq.~(\ref{leg_diagram}) and Tables~\ref{table3} and~\ref{table4},
\beqn
\Delta^{\rm div}_{\rm leg} & = & \frac{g^2}{(4\pi)^2}\left( \frac{\alpha_{\epsilon}}{4} \right) \Bigg[ -\frac{1}{2}|Z_{U}^{1i}|^2 \left( N_{j2} + \frac{1}{3}\tan \theta_W N_{j1} \right)^2 \nonumber \\
&& - \frac{1}{2}|Z_{D}^{1i}|^2 \left( N_{j2} - \frac{1}{3}\tan \theta_W N_{j1} \right)^2 + \frac{1}{2}|Z_{L}^{2i}|^2 (N_{j2} + \tan \theta_W N_{j1})^2  \nonumber \\
&& + \frac{1}{2}|Z_{\nu}^{2J}|^2 (N_{j2} - \tan \theta_W N_{j1})^2  \Bigg] \nonumber \\
&& - \frac{g_{s}^2}{(4\pi)^2}\left( \frac{\alpha_{\epsilon}}{4} \right) \left( \frac{8}{3}|Z_{U}^{1i}|^2 + \frac{8}{3}|Z_{D}^{1i}|^2 \right)~.
\label{leg_div}
\eeqn
By the orthogonality of $N$ and the unitarity of the other mixing matrices,
\beq
\Delta^{\rm div}_{\rm leg} = \frac{g^2 \alpha_{\epsilon}}{(4\pi)^2} \left( \frac{2}{9} \right) \tan^2 \theta_W - \frac{g_{s}^{2} \alpha_{\epsilon}}{(4\pi)^2} \left( \frac{4}{3} \right)~.
\label{leg_div_fin}
\eeq

According to Eq.~(\ref{vert_left}) and Tables~\ref{table5} and~\ref{table6},
\beqn
\Delta^{\rm div}_{\rm vert (a)} & = & \frac{g^2}{(4\pi)^2} \left( \frac{\alpha_{\epsilon}}{2} \right) \Bigg( \frac{1}{2}Z_{D}^{ki*}Z_{D}^{1i}Z_{U}^{ki^{\prime}*}Z_{U}^{1i^{\prime}} \nonumber \\
&& (N_{j2} + \tan \theta_W N_{j1}/3)(-N_{j2} + \tan \theta_W N_{j1}/3) \nonumber \\
&& - \frac{1}{2}Z_{\nu}^{kJ*}Z_{\nu}^{2J}Z_{L}^{ki*}Z_{L}^{2i}(N_{j2} - \tan \theta_W N_{j1})(N_{j2} + \tan \theta_W N_{j1}) \Bigg) \nonumber \\
&& + \frac{g_{s}^{2}}{(4\pi)^2} \left( \frac{\alpha_{\epsilon}}{2} \right) \frac{8}{3}Z_{U}^{*kj}Z_{U}^{1j}Z_{D}^{*ki}Z_{D}^{1i}~.
\label{verta_div}
\eeqn
By the orthogonality of $N$ and the unitarity of the other mixing matrices,
\beq
\Delta^{\rm div}_{\rm vert (a)} = -\frac{g^2 \alpha_{\epsilon}}{(4\pi)^2}\left( \frac{2}{9} \right) \tan^2 \theta_W + \frac{g_{s}^{2} \alpha_{\epsilon}}{(4\pi)^2}\left( \frac{4}{3} \right)~.
\label{verta_div_fin}
\eeq

According to Eq.~(\ref{vert_right}) and Tables~\ref{table7} and~\ref{table9},
\beqn
\Delta^{\rm div}_{\rm vert (b)} & = & -\frac{g^2}{(4\pi)^2}\left( \frac{\alpha_{\epsilon}}{2} \right) \Bigg[ \frac{1}{2}V_{j1}^{*}|Z_{U}^{1i}|^2 (N_{j^{\prime}2} + \tan \theta_W N_{j^{\prime}1}/3) \nonumber \\
&& (-N_{j^{\prime}2}V_{j1} + N_{j^{\prime}4}V_{j2}/\sqrt{2} - N_{j^{\prime}2}U_{j1} - N_{j^{\prime}3}U_{j2}/\sqrt{2} \nonumber \\
&& + N_{j^{\prime}2}U_{j1} + N_{j^{\prime}3}U_{j2}/\sqrt{2} - N_{j^{\prime}2}V_{j1} + N_{j^{\prime}4}V_{j2}/\sqrt{2}) \nonumber \\
&& + \frac{1}{2}U_{j^{\prime}1}|Z_{D}^{1i}|^2(-N_{j2} + \tan \theta_W N_{j1}/3) \nonumber \\
&& (N_{j 2}V^{*}_{j^{\prime} 1} - N_{j 4}V^{*}_{j^{\prime} 2}/\sqrt{2} + N_{j 2}U^{*}_{j^{\prime} 1} + N_{j 3}U^{*}_{j^{\prime} 2}/\sqrt{2} \nonumber \\
&& + N_{j 2}U^{*}_{j^{\prime} 1} + N_{j 3}U^{*}_{j^{\prime} 2}/\sqrt{2} - N_{j 2}V^{*}_{j^{\prime} 1} + N_{j 4}V^{*}_{j^{\prime} 2}/\sqrt{2}) \nonumber \\
&& - \frac{1}{2}V_{j1}^{*}|Z_{\nu}^{2J}|^2 (N_{j^{\prime}2} - \tan \theta_W N_{j^{\prime}1}) \nonumber \\
&& (-N_{j^{\prime}2}V_{j1} + N_{j^{\prime}4}V_{j2}/\sqrt{2} - N_{j^{\prime}2}U_{j1} - N_{j^{\prime}3}U_{j2}/\sqrt{2} \nonumber \\
&& + N_{j^{\prime}2}U_{j1} + N_{j^{\prime}3}U_{j2}/\sqrt{2} - N_{j^{\prime}2}V_{j1} + N_{j^{\prime}4}V_{j2}/\sqrt{2}) \nonumber \\
&& + \frac{1}{2}U_{j^{\prime}1}|Z_{L}^{2i}|^2 (N_{j2} + \tan \theta_W N_{j1}) \nonumber \\
&& (N_{j2}V^{*}_{j^{\prime}1} - N_{j4}V^{*}_{j^{\prime}2}/\sqrt{2} + N_{j2}U^{*}_{j^{\prime}1} + N_{j3}U^{*}_{j^{\prime}2}/\sqrt{2} \nonumber \\
&& + N_{j2}U^{*}_{j^{\prime}1} + N_{j3}U^{*}_{j^{\prime}2}/\sqrt{2} - N_{j2}V^{*}_{j^{\prime}1} + N_{j4}V^{*}_{j^{\prime}2}/\sqrt{2}) \Bigg]~.
\label{vertb_div}
\eeqn
By the orthogonality of $N$ and the unitarity of the other mixing matrices,
\beq
\Delta^{\rm div}_{\rm vert (b)} = 0~.
\label{vertb_div_fin}
\eeq

By Eqs.~(\ref{delt_expand}), (\ref{leg_div_fin}), (\ref{verta_div_fin}) and~(\ref{vertb_div_fin}),
\beq
\Delta^{\rm div} = 0~.
\label{delt_div_fin}
\eeq
Thus, we have demonstrated that the divergent part of $\Delta r_{\beta}^{(V)} - \Delta r_{\mu}$ vanishes.

\newpage
\bibliographystyle{unsrt}

\end{document}